\documentclass[twocolumn]{article}
\usepackage{graphicx}
\usepackage{authblk}
\usepackage{longtable}
\usepackage{booktabs}
\usepackage{tabularx}
\usepackage[normalem]{ulem} 
\title{Validation of a Computational Respiratory System Model for Mechanical Ventilation} 
\author[1,2]{Carlotta Hennigs}
\author[2]{Charlott Danielson}
\author[1,2]{Franziska Bilda} 
\author[1]{Dimitrios Karachalios} 
\author[2]{Niklas Hackelberg} 
\author[3]{Helene Selpien} 
\author[2]{Georg Männel} 
\author[3]{Dirk Schädler} 
\author[5,2]{Folker Spitzenberger}
\author[1,2]{Philipp Rostalski}
\small
\affil[1]{University of Luebeck, Institute of Electrical Engineering in Medicine, Lübeck, Germany}
\affil[2]{Fraunhofer Research Institution for Individualized Medical Technology and Engineering IMTE, Lübeck, Germany}
\affil[3]{Department of Anesthesiology and Intensive Care Medicine, University Medical Center Schleswig-Holstein, Kiel, Germany}
\affil[5]{Centre for Regulatory Affairs in Biomedical Sciences, Technische Hochschule Lübeck, Lübeck, Germany}

\date{}

\usepackage[utf8]{inputenc}
\usepackage[T1]{fontenc}
\usepackage{lmodern}
\usepackage{geometry}
\usepackage{graphicx}
\usepackage{amsmath, amssymb}
\usepackage{siunitx}
\usepackage{booktabs}
\usepackage{caption}
\usepackage{subcaption}
\usepackage{hyperref}
\usepackage{abstract}
\usepackage{xcolor}
\usepackage[T1]{fontenc}
\usepackage[utf8]{inputenc}
\usepackage{layouts}
\usepackage{booktabs}
\usepackage{tabularx}
\usepackage{rotating}
\usepackage{lscape}
\usepackage[ngerman, english]{babel}
\usepackage{csquotes}
\usepackage[utf8]{inputenc}
\usepackage{textgreek}
\usepackage[most]{tcolorbox}
\usepackage{mdframed}
\usepackage{graphicx}
\usepackage{tikz}
\usetikzlibrary{shapes,arrows.meta,positioning}
\usepackage{amsmath}   
\usepackage{xspace}    
\usepackage[european]{circuitikz}
\usepackage[backend=biber,style=numeric]{biblatex}
\usepackage{svg}
\usepackage{colortbl}
\usepackage{longtable}
\usepackage{array}
\usepackage{cases}

\usepackage{etoolbox}

\definecolor{uzlred}{HTML}{7f1519}
\definecolor{UzLred}{rgb}{0.7098, 0.0863, 0.1294}
\definecolor{uzlblue}		{HTML}{004B5A}
\definecolor{uni_color1}	{RGB}{ 0, 75, 90}
\definecolor{uni_color2}	{RGB}{ 0, 70, 114}
\definecolor{uni_color3}	{RGB}{ 0, 88, 119}
\definecolor{uni_color4}	{RGB}{ 0, 97, 122}
\definecolor{uni_darkred}   {RGB}{ 127, 21, 24}
\definecolor{uzl_oceangreen}{cmyk}{1,0,0.2,0.78}
\definecolor{uzl_Grey}		{cmyk}{0,0,0,0.11}
\definecolor{uzl_Gray}		{cmyk}{0,0,0.03,0.1}
\definecolor{blue2}			{cmyk}{1,0,0.2,0.395}
\definecolor{UzLcyan}		{rgb}{0.1412, 0.5608, 0.5216}
\definecolor{UzLyel}		{rgb}{0.9804, 0.7333, 0}
\definecolor{uzl_lightblue}{RGB}{198, 220, 226}

\definecolor{darkblue}{RGB}{0,70,140}

\definecolor{gradeWeak}{HTML}{7A8B99}
\definecolor{gradeMid}{HTML}{6FAED1}
\definecolor{gradeGood}{HTML}{004672}
\definecolor{gradeNA}{HTML}{A0A0A0}

\definecolor{gradeC}{HTML}{7A8B99} 
\definecolor{gradeB}{HTML}{6FAED1} 
\definecolor{gradeA}{HTML}{004672} 
\definecolor{gradeNA}{HTML}{C0C0C0} 

\definecolor{gradeD}{HTML}{A3BFCF} 

\newcommand{\gD}{\cellcolor{gradeD!100}\textbf{d}}
\newcommand{\gC}{\cellcolor{gradeC!100}\textbf{c}}
\newcommand{\gB}{\cellcolor{gradeB!100}\textbf{b}}
\newcommand{\gA}{\cellcolor{gradeA!100}\textbf{a}}

\newcommand{\paw}{\ensuremath{P_\mathrm{aw}}\xspace}

\newcommand{\fio}{\ensuremath{F_{\mathrm{i}}\mathrm{O}_{2}}\xspace} 
\newcommand{\fico}{\ensuremath{F_{\mathrm{i}}\mathrm{CO}_{2}}\xspace} 

\newcommand{\pinsp}{\ensuremath{P_{\mathrm{insp}}}\xspace}

\newcommand{\pmus}{\ensuremath{P_{\mathrm{mus}}}\xspace}
\newcommand{\vcrit}{\ensuremath{V_{\mathrm{crit}}}\xspace}

\newcommand{\vt}{\ensuremath{V_{\mathrm{T}}}\xspace}
\newcommand{\vd}{\ensuremath{V_{\mathrm{D}}}\xspace}

\newcommand{\fs}{\ensuremath{f_s}\xspace}
\newcommand{\pao}{\ensuremath{P_{\mathrm{a}}\mathrm{O}_{2}}\xspace} 
\newcommand{\paco}{\ensuremath{P_{\mathrm{a}}\mathrm{CO}_{2}}\xspace}  
\newcommand{\palvo}{\ensuremath{P_{A}\mathrm{O}_{2}}\xspace}   
\newcommand{\palvco}{\ensuremath{P_{A}\mathrm{CO}_{2}}\xspace}

\newcommand{\pbco}{\ensuremath{P_{\mathrm{b}}\mathrm{CO}_{2}}\xspace} 
\newcommand{\pcsf}{\ensuremath{P_{\mathrm{CSF}}}\xspace}  
\newcommand{\pO}{\ensuremath{p\mathrm{O}_{2}}\xspace} 
\newcommand{\pCO}{\ensuremath{p\mathrm{CO}_{2}}\xspace} 
\newcommand{\fo}{\ensuremath{f_{\mathrm{O}_{2}}}\xspace}
\newcommand{\fco}{\ensuremath{f_{\mathrm{CO}_{2}}}\xspace} 
\newcommand{\pco}{\ensuremath{P_{\mathrm{c}}\mathrm{O}_{2}}\xspace} 
\newcommand{\pcco}{\ensuremath{P_{\mathrm{c}}\mathrm{CO}_{2}}\xspace}
\newcommand{\spo}{\ensuremath{S_{\mathrm{p}}\mathrm{O}_{2}}\xspace}

\newcommand{\etco}{\ensuremath{{\mathrm{e_t}}\mathrm{CO}_{2}}\xspace}

\usepackage{graphicx}
\usepackage{tikz}
\usetikzlibrary{arrows.meta}

\definecolor{vvNavy}{RGB}{0,72,101}
\definecolor{vvBlue}{RGB}{0,125,162}
\definecolor{vvTeal}{RGB}{0,106,132}
\definecolor{vvLightBlue}{RGB}{220,237,242}
\definecolor{vvPale}{RGB}{237,246,248}
\definecolor{vvGreen}{RGB}{0,125,88}
\definecolor{vvOrange}{RGB}{214,133,65}
\definecolor{vvRose}{RGB}{244,232,235}

\definecolor{modelGreen}{HTML}{2F7D68}      
\definecolor{modelGreenLight}{HTML}{DCEDE7} 
\definecolor{modelGreenDark}{HTML}{1E5A4B}  

\newcommand{\workflowlabel}[1]{%
  \parbox[c][0.68cm][c]{1.24cm}{%
    \centering\sffamily\fontsize{5.7}{6.2}\selectfont #1\par%
  }%
}

\tikzset{
  phase frame/.style={
    draw=vvNavy!95!black,
    line width=0.65pt
  },
  phase title/.style={
    text=white,
    font=\sffamily\bfseries\fontsize{6.3}{6.8}\selectfont
  },
  workflow box/.style={
    rectangle,
    rounded corners=1.0pt,
    draw=vvNavy!95!black,
    line width=0.55pt,
    text=white,
    minimum width=1.43cm,
    minimum height=0.90cm,
    inner sep=0pt,
    outer sep=0pt,
    align=center
  },
  isct box/.style={
    workflow box,
    fill=vvBlue
  },
  vv box/.style={
    workflow box,
    fill=vvTeal
  },
  wide isct box/.style={
    isct box,
    minimum width=4.71cm,
    minimum height=0.84cm
  },
  initial model/.style={
    rectangle,
    rounded corners=1.0pt,
    draw=vvNavy!95!black,
    fill=vvGreen,
    line width=0.60pt,
    text=white,
    minimum width=4.71cm,
    minimum height=0.60cm,
    font=\sffamily\fontsize{6.1}{6.6}\selectfont,
    align=center,
    inner sep=1pt
  },
  precheck/.style={
    rectangle,
    rounded corners=1.0pt,
    draw=vvNavy!95!black,
    fill=vvNavy,
    line width=0.60pt,
    text=white,
    minimum width=1.36cm,
    minimum height=1.45cm,
    text width=1.15cm,
    font=\sffamily\fontsize{5.7}{6.3}\selectfont,
    align=center,
    inner sep=1pt
  },
  credibility frame/.style={
    draw=vvBlue!75!vvNavy,
    line width=0.65pt,
    dash pattern=on 1.4pt off 1.4pt
  },
  credibility title/.style={
    text=vvNavy,
    font=\sffamily\bfseries\fontsize{5.1}{5.6}\selectfont,
    align=center
  },
  credibility cell/.style={
    rectangle,
    rounded corners=0.6pt,
    draw=vvBlue!75!vvNavy,
    fill=white,
    text=vvNavy,
    minimum width=1.05cm,
    minimum height=0.36cm,
    text width=0.92cm,
    font=\sffamily\fontsize{4.5}{4.9}\selectfont,
    align=center,
    inner sep=0.5pt
  },
  validated model/.style={
    rectangle,
    rounded corners=1.0pt,
    draw=vvNavy!95!black,
    fill=vvGreen,
    line width=0.60pt,
    text=white,
    minimum width=1.25cm,
    minimum height=0.88cm,
    text width=1.10cm,
    font=\sffamily\fontsize{5.6}{6.1}\selectfont,
    align=center,
    inner sep=1pt
  },
  outcome/.style={
    ellipse,
    draw=vvBlue!75!vvNavy,
    fill=vvLightBlue,
    line width=0.60pt,
    text=vvNavy,
    minimum width=1.52cm,
    minimum height=0.58cm,
    text width=1.25cm,
    font=\sffamily\fontsize{5.6}{6.1}\selectfont,
    align=center,
    inner sep=1pt
  },
  risk outcome/.style={
    ellipse,
    draw=vvGreen!85!black,
    fill=vvRose,
    line width=0.60pt,
    text=vvGreen!70!black,
    minimum width=1.55cm,
    minimum height=0.58cm,
    text width=1.27cm,
    font=\sffamily\fontsize{5.3}{5.9}\selectfont,
    align=center,
    inner sep=1pt
  },
  flow/.style={
    -{Stealth[length=1.5mm,width=1.1mm]},
    draw=vvNavy,
    line width=0.60pt
  },
  subflow/.style={
    -{Stealth[length=1.2mm,width=0.9mm]},
    draw=vvNavy,
    line width=0.45pt
  },
  feedback/.style={
    draw=vvOrange,
    line width=0.75pt,
    densely dashed
  }
}

\begin{document}
\maketitle
\section*{Abstract} 
Computational modeling and simulation have emerged as powerful tools for the assessment of medical device performance and safety, particularly in the context of in silico clinical trial (ISCT)for automated medical systems. In ventilation, where coordinated management of gas exchange, respiratory mechanics, and patient–ventilator interaction is required under evolving pathophysiology, the clinical translation of automated control strategies remains slow and resource-intensive. These challenges are particularly relevant for AI-based therapy-control systems, whose data-driven decision-making must be evaluated across heterogeneous, evolving, and safety-critical patient states that may be sparsely represented in clinical datasets. Mechanistic, physiology-based models provide a complementary and interpretable environment for systematically generating and testing such scenarios. This paper applies a standards-aligned framework for the credibility assessment of an exemplary computational respiratory model, demonstrated using an automated weaning case study. The framework operationalizes ASME V\&V 40 and FDA (U.S. Food and Drug Administration) principles within a structured, guidance-based validation workflow. The illustrative computational physiological model integrates respiratory mechanics, gas exchange, respiratory control, and a ventilator representation, with validation conducted under a clearly defined context of use and explicit questions of interest. Model credibility is assessed across ventilation-relevant factors, including calibration, physiological plausibility, population-based evaluation, and the reproduction of emergent behavior. All model requirements derived from the intended context of use are addressed within the proposed credibility assessment plan, and documented gaps are transparently reported. The resulting credibility argument supports the applicability of the model for its context of use. Strengths are demonstrated in population-based comparison and mechanistic plausibility, while residual limitations relate to the extent of in vivo evidence, population representativeness, and external validation. Overall, the model is considered fit for purpose for medium-low risk preclinical ISCTs of automated weaning strategies. Furthermore, the validation procedure outlined in this article provides a blueprint for the validation of this and similar models in other mechanical ventilation algorithms and related use cases.

\section*{Authors Summary}
Mechanical ventilators support patients who cannot breathe adequately on their own, but choosing and adjusting ventilator settings is complex. Automated ventilator functions could make care more consistent and reduce the workload of intensive-care teams. Before these systems are tested in a clinical setting, they must be assessed safely across a broad range of patient conditions. Computational models that represent breathing, gas exchange, and the interaction between a patient and a ventilator can provide such a testing environment and can be used to conduct ISCTs.\\
We developed and evaluated a computational model of the respiratory system and mechanical ventilation. We also present a structured framework for determining whether such models are sufficiently trustworthy for a particular purpose. The framework adapts the ASME V\&V 40 and a FDA guidance to computational physiological models. We compared our model with published evidence and clinical measurements and tested whether it reproduced expected physiological responses. Our findings support its use in medium-low-risk, preclinical evaluations of automated ventilator functions. Automated weaning was used as an illustrative example. We identify remaining needs, including larger and more diverse patient datasets for comparison. Our approach provides a transparent foundation for developing and assessing computational physiological models.

\section{Introduction}
\label{sec:intro}
\begin{figure*}[t]
\centering
\resizebox{\textwidth}{!}{%
\begin{tikzpicture}[
    x=1cm,
    y=1cm,
    line cap=round,
    line join=round
]

\node[precheck] (prelim) at (0.65,3.3) {
    Automated function under test in ISCT
};

\path[fill=vvPale!55] (1.60,1.10) rectangle (6.88,4.00);
\path[fill=vvNavy] (1.60,3.62) rectangle (6.88,4.00);
\draw[phase frame] (1.60,1.10) rectangle (6.88,4.00);
\draw[phase frame] (1.60,3.62) -- (6.88,4);

\node[phase title] at (4.24,3.81) {Define prospective ISCT use case};

\node[isct box] (qoi) at (2.60,2.98) {
    \workflowlabel{State ?oI}
};

\node[isct box] (cou) at (4.24,2.98) {
    \workflowlabel{Describe COU}
};

\node[isct box] (qois) at (5.88,2.98) {
    \workflowlabel{Define QoI}
};

\node[wide isct box] [fill=modelGreenLight,
    draw=modelGreenDark,
    text=modelGreenDark] (model) at (4.24,1.82) {
    \parbox[c][0.58cm][c]{4.35cm}{%
        \centering\sffamily\fontsize{6.0}{6.6}\selectfont
        Patient--device model
    }
};

\draw[flow] (1.22,3.83) -- (1.62,3.83);

\draw[flow] (qoi.south) -- (qoi.south |- model.north);
\draw[flow] (cou.south) -- (cou.south |- model.north);
\draw[flow] (qois.south) -- (qois.south |- model.north);

\path[fill=vvPale!55] (7.12,1.10) rectangle (15.35,4);
\path[fill=vvNavy] (7.12,3.62) rectangle (15.35,4);
\draw[phase frame] (7.12,1.10) rectangle (15.35,4);
\draw[phase frame] (7.12,3.62) -- (15.35,4);

\node[phase title] at (11.24,3.81) {Patient-device model credibility assessment};

\node[vv box] (?oI) at (8.00,2.98) {
    \workflowlabel{State ?oI}
};

\node[vv box] (COU) at (9.62,2.98) {
    \workflowlabel{Describe COU}
};

\node[vv box] (model risk) at (11.24,2.98) {
    \workflowlabel{Assess model risk}
};

\node[vv box] (cred ev) at (12.86,2.98) {
    \workflowlabel{Identify credibility evidence}
};

\node[vv box] (vvplan) at (14.48,2.98) {
    \workflowlabel{State credibility factors}
};

\node[vv box] (uq) at (8.00,1.82) {
    \workflowlabel{State gradations and goals}
};

\node[vv box] (prospective) at (9.62,1.82) {
    \workflowlabel{Prospective adequacy assessment}
};

\node[vv box] (execute) at (11.24,1.82) {
    \workflowlabel{Execute studies and collect evidence}
};

\node[vv box] (poststudy) at (12.86,1.82) {
    \workflowlabel{Post-study adequacy assessment}
};

\node[vv box] (report) at (14.48,1.82) {
    \workflowlabel{Finalize the credibility report}
};


\draw[flow] (model.east) -- (6.98,1.82) |- (7.15, 3.83);

\draw[subflow] (?oI.east) -- (COU.west);
\draw[subflow] (COU.east) -- (model risk.west);
\draw[subflow] (model risk.east) -- (cred ev.west);
\draw[subflow] (cred ev.east) -- (vvplan.west);

\draw[flow] (vvplan.south) -- (14.48,2.45)
    -- (8,2.45) -- (uq.north);

\draw[flow] (uq.east) -- (prospective.west);
\draw[flow] (prospective.east) -- (execute.west);
\draw[flow] (execute.east) -- (poststudy.west);
\draw[flow] (poststudy.east) -- (report.west);




\node[precheck] [fill=modelGreenLight,
    draw=modelGreenDark,
    text=modelGreenDark] (q model) at (16.4,3.3) {
    \workflowlabel{Assessed Patient--Device \\Model}
};
\draw[flow] (15.3,3.83) -- (15.73,3.83);

\node[vv box] [dashed, color=vvLightBlue, text=vvNavy] (isct) at (16.4,1.8) {
    \workflowlabel{ISCT execution}
};
\draw[flow] [dashed] (q model.south) -- (isct.north);

\end{tikzpicture}%
}
\caption{
Risk-informed framework for assessing the credibility of the patient--device model for its intended context of use. The framework links the setup of an ISCT with the model credibility assessment and the
resulting assessed patient--device model. The dashed box denotes subsequent ISCT execution, which is outside the scope of this study. ?oI: question of interest, COU: context of use, QoI: quantities of interest.
}

\label{fig:credibility_workflow}
\end{figure*}
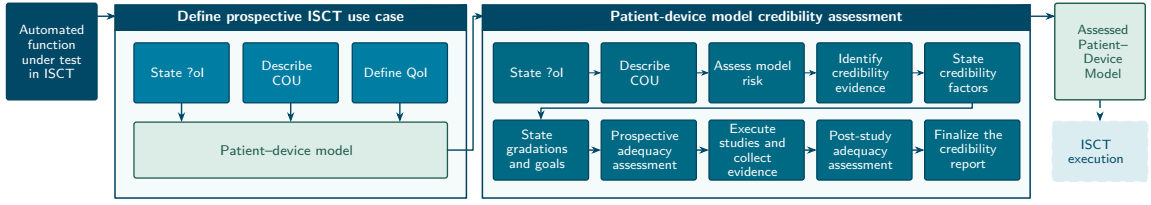

Mechanical ventilation remains indispensable in intensive care, yet tailoring assistance to patient-specific physiology while reducing ventilator-induced harm is challenging. Clinicians must balance gas exchange, respiratory mechanics, and patient–ventilator interaction across dynamic pathophysiology (e.g., acute respiratory distress syndrome (ARDS), chronic obstructive pulmonary disease (COPD), sedation, and weaning)~\cite{lumb_nunns_2017, west_wests_2017}. This challenge is accentuated during weaning, where support is gradually reduced and mismatches between patient effort, respiratory control, and device response can lead to delayed liberation or failure. Weaning therefore represents a high-variability, high-risk phase for decision support and automation. At the same time, there is a growing need to automate key mechanical ventilation functions to improve consistency, reduce clinician workload, and ensure patient safety. Bringing automation to market is expensive and time-consuming, often requiring extensive preclinical testing and, in some cases, animal studies before clinical evaluation is feasible. In silico clinical trials (ISCTs) using computational physiological models (CPMs) offer a complementary pathway to explore decision boundaries, stress-test protocol logic, and quantify safety and performance under controlled variability, thereby de-risking translation when the underlying models are credible and tailored to a well-defined context of use (COU)~\cite{pappalardo_silico_2019, karanasiou_advancing_2025}.\\
Over the past decade, a risk-informed credibility paradigm has emerged: ASME V\&V~40~\cite{asme_2018} prescribes that verification, validation, and uncertainty quantification (V\&V/UQ) be proportionate to model risk and anchored in the intended COU and question of interest (?oI), while a recent FDA (U.S. Food and Drug Administration) guidance operationalizes credibility across evidence categories~\cite{fda_guidance_2023}. ISCT-specific workflows further clarify planning and evidence selection consistent with decision needs~\cite{pathmanathan_credibility_2024, aycock_toward_2024}. A notable precedent outside ventilation is the regulator-accepted UVA/Padova Type-1 Diabetes simulator, which enabled preclinical assessment of closed-loop insulin delivery and set a template for credibility-aware computational evidence in device development~\cite{cobelli_developing_2023, dalla_man_meal_2007}.\\
Within mechanical ventilation, however, the state of the art reveals an incomplete ventilation-specific operationalization of these general frameworks. A systematic review by Warnaar et~al.~\cite{warnaar_computational_2023} found that most computational physiological models address lung mechanics, whereas gas exchange, diaphragm function, and integrative cardiopulmonary control are comparatively underrepresented. Only a minority of studies met comprehensive validation quality criteria, and reporting practices were inconsistent~\cite{warnaar_computational_2023}. As a result, many published efforts validate only subsystems or narrow clinical scenarios, and there is no broadly accepted, regulator-aligned reference simulator for ventilation analogous to other domains. This gap limits the use of ISCT evidence to support preclinical assessment of automation in mechanical ventilation, especially for use cases — such as weaning — where patient–device interaction, gas exchange kinetics, and measurement-chain dynamics are central. Notably, Herrmann et al.~\cite{herrmann_virtual_2025} demonstrated that an integrated lung mechanics and gas exchange model can be validated following ASME V\&V principles for an \spo-based closed-loop controller, illustrating a specific pathway toward comprehensive, regulator-aligned in silico evaluation in ventilation.\\
Against this background, the present work pursues a focused aim: to evaluate the credibility of a coupled patient-device model (PDM) that integrates respiratory mechanics, gas exchange, and respiratory control for the specific COU of assessing automated weaning protocols, using a risk-proportionate plan aligned with ASME V\&V~40 and FDA guidance~\cite{asme_2018, fda_guidance_2023, pathmanathan_credibility_2024}. Consistent with recent extensions to credibility practice, we organize evidence across the FDA’s categories and adapt their implementation to a physiologic PDM in ventilation, with explicit quantities of interest (QoIs) and decision-oriented acceptance criteria for weaning. In this work,  we emphasize model plausibility, model calibration, population-level/literature comparators, uncertainty and sensitivity analyses, and documentation of emergent behaviors and applicability~\cite{fda_guidance_2023}.\\
To operationalize this framework, the present work proceeds in two complementary steps. First, we define a precise context of an prospective ISCT for the automated weaning protocol (AWP), establishing the decision-relevant questions and clinical constraints that will guide subsequent model validation. Second, we apply a structured, risk-informed credibility assessment workflow to the underlying patient–device model, demonstrating how ASME V\&V 40 and FDA guidance can be systematically adapted to mechanical ventilation and ISCTs with focus on model validation (see Figure~\ref{fig:credibility_workflow}). It should be noted that conducting the ISCT is not part of this work. However, it is necessary to place the prospective ISCT set up of this trial within the context of use as part of the model credibility assessment. Together, these sections bridge the gap identified above: from general credibility principles to practical, ventilation-focused implementation.

\section{Definition of a Prospective ISCT Use Case}
\label{sec:case_study}
\begin{figure}[t]
\centering
\resizebox{0.45\textwidth}{!}{%
    \begin{tikzpicture}[
        scale=0.85,
        transform shape,
        line cap=round,
        line join=round,
        ventilatorpanel/.style={
            draw=vvNavy,
            fill=vvPale!55,
            minimum width=5cm,
            minimum height=4.3cm,
            rounded corners=2pt,
            line width=0.45pt,
            align=center
        },
        devicebox/.style={
            draw=vvNavy,
            fill=white,
            text=vvNavy,
            minimum width=4.5cm,
            minimum height=1cm,
            rounded corners=2pt,
            line width=0.45pt,
            align=center,
            font=\sffamily\small
        },
        patientbox/.style={
            draw=modelGreenDark,
            fill=modelGreenLight,
            text=modelGreenDark,
            minimum width=1.6cm,
            minimum height=4.3cm,
            rounded corners=2pt,
            line width=0.45pt,
            align=center,
            font=\sffamily\small
        },
        signallabel/.style={
            font=\sffamily\small,
            text=vvNavy
        },
        paneltitle/.style={
            font=\sffamily\small\bfseries,
            text=vvNavy
        }
    ]

    \node[patientbox] (pat) at (4.8,0.3) {\textbf{Patient}};

    \node[ventilatorpanel] (Vent) at (0,0.3) {};

    \node[devicebox] (DP) at (0,1) {Data Processing};
    \node[devicebox] (AWP) at (0,-1) {AWP};

    \draw[flow, draw=vvNavy] (-1.3,0.5) to (-1.3,-0.5);
    \draw[flow, draw=vvNavy] (0,0.5) to (0,-0.5);
    \draw[flow, draw=vvNavy] (1.3,0.5) to (1.3,-0.5);
    \draw[flow, draw=vvNavy] (2.25,-1) to (4,-1);
    \draw[flow, draw=vvNavy] (4,1) to (2.25,1);

    \node[paneltitle, above left=0.25cm and -3.25cm of DP] {Ventilator};

    \node[signallabel, above right=-0.6cm and 0.4cm of DP]
        {$F_\mathrm{CO_2}$};

    \node[signallabel, above right=-1.2cm and 0.4cm of DP]
        {$\dot{V}$};

    \node[signallabel, above right=-0.8cm and -1cm of AWP]
        {\pinsp};

    \node[signallabel, above right=-0.6cm and 0.4cm of AWP]
        {\paw};

    \node[signallabel, above right=0.2cm and -0.9cm of AWP]
        {\etco};

    \node[signallabel, above right=0.2cm and -2.2cm of AWP]
        {RR};

    \node[signallabel, above right=0.2cm and -3.5cm of AWP]
        {\vt};

    \end{tikzpicture}%
    }

    \caption{High-level functional representation of patient-device interaction with automated weaning protocol (AWP) and its data pathways. \vt: Tidal volume, RR: Respiratory rate, \etco: End-tidal CO$_2$, \pinsp: Inspiratory pressure, \paw: Airway pressure, $F_\mathrm{CO_2}$: Mainstream capnogram, $\dot{V}$: Airway flow.}
    \label{fig:ISCT_overview}
\end{figure}
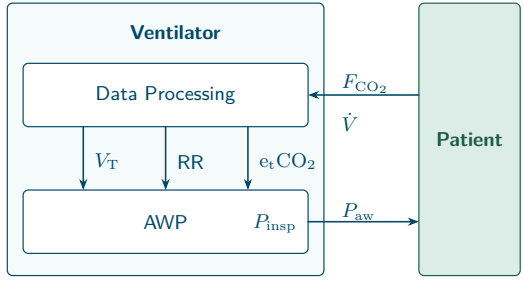

ISCTs for automated weaning represent an application where credibility assessment is essential. In this section, the clinical scenario, decision-relevant variables, and questions of interest are specified that will structure the subsequent validation plan. This explicit framing of context of use and questions of interest — cornerstones of ASME~V\&V~40 — ensures that all evidence collected downstream targets the right outcomes under the right conditions. This approach avoids the pursuit of generic validation that is separated from clinical reality. As a hypothetical case study, a representative automated weaning protocol based on the publicly available SmartCare/PS algorithm (Drägerwerk AG \& Co. KGaA, Lübeck, Germany) is selected~\cite{neumann_smartcareps_2015}. The implementation used in this work is derived exclusively from publicly available descriptions of SmartCare/PS and may therefore differ from the proprietary algorithm implemented in commercial systems by Dräger. SmartCare/PS has been in widespread clinical use for more than two decades and is among the most established commercially available closed-loop weaning systems in mechanical ventilation. Its protocol logic explicitly integrates indices of gas exchange, respiratory mechanics, and spontaneous breathing activity to guide stepwise reduction of pressure support and assess readiness for extubation. These characteristics make SmartCare/PS a well-suited and clinically relevant reference application for demonstrating model credibility in the context of automated weaning. Although the algorithm itself is relatively mature and thus less “innovative” from an ISCT perspective, it is supported by extensive clinical experience and published trial data. This provides a well-characterized, reproducible benchmark against which model behavior can be compared, making SmartCare/PS an attractive choice for a first credibility-focused case study. To support preclinical in silico evaluation of this protocol, a PDM is defined, establishing the functional, physiological, and clinical requirements that must be addressed and validated for the context of use. The PDM captures key aspects of respiratory mechanics, gas exchange, and respiratory center. The case study focuses on an AWP designed to adjust inspiratory pressure \pinsp in CPAP/CSV-PS mode based on real-time patient responses, similar to systems such as SmartCare/PS. By continuously monitoring variables such as respiratory rate~RR, tidal volume~\vt, and end-tidal CO$_2$~\etco, the trial explores how the AWP interacts with patient physiology to tailor ventilatory support. \\
The overall data flow and functional interaction between the patient model and the AWP and device model are illustrated in Figure~\ref{fig:ISCT_overview}. Based on this setup, an ISCT is defined to systematically evaluate the behavior and credibility of the automated weaning protocol. In the following, the trial design is structured around clearly specified questions of interest, context of use, and quantities of interest, which then inform the derivation of model requirements and a risk-informed validation strategy. This provides a coherent link from the clinical use case to concrete modeling assumptions, required model performance, and the subsequent credibility assessment.

\subsection{Device-Level Question of Interest, Context of Use and Quantities of Interest}
\label{sec:Qoi_COI}
\subsubsection{Question of Interest} For the use case of evaluating the AWP in an ISCT, the following questions of interest can be formulated:
\begin{enumerate}
	\item Does the automated weaning protocol maintain tidal volume \vt, respiratory rate RR, and end-tidal carbon dioxide \etco within the predefined acceptable working ranges over time?
	\item Do the resulting of tidal volume \vt and plateau pressure $P_\mathrm{plat}$ satisfy guideline-based criteria for lung-protective ventilation?
	\item  Does patient-ventilator asynchrony occur during operation with the automated weaning protocol? If so, what types and how often?
	\item Does inspiratory pressure oscillations occur during operation with the automated weaning protocol? If so, how often?
\end{enumerate}

\subsubsection{Context of Use}
The aim of the prospective ISCT is to evaluate in silico whether the automated weaning protocol can autonomously implement a clinical therapeutic strategy under predefined conditions and to assess its impact on ventilatory performance and patient-ventilator interaction. In particular, the prospective ISCT is used to quantify how well AWP maintains tidal volume \vt, respiratory rate RR, and end-tidal CO$_2$ \etco within the predefined acceptable working point ranges, whether the resulting combinations of \vt and plateau pressure $P_\mathrm{plat}$ are consistent with guideline-based lung-protective ventilation, and how often patient-ventilator asynchrony and \pinsp oscillations occur across different scenarios. It enables investigation of device behavior in a physiologically and pathophysiologically plausible environment prior to clinical testing. \\
This COU describes a preclinical modeling and simulation workflow supporting preclinical evaluation and iterative design improvement of AWP. It does not demonstrate clinical effectiveness or safety, nor should results be used to guide patient care. Model assumptions, uncertainties, and boundary conditions must be explicitly documented and justified within the associated credibility plan.

\subsubsection{Quantities of Interest}
Tidal volume \vt, respiratory rate RR, and end-tidal CO$_2$ \etco, plateau pressure $P_\mathrm{plat}$, and quantitative descriptors of patient-ventilator asynchrony, including the incidence (number of events) and relative frequency of specific asynchrony types and quantitative descriptors of \pinsp oscillations.

\subsection{Model under assessment}
For the ISCT, a coupled patient–device model is used that simulates respiratory mechanics, gas exchange, respiratory control, and interaction with the ventilator. The model generates the signals relevant for automated weaning, including respiratory rate, tidal volume, end-tidal CO$_2$, and pressure-related parameters. The AWP is clearly distinguished from the patient–device model as the external control logic under test. The complete model architecture, assumptions, parameterization, and numerical implementation are presented in Appendix~\ref{app:pdm}. In the main paper, the model is evaluated exclusively with regard to its suitability for the defined COU.
\\ \\
The prospective ISCT setup and the patient--device model credibility assessment are not independent activities. The prospective ISCT setup defines the device-level decision question that a future ISCT would address, including the clinical scenarios, relevant outputs, and safety-relevant behaviours of the
automated weaning protocol. These elements determine the capabilities required from the patient--device model and are therefore translated into the derived model question of interest, model context of use, model risk, and credibility requirements assessed in the following section. This study assesses model
credibility for this derived context. It does not execute the prospective ISCT or draw conclusions regarding the performance of the automated weaning protocol.


\section{Risk-Informed Patient-Device Model Credibility Assessment}
\label{sec:cred_asses_framework}
Building on the ISCT setup described in Section~\ref{sec:case_study} — including the SmartCare/PS-based automated weaning protocol, the defined context of use, questions and quantities of interest, and the coupled PDM — this section focuses on a risk-informed credibility assessment of the PDM. The objective is to determine whether the model is sufficiently credible for its specific context of use in the preclinical in silico evaluation of automated weaning.\\
To this end, a dedicated ?oI/COU-based framework (see Figure~\ref{fig:credibility_workflow} and Section~\ref{sec:cred_assessment}) is applied, consistent with ASME V\&V~40~\cite{asme_2018} and recent FDA guidance~\cite{fda_guidance_2023} on model credibility, as well as ISCT-specific workflows~\cite{pathmanathan_credibility_2024}. The following sections implement the credibility assessment framework, covering model risk assessment, selection and execution of credibility activities, and synthesis of evidence in relation to the defined context of use.


\subsection{Credibility Assessment with Focus on Model Validation}
The nine-step framework outlined in Section~\ref{sec:cred_assessment} is now applied to the PDM that can be used in an ISCT for an AWP. First, the question of interest and context of use is stated. Then, the model risk is assessed, evidence mapped to applicable credibility evidence categories, validation studies designed, validation tests executed, and a post-study adequacy assessment conducted. Throughout, each activity is tied to the ASME V\&V~40 and FDA framework, ensuring traceability and compliance with regulatory and scientific best practices. This section is the substantive core of the credibility assessment results: it documents what is tested, against which comparator, with what acceptance criteria, and what results are obtained.
\subsubsection{Derived Model Question of Interest} 
The device-level question previously defined cannot be addressed unless the patient--device model can represent the physiological processes and observable outputs on which the prospective ISCT depends. Accordingly, the derived model question of interest is:\\
Can the patient--device model provide sufficiently credible representations of the relationships between ventilator settings, patient physiology, and the required outputs---including tidal volume, respiratory
rate, end-tidal carbon dioxide, pressure-related variables, and specified patient--ventilator asynchronies---within the defined physiological and pathophysiological scenarios, such that it can support the prospective ISCT use case?
\subsubsection{Derived Model Context of Use}
The model is designed for the pre-clinical evaluation of automated ventilator functions in adult patients. It simulates respiratory mechanics, gas exchange, and respiratory control based on literature-derived or clinically fitted patient parameters. Realistic clinical scenarios, including ARDS and COPD, are reproduced, and model outputs are comparable to standard physiological monitoring. The model serves as an alternative to pre-clinical animal testing for the assessment of automated ventilator functions, such as automated weaning. Regulatory and clinical decision-making additionally relies on bench testing, in vivo studies, and clinical data. Within the scope of this study, the model is used exclusively for device assessment, including the AWP and regulatory documentation, and is not intended for integration into clinical devices or real-time patient care.

\subsubsection{Model risk} 
Following the ASME V\&V 40 framework, model risk is determined by assessing Model Influence and Decision Consequence, both evaluated using a qualitative gradation scale of \textit{low}, \textit{medium}, or \textit{high}. 
Based on the COU, in this hypothetical scenario the model’s influence on decision-making is considered \textbf{medium} following the gradations proposed in ASME V\&V 40, as the simulation outputs of the computational model (and the overall ISCT) are a moderate factor in the decision whether the AWP is safe for patient use. It is not the sole or decisive source of information for final regulatory and subsequent clinical decision-making. The overall hypothetical evaluation of the AWP additionally relies on orthogonal methods and approaches, including bench testing, in vivo studies, and clinical data. Because these complementary sources of evidence exist and the model is used exclusively for initial pre-clinical evaluation without patient contact, its influence in making a decision is kept at the medium level rather than high.\\
The significance of an adverse outcome resulting from an incorrect decision is considered \textbf{low}, corresponding to the gradations of ASME V\&V 40 for decision consequence where an incorrect decision would not adversely affect patient safety or health or may only result in nuisance to the physician (user) or have other minor impacts. The model informs pre-clinical evaluation without direct clinical impact. Any incorrect decisions would not lead to patient harm, as multiple safety mechanisms are in place, including ventilator alarms, clinician oversight, and downstream validation steps as described above. \\
Following the ASME V\&V 40 risk matrix, model influence is classified as medium and decision consequence as low, resulting in an overall \textbf{medium-low} model risk.

\subsubsection{Credibility Evidence}
The credibility goals define the required level of confidence in the model’s predictions and guide the verification, validation, and uncertainty assessment activities. The primary quantities of interest are tidal volume \vt, respiratory rate RR, end-tidal carbon dioxide \etco and emergence of asynchrony types. 
Following the ASME~V\&V~40 framework, the model is classified as medium-low risk, implying that credibility must be established through rigorous procedures. Accordingly, the steps are:		
\begin{enumerate}
	\item \textbf{Uncertainty Assessment:} Quantify how input parameter variability affects the QoIs and assess the robustness of model predictions under different physiological scenarios. Sensitivity analyses identify the parameters that most strongly influence the outputs.  
	\item  \textbf{Verification:} Ensure that the model is implemented correctly, free of numerical errors, and internally consistent with the underlying physiological assumptions. This includes code verification, convergence tests, and cross-checks of model equations.			
	\item \textbf{Validation:} Demonstrate that the model accurately predicts the selected QoIs under some clinically relevant conditions. This requires quantitative comparison with literature and clinical data and statistical evaluation of agreement, for instance using metrics such as bias, RMSE, and correlation.
    \item \textbf{Supplementary credibility evidence:}
 In the field of physiological modeling, computational model assessment may include additional credibility factors beyond traditional verification and validation. These comprise \textit{computational model calibration} to ensure accurate parameterization against observed (clinical) data, \textit{computational model plausibility} to verify underlying equations and assumptions against physiological principles and expert knowledge, and \textit{emergent model behavior} to confirm the model's ability to reproduce known, non-programmed physiological phenomena that arise naturally from subsystem interactions.
\end{enumerate} 
\noindent	

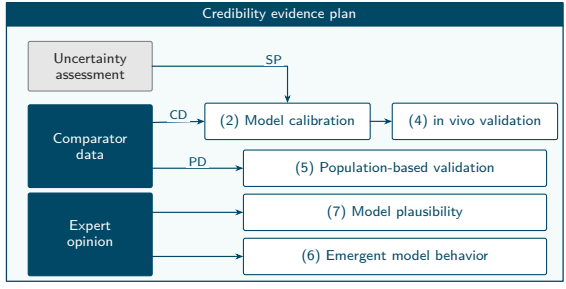
\begin{figure}[t]
\centering
\resizebox{\columnwidth}{!}{%
\begin{tikzpicture}[
    x=1cm,
    y=1cm,
    line cap=round,
    line join=round,
    validation category/.style={
        vv box,
        text width=4.05cm,
        minimum width=5.5cm,
        minimum height=0.66cm,
        align=center,
        fill=white,
        draw=vvNavy,
        text=vvNavy,
        font=\sffamily\fontsize{7.4}{8.0}\selectfont
    },
    source box/.style={
        vv box,
        text width=1.85cm,
        minimum width=2.25cm,
        align=center,
        fill=vvNavy,
        draw=vvNavy,
        text=white,
        font=\sffamily\fontsize{7.4}{8.0}\selectfont
    },
    uncertainty box/.style={
        vv box,
        text width=1.85cm,
        minimum width=2.25cm,
        align=center,
        fill=black!10,
        draw=black!50,
        text=black!85,
        font=\sffamily\fontsize{7.4}{8.0}\selectfont
    },
    evidence label/.style={
        font=\sffamily\fontsize{6.5}{7.1}\selectfont,
        text=vvNavy,
        fill=vvPale!55,
        inner sep=1.0pt,
        align=center
    }
]

\path[fill=vvPale!55] (0.20,0.10) rectangle (10.30,5.15);
\path[fill=vvNavy]     (0.20,4.73) rectangle (10.30,5.15);

\draw[phase frame] (0.20,0.10) rectangle (10.30,5.15);
\draw[phase frame] (0.20,4.73) -- (10.30,4.73);

\node[phase title] [font=\sffamily\fontsize{7.4}{8.0}\selectfont] at (5.15,4.94) {
    Credibility evidence plan
};

\node[uncertainty box] (uq) at (1.72,4.) {
    Uncertainty\\assessment
};

\node[source box, minimum height=1.50cm] (comparator) at (1.72,2.55) {
    Comparator\\data
};

\node[source box, minimum height=1.50cm] (expert) at (1.72,0.95) {
    Expert\\opinion
};

\node[
    validation category,
    text width=2.7cm,
    minimum width=3cm,
    align=center
] (cal) at (5.3,3.0) {
    (2) Model calibration
};

\node[
    validation category,
    text width=2.7cm,
    minimum width=3cm,
    align=center
] (vivo) at (8.7,3.0) {
    (4) in vivo validation
};

\node[validation category] (pop) at (7.25,2.15) {
    (5) Population-based validation
};

\node[validation category] (plaus) at (7.25,1.35) {
    (7) Model plausibility
};

\node[validation category] (emergent) at (7.25,0.55) {
    (6) Emergent model behavior
};


\draw[flow]
    (uq.east) -|
    node[evidence label, above, pos=0.45] {SP}
    (cal.north);

\draw[flow]
    ([yshift=0.43cm]comparator.east)
    -- node[evidence label, above, pos=0.50] {CD}
    (cal.west);

\draw[flow]
    ([yshift=-0.40cm]comparator.east)
    -- node[evidence label, above, pos=0.50] {PD}
    (pop.west);

\draw[flow]
    ([yshift=0.40cm]expert.east)
    -- (plaus.west);

\draw[flow]
    ([yshift=-0.40cm]expert.east)
    -- (emergent.west);

    \draw[flow]
    ([yshift=-0.40cm]cal)
    -- (vivo.west);


\end{tikzpicture}%
}
\caption{
Selected credibility evidence categories and associated evidence inputs for
the validation plan. Category numbers correspond to Table~2 of the FDA
guidance~\cite{fda_guidance_2023}. Navy-framed boxes denote selected
validation evidence categories, the gray box denotes uncertainty assessment,
and navy-filled boxes denote evidence inputs. SP: sensitive model parameters;
CD: clinical data; PD: population-based or literature data.
}
\label{fig:plan_VV}
\end{figure}

\paragraph{Credibility Assessment Plan} 
Credibility‐evidence categories for validation were selected based on Table~2 in the FDA guidance~\cite{fda_guidance_2023}, and their dependencies were determined (see Figure \ref{fig:plan_VV}).  Verification steps such as numerical code verification and calculation verification are critical processes in scientific computing~\cite{oberkampf_verification_2025} and vital for the model's credibility. However, while comprehensive verification is an essential prerequisite to ensure the numerical correctness of any validation results, it is outside the scope of this work, which will focus on demonstrating the validation framework in the following sections. Owing to limited data availability, only parts of the model can be assessed using clinical data. For Test~S--2, patient-specific calibration and validation are temporally separated: parameters are identified only during the calibration interval and are subsequently kept fixed during a later validation interval
that is not used for parameter estimation. The remaining model components are evaluated against literature data using population-based methods. Before starting the validation process, the comparator data (clinical and literature based) and expert opinions have to be collected and processed. Furthermore, a sensitivity analysis and uncertainty quantification of the baseline patient model must be performed before validation can begin. Both are combined depicted in the \textit{Uncertainty assessment} block. These results are particularly important for calibrating the model, as they help to identify the most sensitive parameters.  In addition to the data-based validation steps, the categories of \textit{Model plausibility} and \textit{Emergent phenomena}, which provide credibility evidence, are also included here. These should primarily be described and evaluated on the basis of clinical expert knowledge and partially supported by literature. They serve to supplement the description of model properties and create an additional source of confidence in the model's overall behavior, so their importance should not be underestimated.\\
Based on the available data, the individual validation tests for the patient model, device model, and PDM are summarized in Table~\ref{tab:assesment_plan}. Tests V--1 to V--3 evaluate gas exchange and respiratory control in the patient model to ensure physiological plausibility. Tests S--1 to S--2 test the model under representative clinical conditions, including lung mechanics, pressure support ventilation, and patient-ventilator asynchrony. Tests S--1 and S--2 assess overall model plausibility and emergent system behaviors. The comparators, ranging from literature and clinical datasets to expert opinion, provide a rationale for each validation step and link it to credible reference data. \\
The model predictions are considered acceptable if they meet one of the following general criteria: the results lie within a 90\,\% confidence interval, the mean absolute percentage error is less than or equal to 10\,\% compared to the comparator data, or, if these criteria are not applicable, a well-reasoned justification is provided.  \\ 
In this work, these factor definitions and gradation criteria are adopted to rate the strength of the evidence collected for the PDM and to derive acceptance decisions for the specified context of use. The detailed mapping of credibility factors, gradations, and acceptance rules applied in this case study is provided in the Appendix~\ref{appen:model_val}.

\begin{table*}[t]
	\centering
	\small
	\caption[Assessment plan for model validation and supplementary credibility evidence.]{Assessment plan for model validation (V) and supplementary credibility evidence (S). Category (Cat.) based on~\cite{fda_guidance_2023}, BPM: Baseline patient model, PDM: Patient-device model.}
	\label{tab:assesment_plan}
	\begin{tabularx}{\textwidth}{>{\centering\arraybackslash}p{0.7cm} p{1cm} p{4.7cm} p{3.8cm} >{\centering\arraybackslash}p{0.9cm}>{\centering\arraybackslash}p{1.6cm}}  
		\toprule
		\textbf{Test} & \textbf{Model} & \textbf{Description} & \textbf{Comparator}  & \textbf{Cat.} & \textbf{Req.}\\ \midrule
		V--1  & BPM & Gas exchange, minute ventilation to alveolar partial pressure & Literature data~\cite{lumb_nunns_2017} & 5 & AWP--P4\\
		V--2  & BPM & Gas exchange, iso shunt & Literature data~\cite{lumb_nunns_2017} & 5 & AWP--P4\\
		V--3  & BPM & Resp. center, chemical feedback & Literature data~\cite{reynolds1973transient, reynolds_transient_1972} & 5 & AWP--P5 \\
		
		V--4  & PDM & Resp. center, PSV levels & Literature data~\cite{vitacca_assessment_2004} & 5 & AWP--P5, P6, C1\\
		V--5  & PDM & Resp. center, asynchrony & Literature data~\cite{sauer_automated_2024, thille_patient-ventilator_2006, mulqueeny_automatic_2007} & 5 & AWP--C2\\
		S--1  & PDM & Lung mechanics, COPD & Clinical data set of ventilated COPD patients~\cite{grashoff_surface_2021} & 2 & AWP--P1, P2\\
        S--2  & PDM & Lung mechanics, gas exchange & Clinical data of ventilated patients~\cite{niras_data} & 2, 4 & AWP--P1 -- P4\\

		S--3  & PDM & Model plausibility & Literature data~\cite{lumb_nunns_2017}, expert opinion & 7 \\
		S--4  & PDM & Emergent phenomena & Literature data~\cite{lumb_nunns_2017}, expert opinion & 6 \\
		\bottomrule
	\end{tabularx}
\end{table*}

\subsubsection{Adequacy Assessment}
A prospective adequacy assessment is conducted to align the credibility plan with the defined questions of interest and context of use. Before execution, each QoI (\vt, RR, \etco, asynchrony, $P_\mathrm{plat}$) was mapped to explicit acceptance criteria and to suitable evidence categories (calibration, population-based, in vivo validation, model plausibility, emergent behavior). The suitability of each comparator for the defined COU was assessed using pre-specified criteria: alignment of input conditions, output/QoI definitions, physiological and pathophysiological ranges, and clinical scenario conditions. Data sources, responsibilities for data handling, protocols, software environments, and versioning were documented to support traceability and reproducibility. 
Applicability is supported by population-/clinical comparators, literature envelopes, and targeted measurement-chain checks. Given a medium-low model-risk assessment, the planned scope, depth (statistical calibration, uncertainty quantification and sensitivity analysis, plausibility, emergent behavior), and documentation were deemed proportionate. Residual risks (e.g.\ limited clinical data span, cohort representativeness) are captured as mitigation actions (expanded datasets, external validation, identifiability analysis) and reflected in the post-study assessment and final credibility report.

\subsubsection{Execute Studies and Collecting Evidence} 
\begin{figure*}[t]
	\centering
	\begin{subfigure}[t]{0.48\textwidth}
		\centering
	\includegraphics[width=0.95\textwidth]{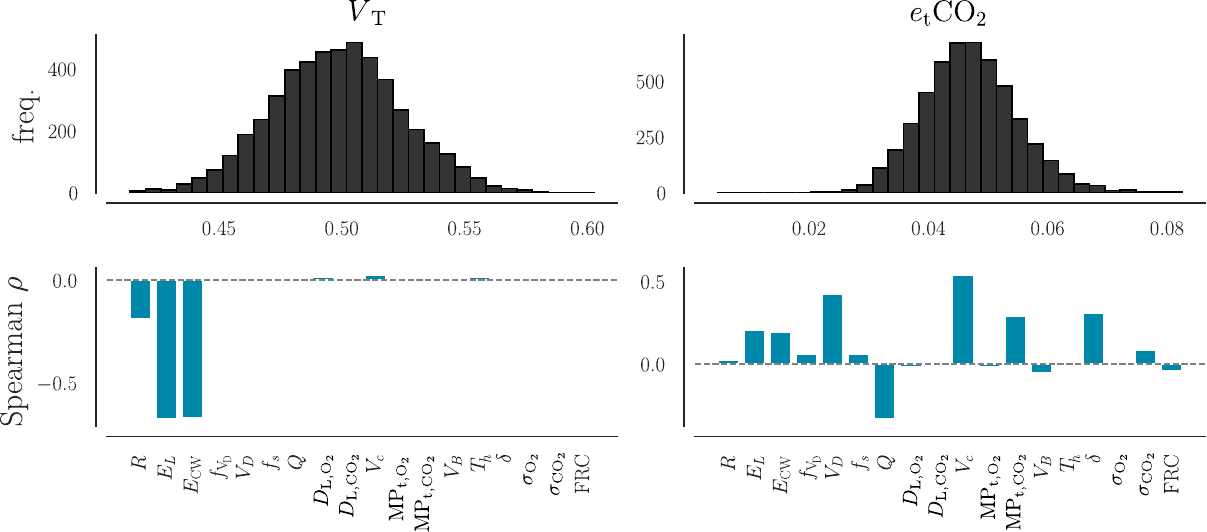}
	\caption{Results of the uncertainty quantification and sensitivity analysis of the lung mechanics and gas exchange submodels. \vt: Tidal volume [L], \etco: End-tidal CO$_2$.}
	\label{fig:res_SA_UQ_LM_GE}
	\end{subfigure}
	\hfill
	\begin{subfigure}[t]{0.48\textwidth}
		\centering
    	\includegraphics[width=0.95\textwidth]{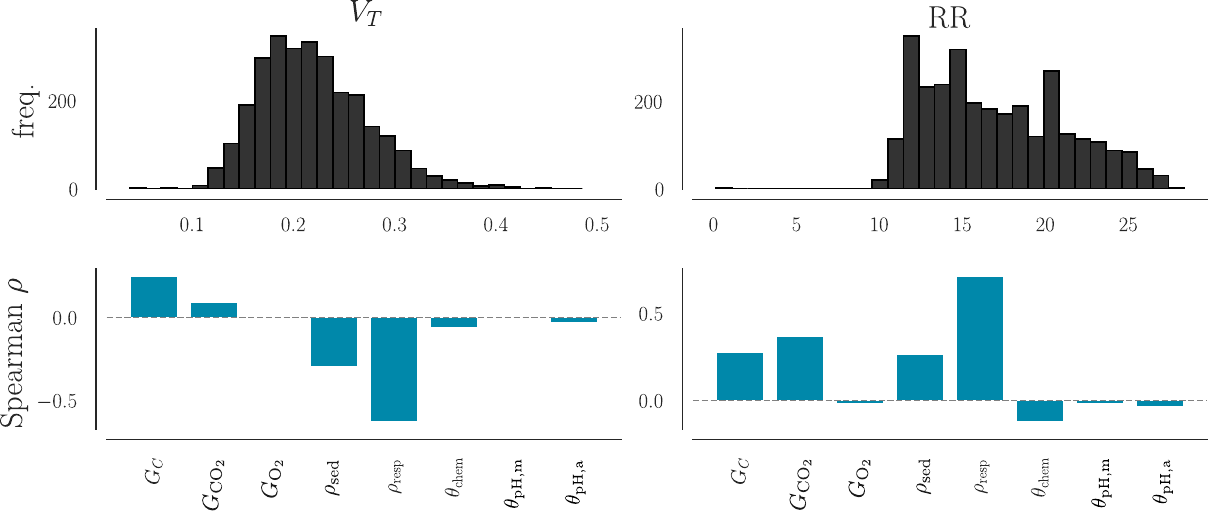}
    	\caption{Results of the uncertainty quantification and sensitivity analysis of the respiratory center submodels. \vt: Tidal volume [L], RR: Respiratory rate [1/min].}
    	\label{fig:res_SA_UQ_RC}
	\end{subfigure}
	\hfill
	\caption{Results of the uncertainty quantification and sensitivity analysis.}
	\label{fig:step1_2}
\end{figure*}

\paragraph{Uncertainty Assessment} 
Global uncertainty and sensitivity analyses were performed on the respiratory system model divided in two sub groups: Lung mechanics + Gas Exchange submodel and respiratory center submodel. In the following, the input ($\mathbf{x}_i$) and output ($\mathbf{y}_i$) with $i =$ LM, GE and RC parameter are defined:
\begin{align*}
	\nonumber\mathbf{x}_\mathrm{LM, GE} =& [R, E_L, E_\mathrm{CW}, f_\mathrm{V_D}, \vd, \fs, Q, D_\mathrm{L,O_2}, \\ &D_\mathrm{L,CO_2}, V_c, \mathrm{MP_{t, O_2}}, \mathrm{MP_{t, CO_2}}, V_B, T_\mathrm{h}, \delta, \\ & \sigma_\mathrm{O_2}, \sigma_\mathrm{CO_2}, \mathrm{FRC}] \\[2mm]
	\nonumber\mathbf{x}_\mathrm{RC} =& [G_C, G_\mathrm{CO_2}, G_\mathrm{O_2}, \rho_{\mathrm{sed}}, \rho_{\mathrm{resp}}, \theta_{\mathrm{chem}}, \\& \theta_\mathrm{pH,m}, \theta_\mathrm{pH,a}] 
\end{align*}
and 
\begin{align*}
	\nonumber\mathbf{y}_\mathrm{LM, GE} &= [\vt, \etco] \\[2mm]
	\nonumber\mathbf{y}_\mathrm{RC} &= [\vt, \mathrm{RR}] 	
\end{align*}
All model parameter with description, mean value and standard deviation are listed in Table~\ref{tab:sa_uq_input} for the SA and UQ. Literature based mean standard values are listed in Table~\ref{tab:model_input_parameter} of the appendix.\\
Parameter sampling is done using the Latin Hypercube Sampling (LHS)~\cite{loh_latin_1996, helton_latin_2003}. For sensitivity analysis the Spearman rank correlation~\cite{armitage_spearman_2005} is calculated. \\
The results are visualized in Figure~\ref{fig:res_SA_UQ_LM_GE} and ~\ref{fig:res_SA_UQ_RC}. Overall, \vt~was comparatively robust to the sampled parameter variability, whereas \etco~in the lung mechanics and gas-exchange analysis and RR in the respiratory-center analysis showed substantial propagated variability. \\
UQ/SA for the lung mechanics submodel showed very low uncertainty for \vt. \etco~was more variable due to V/Q distributions, diffusion capacity, and blood transport. \vt~was chiefly negatively associated with lung and chest-wall elastance $E_L$, $E_{\mathrm{CW}}$, while airway resistance $R$ became limiting when the time constant $\tau = R \cdot C$ approached $T_{\mathrm{insp}}$. \\
In the model, gas exchange depends on how well air reaches the lungs and is matched to blood flow (V/Q matching; Figure~\ref{fig:res_SA_UQ_LM_GE}). If part of the lung does not participate in gas exchange (alveolar dead space \vd), end-tidal CO$_2$~levels rise. Conversely, higher shunt fraction \fs~or cardiac output $Q$ reduce \etco.\\
Breathing is primarily controlled by CO$_2$ levels in the blood (Figure~\ref{fig:res_SA_UQ_RC}). The amount of air inhaled per breath (tidal volume, \vt) is fairly stable, while the breathing rate (RR) can vary substantially. Central CO$_2$ sensitivity ($G_C$) is the main driver, peripheral CO$_2$ sensitivity ($G_\mathrm{CO_2}$) has a smaller effect, and oxygen sensitivity ($G_\mathrm{O_2}$) has minimal impact under normal oxygen conditions. Sedation reduces tidal volume slightly and increases respiratory rate. Increasing the severity of respiratory failure ($\rho_\mathrm{resp}$) leads to faster, more effortful breathing but smaller breaths. Higher CO$_2$ or pH set points ($\theta_\mathrm{chem}$) dampen the overall drive to breathe, with minimal effect on tidal volume.
In summary, $E_L$ and $E_{\mathrm{CW}}$ were the dominant contributors to the remaining variability in \vt, with $R$ becoming relevant under high-resistance conditions. For \etco, variability in dead-space and shunt parameters, cardiac output $Q$, diffusion capacity, and blood-transport parameters should be considered. \\
Among the sampled respiratory-center parameters, the respiratory-failure scaling factor $\rho_\mathrm{resp}$ had the largest influence on the respiratory-center outputs, reflecting its direct modulation of respiratory drive in the simulated respiratory-failure scenarios. Central CO$_2$ sensitivity ($G_C$) was the most influential chemoreflex-gain parameter, whereas peripheral CO$_2$ sensitivity ($G_\mathrm{CO_2}$) had a smaller effect and oxygen sensitivity ($G_\mathrm{O_2}$) had minimal impact under normal oxygen conditions. In summary, $\rho_\mathrm{resp}$ was the dominant contributor to variability in the respiratory-center outputs. $G_\mathrm{CO_2}$, $\rho_\mathrm{sed}$, and the chemical set points had secondary or modulating effects. In contrast, $G_\mathrm{O_2}$ had minimal influence under the investigated normoxic conditions. These rankings apply to the sampled parameter ranges and simulation conditions.

\paragraph{Verification} 
In line with ASME V\&V~40, numerical code and calculation verification, alongside software quality assurance, are essential prerequisites for any validation exercise. In the present work, basic implementation checks (e.g.\ consistency of units, conservation properties, regression tests of key submodels) have been performed to ensure that the code behaves as intended under the tested conditions. However, a comprehensive verification of the model implementation -- including structured code review, formal unit tests for all submodels, and calculation verification through step-size convergence studies -- has not yet been fully executed and documented. As a consequence, the validation results reported in this work should be interpreted as conditional on the correctness of the current implementation.Therefore, all presented validation work will need to be reevaluated once comprehensive verification is properly executed. Such a verification campaign must be completed prior to any use of the model in a regulatory submission or for direct clinical decision support.

\begin{figure*}[t]
	\centering
	\begin{subfigure}[t]{0.48\textwidth}
		\centering
		\includegraphics[width=\textwidth]{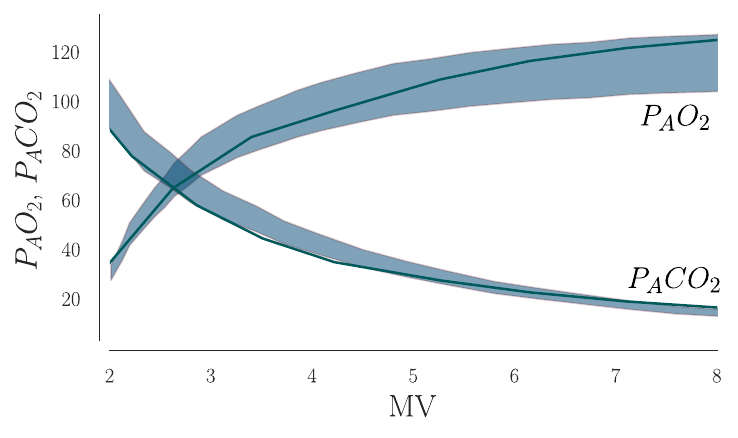}
		\caption{Test V--1: Simulation results (blue line) of minute ventilation MV [L/min] as a function of alveolar partial pressure of O$_2$ and CO$_2$ \palvo and \palvco [mmHg] in comparison to literature date (90\% confidence interval)~\cite{lumb_nunns_2017}.}
		\label{fig: MV_gasexchange}
	\end{subfigure}
	\hfill
	\begin{subfigure}[t]{0.48\textwidth}
		\centering
		\includegraphics[width=\textwidth]{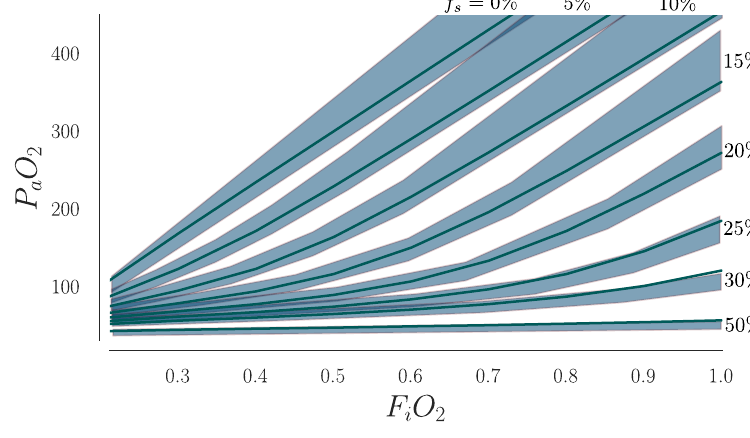}
		\caption{Test V--2: Simulation results (blue line) of arterial partial pressure \pao [mmHg] as a function of pulmonary shunt \fs~and inspiratory oxygen fraction \fio in comparison to literature date (90\% confidence interval)~\cite{lumb_nunns_2017}.}
		\label{fig: iso_shunt}
	\end{subfigure}
	\hfill
	\caption{Population based evidence: Simulation results of validation Test V--1 (a) and Test V--2 (b) for the baseline patient model.}
	\label{fig:step1_2}
\end{figure*}

\paragraph{Validation}
Validation is carried out based on the assessment plan set out in Table~\ref{tab:assesment_plan}. The methods and results of all five tests are explained in detail in the following: \\ \\
\noindent
\textcolor{darkblue}{\textbf{Test V--1: Gas Exchange, MV to \palvo, \palvco}} The model’s ability to reproduce alveolar oxygen and carbon dioxide partial pressures under varying minute ventilation is evaluated. Literature data~\cite{lumb_nunns_2017} provide reference values for comparison. For results see Figure~\ref{fig: MV_gasexchange}, simulation results are within 90\,\% confidence interval.\\ \\
\noindent
\textcolor{darkblue}{\textbf{Test V--2: Gas Exchange, Iso shunt}} The impact of shunt fraction on gas exchange is assessed by simulating scenarios with fixed shunt levels. Reference values are taken from literature~\cite{lumb_nunns_2017}. For results see Figure~\ref{fig: iso_shunt}, simulation results are within 90\,\% confidence interval.\\ \\
\noindent
\textcolor{darkblue}{\textbf{Test V--3: Respiratory Center, Chemical Feedback}} The response of the respiratory controller to changes in CO$_2$ and O$_2$ blood levels is tested. Literature data on transient chemical feedback responses~\cite{reynolds1973transient, reynolds_transient_1972} are used for validation. For results see reference~\cite{hennigs_respcenter_2025}, simulation results are within 90\,\% confidence interval and MAPE of simulation results is less than 10\,\%.  \\ \\
\noindent
\textcolor{darkblue}{\textbf{Test V--4: Respiratory Center, PSV Levels}}
The effect of different pressure support ventilation levels PSV on respiratory control is assessed. Literature data~\cite{vitacca_assessment_2004} serve as reference. For results see reference~\cite{hennigs_respcenter_2025}, MAPE of simulation results is less than 10\,\%.\\ \\
\noindent
\textcolor{darkblue}{\textbf{Test V--5: Respiratory Center, Asynchrony}} Patient-ventilator interaction and asynchrony are examined. Data from literature~\cite{sauer_automated_2024, thille_patient-ventilator_2006, mulqueeny_automatic_2007} provide benchmark observations. For results see reference~\cite{hennigs_respcenter_2025}, simulation results differ less than 5\,\% compared to literature data.
\paragraph{Supplementary credibility evidence} Supplement credibility evidence is created based on the assessment plan set out in Table~\ref{tab:assesment_plan}. The methods and results of the four tests are explained in detail in the following: \\ \\
\textcolor{darkblue}{\textbf{Test S--1: Lung Mechanics, COPD}} The model’s representation of lung mechanics in patients with COPD is evaluated. Clinical data from ventilated COPD patients~\cite{grashoff_surface_2021} provide the comparison. For results see reference~\cite{Hennigs_COPD_2024}.\\ 
\noindent
\textcolor{darkblue}{\textbf{Test S--2: Lung mechanics, gas exchange}} Model calibration employed nonlinear optimization to fit the ODE-based patient model to clinical data. The clinical data was obtained from a study by Schädler~\cite{niras_data}. The study was reviewed and approved by the Ethics Committee of the Medical Faculty of the Christian-Albrechts-University of Kiel (approval number D411/13). Four patients were investigated. The data were recorded using the automated weaning protocol SmartCare/PS by Dräger (Drägerwerk AG \& Co. KGaA, Lübeck, Germany). Only two measurements were available: airway flow $\dot{V}$ and capnogram~$F_\mathrm{CO_2}$. Lung volume was reconstructed by temporal integration of the measured airway-flow signal after leakage correction and was subsequently used for comparison with the simulated volume trajectory. The model was driven by synchronized clinical input signals comprising airway pressure relative to PEEP, external PEEP, inspired oxygen and carbon-dioxide fractions, and heart rate. Non-smooth physiological operators within the model (e.g. cardiac resetting, flow switching) were reformulated as smooth sigmoid functions to enable robust parameter identification while maintaining differentiability. An iterative sliding-window formulation can subsequently be used to update patient-specific parameters online as physiology evolves over time. Further methodological details and patient-specific results are provided in Section~\ref{append:model_cal} of the Appendix. \\
For the exemplary reference Patient~3, patient-specific identification was evaluated using temporally separated calibration and validation intervals. Agreement was quantified over the complete lung-volume and capnogram signals using the coefficient of determination ($R^2$), root mean square error (RMSE), and mean absolute error (MAE). RMSE and MAE provide absolute errors in the
physical units of the respective signals and remain well defined throughout
the complete capnogram, including intervals in which CO$_2$ approaches zero.
Table~\ref{tab:model_cal} compares the physiologically informed global prior,
the patient-specific calibration, and the subsequent frozen temporal
validation. Results for the additional investigated patients are reported in
Appendix~B.2.
\begin{figure*}[t]
    \centering
    \includegraphics[width=\textwidth]{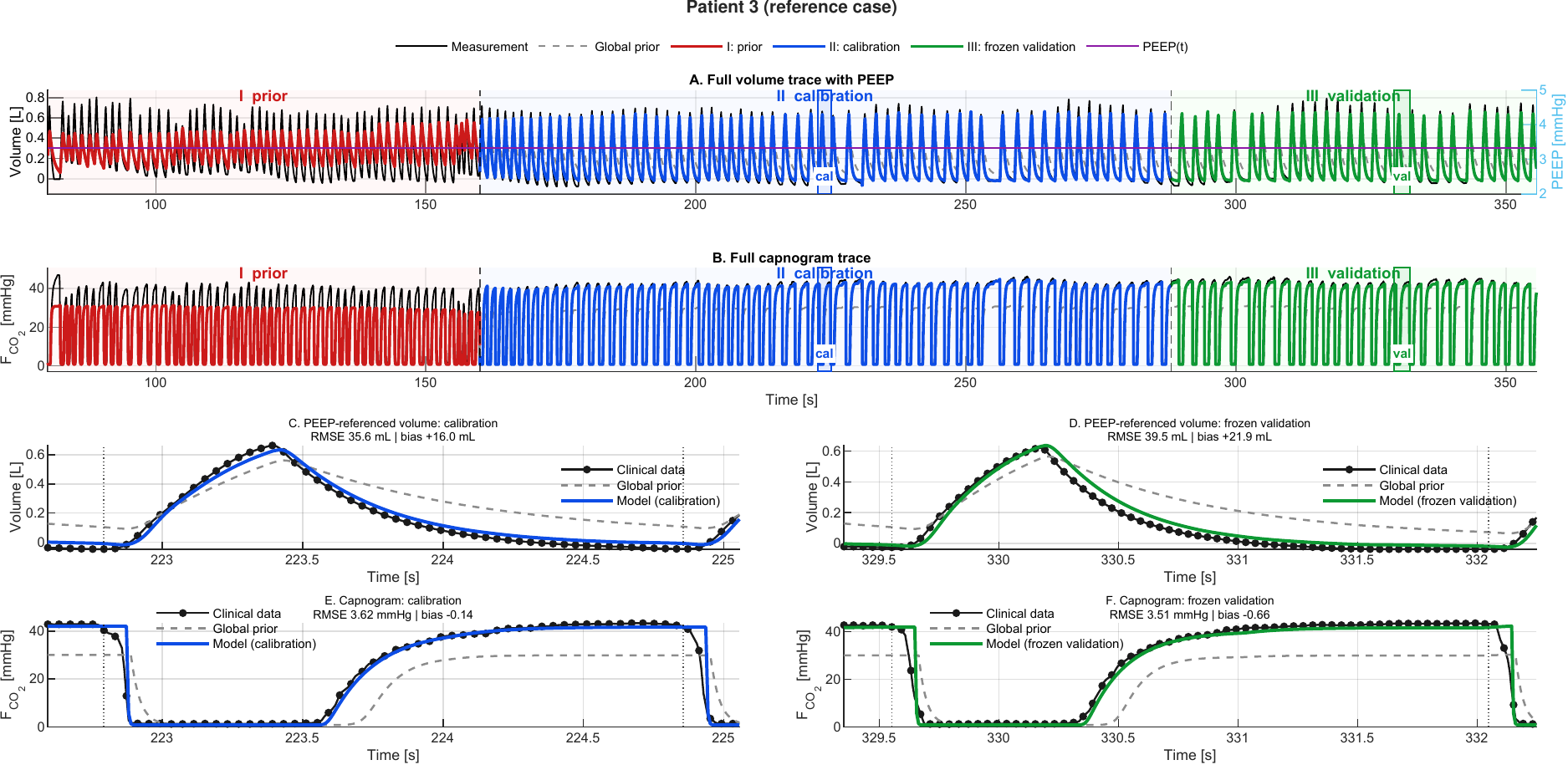}
    \caption{Test S--2: Patient-specific calibration and frozen temporal validation for the exemplary
reference Patient~3. (A,B) Full time courses of lung volume and capnogram, respectively, showing
the physiologically informed global prior and the three evaluation regimes:
I, prior; II, patient-specific calibration; and III, frozen temporal validation,
i.e. prediction over a later unseen interval with all patient-specific parameters
fixed at their calibrated values. The validation interval is not used for
parameter estimation.
(C,D) Representative single-breath comparisons of lung volume during calibration
and frozen validation, respectively.
(E,F) Corresponding capnogram comparisons using the same breath windows.
Clinical data are shown in black, the global prior as a gray dashed line, and
the regime-specific model trajectories in red (prior), blue (calibration), and
green (frozen validation).
RMSE and bias reported in panels C--F quantify the local agreement within the
displayed breath windows. The representative breaths are selected
deterministically from the center of the respective intervals.}
    \label{fig:model_calibration}
\end{figure*}
\begin{table*}[t]
	\centering
	\small
	\setlength{\tabcolsep}{5pt}
	\caption{
Test S--2: Patient-specific calibration and frozen temporal validation
for the reference case (patient~3) shown in Figure~\ref{fig:model_calibration}.
The global prior denotes the model before patient-specific adaptation.
Parameters are identified only during calibration and remain fixed during
the subsequent unseen validation interval. Metrics are evaluated over the
complete calibration and validation intervals, whereas the RMSE and bias
values shown in Figure~\ref{fig:model_calibration} refer to the displayed
representative windows. Results for the additional investigated patients
are reported in Appendix~B.2. $R^2$: coefficient of determination;
RMSE: root mean square error; MAE: mean absolute error.
}
	\label{tab:model_cal}
	\begin{tabular}{@{}lcccccc@{}}
		\toprule
		& \multicolumn{3}{c}{\textbf{Lung volume $V$}}
		& \multicolumn{3}{c}{\textbf{Capnogram $F_{\mathrm{CO_2}}$}} \\
		\cmidrule(lr){2-4}
		\cmidrule(l){5-7}
		\textbf{Stage}
		& \textbf{$R^2$}
		& \textbf{RMSE [mL]}
		& \textbf{MAE [mL]}
		& \textbf{$R^2$}
		& \textbf{RMSE [mmHg]}
		& \textbf{MAE [mmHg]} \\
		\midrule
		Prior -- calibration
		& 0.616 & 140.4 & 129.0
		& 0.679 & 10.32 & 8.50 \\
		\textbf{Calibration}
		& \textbf{0.964} & \textbf{42.7} & \textbf{34.5}
		& \textbf{0.966} & \textbf{3.34} & \textbf{1.22} \\
		\addlinespace[2pt]
		Prior -- validation
		& 0.671 & 130.9 & 121.1
		& 0.634 & 11.22 & 9.45 \\
		\textbf{Frozen validation}
		& \textbf{0.967} & \textbf{41.5} & \textbf{31.7}
		& \textbf{0.977} & \textbf{2.81} & \textbf{1.36} \\
		\bottomrule
	\end{tabular}
\end{table*}
Across the four investigated patients, frozen temporal validation showed consistent agreement for lung-volume and capnogram dynamics under distinct patient-specific conditions (Table~\ref{tab:model_cal_all_patients}). Patient~3 served as the reference case, Patient~1 included a PEEP intervention during validation, Patient~2 provided a long-duration capnogram record with additional adaptation of the effective metabolic CO$_2$ production parameter, and Patient~4 was used to assess robustness in the presence of a slow capnogram artefact. Further patient-specific calibration, validation, parameter adaptation, and diagnostic results are provided in Appendix~B.2.
\begin{table}[t]
    \centering
    \scriptsize
    \setlength{\tabcolsep}{4pt}
    \renewcommand{\arraystretch}{1.08}
\caption{
Test S--2: Frozen temporal validation across all investigated patients.
Metrics are evaluated over the complete frozen-validation intervals,
whereas the RMSE values shown in the corresponding patient-specific
figures refer to the displayed representative windows.
Further patient-specific details are provided in Appendix~B.2.
}
    \label{tab:model_cal_all_patients}

    \begin{tabular}{@{}lcccc@{}}
        \toprule
        &
        \multicolumn{2}{c}{\textbf{Volume $V$}} &
        \multicolumn{2}{c}{\textbf{Capnogram $F_{\mathrm{CO_2}}$}} \\
        \cmidrule(lr){2-3}
        \cmidrule(l){4-5}

        \textbf{Patient / Case} &
        $\mathbf{R^2}$ &
        \shortstack{\textbf{RMSE}\\\textbf{[mL]}} &
        $\mathbf{R^2}$ &
        \shortstack{\textbf{RMSE}\\\textbf{[mmHg]}} \\
        \midrule

        1 -- \textit{PEEP change}      & 0.811 & 117.6 & 0.950 & 3.93 \\
        2 -- \textit{Long-duration} & 0.913 &  81.2 & 0.932 & 4.51 \\
        3 -- \textit{Reference} & 0.967 &  41.5 & 0.977 & 2.81 \\
        4 -- \textit{Artefact}  & 0.975 &  21.4 & 0.921 & 8.13 \\

        \bottomrule
    \end{tabular}
\end{table}

The global prior represents the physiological model before patient-specific adaptation. During the calibration interval, a selected subset of physiologically meaningful parameters is identified from the available clinical measurements. The resulting parameterization therefore reflects
individual respiratory mechanics and gas-exchange characteristics supported
by the available measurements, while the mechanistic state-space formulation
preserves an interpretable internal representation of the underlying
physiological dynamics.\\
\noindent
Following calibration, the identified parameters are kept fixed and the model is propagated forward over a temporally separate validation interval that was not used for parameter estimation. The preserved agreement for both lung
volume and capnography in this unseen interval provides evidence that, over
the investigated time horizon, the relevant patient-specific parameters can
be approximated as quasi-static and that the individualized model captures
sufficient physiological dynamics for short-term prediction. Figure~\ref{fig:model_calibration} illustrates this calibration--prediction
sequence over the complete trajectories and representative individual
breaths.\\

\par\noindent\textcolor{darkblue}{\textbf{Test S--3: Model plausibility}} This test re-examines the sensitivity analysis and the correlations between input and output parameters to assess whether parameter-induced behavior is plausible and consistent with clinical expert opinion. The predicted increases and decreases agree with expert judgment. The governing equations (Table~\ref{tab:governing_eq_results}) of the physiological model were then systematically evaluated to ensure scientific rigor and transparency. For each equation, literature quality (LQ), data quality (DQ), and validation evidence (V) were graded. Literature quality was rated \emph{LOW}, \emph{MEDIUM}, or \emph{HIGH} based on citation count and author prominence (gradations in Table~\ref{tab:lit_grading} in the Appendix). Equations were classified as physics-based or study-based. Physics-based DQ reflected adherence to fundamental physical principles or physiological mechanisms, whereas study-based DQ reflected cohort size, population coverage, availability of patient-level data, and study design. Model validation (V) was rated \emph{LOW} for plausibility checks only, \emph{MEDIUM} for comparison with other models or literature, and \emph{HIGH} when clinical data were available. Each dimension was scored $1$--$3$, and an overall quality grade was computed as 
\begin{equation}
	\text{Total Grade} = \text{LQ} + \text{DQ} + \text{V} - 2,  
\end{equation}
yielding scores from 1 (lowest) to 7 (highest). The results of this analysis are summarized in Table~\ref{tab:governing_eq_ana_short}. \\ \\
\begin{table}[t]
	\centering
	\small
	\caption[Model governing equation analysis results.]{Model governing equation analysis results, for the list and naming of the models governing equations see Table~\ref{tab:governing_eq_results} in the Appendix. LM: Lung mechanics, GE: Gas exchange, RC: Respiratory center, LQ: Literature quality, DQP: Data quality physics based, DQS: Data quality study based, V: Model validation. }
	\label{tab:governing_eq_ana_short}
	\begin{tabular}{p{1.1cm}>{\centering\arraybackslash}p{0.5cm}>{\centering\arraybackslash}p{0.5cm}>{\centering\arraybackslash}p{0.5cm}>{\centering\arraybackslash}p{0.5cm}>{\centering\arraybackslash}p{0.5cm}>{\centering\arraybackslash}p{0.8cm}}  
		\toprule
		\textbf{Eq. ID}  & \textbf{Ref.} & \textbf{LQ} & \textbf{DQP} & \textbf{DQS} & \textbf{V} & \textbf{Total}\\ 
		\midrule
		LM1--a & \cite{Hennigs_COPD_2024}  & 1 &  & 2 & 3 & \textbf{4} \\
		LM1--b & \cite{suki_mathematical_2025}  & 3 & 3 &  & 3 & \textbf{7} \\ 
		GE1--5 & \cite{ben-tal_simplified_2006}  & 3 & 3 &  &  1 & \textbf{5}\\
		GE6--9, BL4& \cite{chiari_comprehensive_1997}  & 3 & 3 &  & 3 & \textbf{7} \\ 
		BL2& \cite{hills_ph_1973}  & 3 & 3 & & 3 & \textbf{7} \\ 
		BL3 &\cite{loeppky_relationship_1993}  &  3 & & 2 & 3 & \textbf{6} \\ 
		RC1--2, RC4, RC8 &\cite{JAWORSKI2019148} & 3 & 2 &  & 2 & \textbf{5} \\ 
		RC1--2, RC8 & \cite{botros_neural_1990}  &3 & 2 &  & 2 & \textbf{5} \\ 
		RC3 &\cite{reynolds_transient_1972}  & 3 &  & 1 & 2 & \textbf{4} \\ 
		RC3 &\cite{reynolds1973transient}  & 3 &  & 1 & 1 & \textbf{3} 		\\
		\bottomrule
	\end{tabular}
\end{table}%
\noindent\textcolor{darkblue}{\textbf{Test S--4: Emergent phenomena}} Emergent model behavior is evidence that demonstrates that the finalized PDM reproduces known physiological phenomena under specified conditions, even though these behaviors were not explicitly implemented or programmed in the governing equations. Instead, such behaviors arise from the interaction of physiological submodels — such as chemoreflex control, neuronal oscillators, lung mechanics, and gas exchange — and are assessed based on established scientific knowledge and qualitative consistency with clinical observations. These emergent behaviors strengthen the model’s credibility, particularly in simulating complex physiological and pathophysiological states. Below two examples of emergent behavior of the proposed model are described in detail. A more comprehensive list can be found in Section~\ref{sec:Emergent model behavior} of the Appendix. \\
\textbf{Lung mechanics -- Auto-PEEP:} \\
The lung mechanic submodel demonstrated that under conditions of high respiratory rate and increased resistance, the model produces intrinsic positive end-expiratory pressure (auto-PEEP) due to incomplete exhalation. This behavior is not explicitly coded but emerges from the interaction of flow resistance and timing. \\ 	
\textbf{Respiratory center -- Asynchrony Emergence:}
Evidence (see test V--5) is collected to demonstrate that, depending on ventilator settings and patient model parameters, the model exhibits patient-ventilator asynchrony events. These asynchronies are not explicitly coded but emerge from the interaction between ventilator support and simulated respiratory drive. This can be included as credibility evidence for model realism under clinical conditions. These emergent behaviors are not pre-specified in the model logic, but arise naturally from physiologically motivated interactions between components. Their alignment with observed clinical or experimental phenomena provides qualitative evidence supporting the credibility of the model, particularly in simulations involving disease states, therapeutic interventions, or patient-specific adaptation.

\subsubsection{Post-Study Adequacy Assessment}
\paragraph{Applicability}
Applicability (fit-to-COU) was quantified from Table~\ref{tab:applicability_summary} by mapping the reported gradations to numeric scores and aggregating along ASME V\&V~40 and FDA guidance aligned evidence categories. Gradations were mapped as $\gA=1$ (weak), $\gB=2$ (moderate), $\gC=3$ (good) to a subscore $\overline{S}_k$ for each evidence category. For each evidence category, only factors that directly reflect the fit to the COU were included:
\begin{enumerate}
	\item Population-based validation: \emph{Equivalency}, \emph{Relevance of QoI}, \emph{Relevance to COU}
	\item Model calibration: \emph{Relevance of QoI}, \emph{Relevance to COU}
	\item Model plausibility: \emph{Consistency}
	\item Emergent model behavior: \emph{Relevance to COU}, \emph{Impact}
\end{enumerate}    
\noindent
Within each category, per-row averages of the selected factors were computed. The composite applicability score $S_{\mathrm{app}}$ was calculated as the unweighted arithmetic mean of the category subscores and classified into three bands: weak (a) $\le 1.5$, moderate (b) $1.5~{<}~S_{\mathrm{app}}~\le~2.3$, and good (c) $>2.3$. In Table~\ref{tab:applicability_summary} the applicability-relevant subscores are listed. All categories contribute equally to the composite. Strong calibration and plausibility (both 3.0), while population-based validation (2.0) and emergent behavior coverage (2.0) remain solid. In the preclinical ISCT context of ventilation this indicates a robust fit-to-COU, with potential future gains from expanded clinical comparators and broader demonstrations of weaning-specific emergent behaviors.
\begin{table}[t]
	\centering
	\small
	\caption{Applicability (fit-to-COU) summary.  $\overline{S}_k$: Category subscore, $S_{\mathrm{app}}$: Composite application score.}
	\label{tab:applicability_summary}
	\begin{tabular}{lc}
		\toprule
		\textbf{Evidence category} & \textbf{Subscore} $\overline{S}_k$ \\
		\midrule
		Population-based validation     & 2.00 \\
		Model calibration               & 3.00 \\
		Model plausibility              & 3.00 \\
		Emergent model behavior         & 2.00 \\
		\midrule
		\textbf{Composite $S_{\mathrm{app}}$ }   & \textbf{2.5 (good, c)} \\
		\bottomrule
	\end{tabular}
\end{table}

\paragraph{Credibility Assessment}
All baseline requirements — including functional, physiological, and clinical aspects — are addressed by targeted validation tests, with complete coverage documented. Each requirement is verified through at least one test, supported either by direct comparison with literature or clinical data, or by demonstrating plausible and emergent behaviors. For example, lung mechanics in COPD patients are assessed using clinical data (Test S--1), shunt effects on gas exchange are tested under controlled conditions (Test V--2), and chemical feedback in the respiratory center is examined through literature-based CO$_2$ and O$_2$ perturbations (Tests V--1 and V--3). Functional requirements, such as model calibration, forward prediction, and closed-loop ventilator triggering, are validated through targeted simulations. Calibration and predictive performance are assessed by comparing model outputs with patient data (Test S--2), while closed-loop triggering of the ventilator is demonstrated in pressure support scenarios (Test V--4). Clinical requirements, including maintenance of ventilation within safe ranges and the correct capture of common patient-ventilator asynchronies, are evaluated in scenarios mimicking real clinical conditions (Tests V--4 and V--5). Finally, overall model credibility is strengthened by cross-cutting evaluations: general plausibility is confirmed through literature and expert review (Test S--3), and emergent phenomena, such as auto-PEEP and spontaneous asynchronies, are reproduced to demonstrate that the model captures non-programmed, physiologically relevant behaviors (Test S--4). \\
For model plausibility, Table~\ref{tab:governing_eq_ana_short} synthesizes ten equation-source pairs and shows total grading spanning $3-7$ (median~5), reflecting a balanced evidence base with clear clusters. A high‑evidence tier (Total:~7) comprises core gas‑exchange chemistry and buffering \cite{chiari_comprehensive_1997,hills_ph_1973} and the one-compartment lung mechanics model~\cite{suki_mathematical_2025, bates_lung_2009}. These entries combine strong literature (LQ:~3), a robust physics foundation (DQP:~3), and direct validation (V:~3), indicating mature, well‑corroborated building blocks for oxygen and CO$_2$ transport, acid-base balance, and pressure-volume mechanics. A moderate tier (Total: ~$5-6$) covers respiratory‑control relations~\cite{JAWORSKI2019148,botros_neural_1990}, simplified gas‑exchange~\cite{ben-tal_simplified_2006}, and transport proxies~\cite{loeppky_relationship_1993}. Conceptually, these are well‑grounded (high LQ, physics or study basis), but are limited by the depth or directness of clinical validation (V:~$1-2$). In practice, they are fit‑for‑purpose in a low-medium‑risk context, yet offer clear headroom: adding targeted clinical comparisons (e.g.\ ventilatory responses to chemo‑stimuli, shunt/dead‑space effects on \pao and \paco) would lift validation grading and therefore total grading. Low scores (Total:~$3-4$) are confined to legacy or study‑limited items (RC3~\cite{reynolds_transient_1972,reynolds1973transient}). These entries mainly reflect narrower cohorts (low DQS) or limited validation rather than fundamental conceptual issues, and are either replaceable (via the high‑grade physics alternative for LM1) or upgradeable with contemporary datasets and tests.\\
In summary, 79 criterion ratings across nine tests were synthesized using the scale ($a$ = weak = 1, $b$ = moderate = 2, $c$ = good = 3, $d$ = very good). A detailed view is listed in Table~\ref{tab:assess_credibility} in the Appendix. A summary is provided in Table~\ref{tab:assess_credibility_sum}. Overall, the evidence profile shifts toward moderate-to-good, with concentrated strengths and fewer weak areas. The total score is 2.15.\\
While these quantitative scores provide a structured summary of the collected evidence, we acknowledge that a purely quantitative approach can be subjective in assigning weights to credibility levels. Therefore, these scores are utilized as a supplementary tracking tool rather than the sole determinant. The overall adequacy assessment relies on a holistic approach of the totality of the evidence, applying engineering judgment and providing rationale for why the combined credibility evidence is sufficient to support using the model for the defined COU, given the medium-low risk assessment.\\
\textbf{Population-based validation} (Tests V--1, V--2, V--3, V--4, V--5) shows consistent strengths in output equivalency and agreement with generally moderate rigor of comparison and data quantity. Gaps remain in subject-range coverage and completeness of patient-level data, indicating the need for broader cohorts and richer metadata to fully represent the intended population.\\
\textbf{Model calibration} (Test S--1 and S--2) Test S--2 demonstrates strong fit, confirming accurate reproduction of respiratory mechanics and gas exchange dynamics under clinical conditions. Test S--2 is moderate-to-good on inputs analyzed and rigor (both c), with moderate output span (b), but limited sample size (a), yielding an average of 1.87 at the category level.\\
\textbf{Model plausibility} (Test S--3) is moderate-to-good across justification, assumptions, and consistency, and is reinforced by expert endorsement and overall plausibility (both c), with a minor residual gap in the depth of justification and input-parameter rationale for selected submodels.\\
\textbf{Emergent behavior} (Test S--4) is relevant to the COU (b), with good reproducibility and consistency (both c), good identification (c), moderate confirmation (b), and moderate impact (b).\\
Given that the model risk for the COU is classified as medium-low, the attained validation level is sufficient per a risk-informed interpretation of ASME~V\&V~40 and FDA credibility guidance. The current body of evidence (moderate-to-good on average) with specific strong elements (agreement/equivalency, model calibration, plausibility with expert endorsement) — meets the credibility goals appropriate for medium-low decision impact. To elevate confidence and enable higher‑impact use cases, it is recommended to broadening the population based datasets and strengthening calibration with increased patient numbers and independent external validation. 

\definecolor{gradeD}{HTML}{B8D4E8}
\colorlet{gradeVeryGood}{gradeD}
\newcommand{\distbarScore}[6]{%
  \begin{tikzpicture}[x=#6, y=1cm, baseline=0.6ex]
    \pgfmathsetmacro{\n}{#5}
    \pgfmathsetmacro{\score}{#1*1 + #2*2 + #3*3 + #4*4}
    \pgfmathsetmacro{\scale}{(\n>0)?(\score/(4*\n)):0}
    \pgfmathsetmacro{\fa}{(\n>0)?(#1/\n*\scale):0}
    \pgfmathsetmacro{\fb}{(\n>0)?(#2/\n*\scale):0}
    \pgfmathsetmacro{\fc}{(\n>0)?(#3/\n*\scale):0}
    \pgfmathsetmacro{\fd}{(\n>0)?(#4/\n*\scale):0}
    \fill[gradeWeak]     (0,0)
        rectangle (\fa,0.18);
    \fill[gradeMid]      (\fa,0)
        rectangle (\fa+\fb,0.18);
    \fill[gradeGood]     (\fa+\fb,0)
        rectangle (\fa+\fb+\fc,0.18);
    \fill[gradeVeryGood] (\fa+\fb+\fc,0)
        rectangle (\fa+\fb+\fc+\fd,0.18);
    \draw[black, line width=0.2pt] (0,0) rectangle (1,0.18);
  \end{tikzpicture}%
}

\begin{table*}[t!]
  \small
  \setlength{\tabcolsep}{3pt}
  \renewcommand{\arraystretch}{1.15}
  \centering
  \caption[Credibility summary.]{Credibility summary
    (a = weak, b = moderate, c = good, d = very good)
    based on Table~\ref{tab:assess_credibility}.
    Note that for selected categories, a rating of (d) is applicable,
    representing the highest achievable evidence level.
    The mean score is computed by assigning numerical values
    a = 1, b = 2, c = 3, d = 4 and using
    $\text{score} = (1 \cdot \#a + 2 \cdot \#b + 3 \cdot \#c
    + 4 \cdot \#d)/n$,
    where $\#a$, $\#b$, $\#c$, and $\#d$ are the counts of criteria
    rated a, b, c, and d, respectively, and $n$ is the total number
    of criteria. Bar length is scaled by this score.}
  \label{tab:assess_credibility_sum}
  \begin{tabular}{p{2cm}
      >{\centering\arraybackslash}p{1cm}
      >{\centering\arraybackslash}p{1cm}
      >{\centering\arraybackslash}p{1cm}
      >{\centering\arraybackslash}p{1cm}
      >{\centering\arraybackslash}p{1cm}
      >{\centering\arraybackslash}p{1cm}
      >{\centering\arraybackslash}p{4.5cm}}
    \toprule
     & \textbf{n} & \textbf{a} & \textbf{b} & \textbf{c}
     & \textbf{d} & \textbf{Score} & \textbf{Distribution} \\
    \midrule
    Test V--1 & 10 & 3 & 4 & 3 & 0 & 2.00
      & \distbarScore{3}{4}{3}{0}{10}{4cm} \\
    Test V--2 & 10 & 4 & 3 & 3 & 0 & 1.90
      & \distbarScore{4}{3}{3}{0}{10}{4cm} \\
    Test V--3 & 10 & 4 & 4 & 2 & 0 & 1.80
      & \distbarScore{4}{4}{2}{0}{10}{4cm} \\
    Test V--4 & 10 & 1 & 9 & 0 & 0 & 1.90
      & \distbarScore{1}{9}{0}{0}{10}{4cm} \\
    Test V--5 & 10 & 2 & 6 & 2 & 0 & 2.00
      & \distbarScore{2}{6}{2}{0}{10}{4cm} \\
    Test S--1 &  8 & 1 & 4 & 3 & 0 & 2.25
      & \distbarScore{1}{4}{3}{0}{8}{4cm} \\
    Test S--2 &  12 & 5 & 4 & 2 & 0 & 1.58
      & \distbarScore{3}{3}{2}{0}{8}{4cm} \\
    Test S--3 &  7 & 0 & 1 & 3 & 3 & 3.29
      & \distbarScore{0}{1}{3}{3}{7}{4cm} \\
    Test S--4 &  6 & 0 & 3 & 2 & 1 & 2.67
      & \distbarScore{0}{3}{2}{1}{6}{4cm} \\
    \textbf{Total} & \textbf{79} & \textbf{18} & \textbf{37}
      & \textbf{20} & \textbf{4} & \textbf{2.15}
      & \distbarScore{22}{39}{21}{4}{86}{4cm} \\
    \bottomrule
  \end{tabular}
\end{table*}

\paragraph{Credibility Gaps and Recommendations}
The post-study adequacy assessment synthesized evidence across categories and compared outcomes against predefined acceptance criteria for each QoI (\vt~and RR within protective targets, acceptable \etco, stable patient-ventilator interaction). All baseline requirements (AWP--F1 -- F5, AWP--P1 -- P3, AWP--C1 -- C2) are covered and supported by plausibility or emergent-behavior evidence. With the updated results, the residual gaps are:
\begin{itemize}
	\item \textbf{Population-based validation:} Subject-range coverage and completeness of patient-level data remain limited. Broader, stratified, multi-center cohorts with richer metadata are needed to strengthen representativeness and coverage for the intended population. While this gap reduces confidence in the model's broad statistical generalizability across highly heterogeneous populations, it does not invalidate the model for the defined COU. 
	\item \textbf{Model calibration:} The model calibration in Test S--2 is based on a small number of patients and only two available measurements. To improve robustness and generalizability, calibration should be performed on a larger cohort with more comprehensive and advanced measurements. The limited sample size restricts the statistical power of the parameter identification. However, because the model is used exclusively for initial preclinical evaluation and is not the sole source of regulatory decision-making (as orthogonal bench and clinical data will be used later), this limitation is acceptable for the current risk level. The strong goodness-of-fit (R$^2 \approx$ 95\%) on the available data demonstrates the model's capability to accurately reproduce physiological processes and dynamics.
	\item \textbf{in vivo validation:} The availability of in vivo validation is currently limited. Only partial clinical comparators were used — for example, Test S--2 model calibration and a followed up in vivo validation was performed on four patients. Validation should be expanded by including a larger, independent patient cohort that aligns with the defined context of use. The small scope of independent in vivo validation is a significant limitation for higher-risk applications, as it means the model's predictive power in unseen clinical populations remains unproven. For this medium-low risk COU, the gap is mitigated by the fact that downstream safety mechanisms and additional clinical evaluations are mandatory before any patient use of the algorithm under test, meaning an incorrect preclinical decision will not lead to direct patient harm.
	\item \textbf{Emergent behavior:} Relevance and qualitative consistency are moderate-to-good, but reproducibility remains moderate. Standardized seeds, protocols, and openly shared scenario scripts and datasets would consolidate this category and enable quantitative impact assessment on QoIs. Because emergent behaviors serve as supplementary credibility evidence rather than primary quantitative validation, their moderate reproducibility does not critically undermine the core validation arguments. However, the mere appearance of these non-programmed, complex physiological phenomena already indicates a close model representation of the real underlying physiological processes, providing additional qualitative confidence in the model's mechanistic foundations.
\end{itemize}

\paragraph{Conclusion and Risk Alignment}
In line with ASME~V\&V~40~\cite{asme_2018} and FDA guidance~\cite{fda_guidance_2023}, the credibility activities are proportionate to a medium-low model-risk COU for preclinical ISCT in automated weaning function. The evidence bundle — model plausibility (3.0), model calibration (3.0), population-based validation (2.0), and emergent behavior (2.0) — supports that the model is fit-for-purpose for the stated COU, with an overall good applicability ($S_{\mathrm{app}}=2.5$). The remaining limitations are traceable and documented: comparator span and depth, representativeness of population datasets, and the need for explicit external validation in calibration. \\
For decisions of greater consequence or higher model influence, the evidence should be escalated accordingly: expanded in vivo comparators with statistical power, external validation and stricter calibration statistics, explicit validation of measurement-chain effects, and controller timing (e.g.\ hardware-/software-in-the-loop). 

\subsubsection{Credibility Report}
This paper and the Appendix~\ref{appen:credibility} constitutes the credibility report, providing a traceable account of methods, datasets, protocols, analyses, acceptance criteria, and conclusions aligned with current standards and guidance.

\section{Discussion}
The credibility assessment with a focus on validation activities has produced an evidence bundle that supports the model's fitness-for-purpose for preclinical ISCT in automated weaning function, subject to completion of systematic verification activities. These results are now placed in a broader context: How do they compare to prior credibility efforts in computational ventilation models? What do the residual gaps imply for practical translation and regulatory approval pathways? And what additional evidence would be required if the model were to support higher-stakes decisions or broader clinical populations? This discussion anchors the technical findings in the landscape of ventilation modeling, ISCT standards, and regulatory practice, and provides an honest of both strengths and limitations.\\
\subsection{Credibility Assessment with Focus on Model Validation}
This work applied a structured, risk-informed credibility assessment workflow to a PDM for automated weaning, closely following ASME~V\&V~40~\cite{asme_2018} and FDA guidance~\cite{fda_guidance_2023} on computational modeling and simulation (CM\&S). The methodology articulated a clear questions of interest and context of use, assessed model risk (medium-low influence and consequence), and mapped credibility evidence categories to ventilation-relevant activities across validation with documentation and traceability~\cite{asme_2018,fda_guidance_2023}. In addition, ISCT-specific guidance was incorporated to plan, execute, and assess adequacy in a cohort-based setting, consistent with Pathmanathan et~al.~\cite{pathmanathan_credibility_2024}. \\
The UQ/SA results indicate predictable and interpretable model behavior: tidal volume \vt exhibits low uncertainty and is dominated by elastance (lung and chest wall), whereas \etco reflect V/Q matching, shunt and dead space, diffusion, solubility, and transport~\cite{lumb_nunns_2017}. Respiratory control is primarily CO$_2$-driven, with central chemosensitivity governing RR and effort, and sedation reducing inspiratory muscle pressure amplitude as expected~\cite{JAWORSKI2019148,botros_neural_1990}. \\
Validation tests meet the acceptance criteria for the stated QoIs and comparators (literature-based envelopes for gas exchange and PSV responses, asynchrony incidence, and calibration/forward prediction against clinical data), and model plausibility is rated as good, supported by expert review and alignment with published evidence~\cite{lumb_nunns_2017,vitacca_assessment_2004,sauer_automated_2024,thille_patient-ventilator_2006,mulqueeny_automatic_2007}. Emergent phenomena such as auto-PEEP and patient-ventilator asynchronies arise naturally from submodel interactions and align qualitatively with clinical observation, providing additional confidence in complex regimes~\cite{sauer_automated_2024, thille_patient-ventilator_2006}.\\
ASME~V\&V~40 mandates that verification and validation are co-equal and inseparable credibility prerequisites. Verification is not optional but required as a foundation for validation results. Without verified code, validation comparisons cannot be interpreted credibly, as observed discrepancies between model predictions and comparator could reflect either model inadequacy or implementation errors. For the present work, verification activities are currently ongoing but have not yet been fully and systematically executed and documented. The model verification gap includes e.g.\ code correctness checks, numerical convergence tests for the RK23 solver across the operational time step range, cross-checks of conservation laws, and structured code review of submodel implementations against published equations and parameter definitions. This gap does not invalidate the validation results presented here, but it does indicate that the credibility argument is provisionally complete pending verification closure.\\ 
In ventilation, independent bench comparators are limited because common lung simulators are mostly restricted to lung mechanics or model-based themselves. Within these constraints, the credibility plan and execution are consistent with ASME V\&V~40 and FDA guidance expectations and ISCT workflows.\\
The applicability (fit-to-COU) score is good ($S_{\mathrm{app}}{=}2.5$), with strong calibration and plausibility (3.0 each), and solid population-based validation (2.0), while emergent behavior coverage is moderate (2.0). These results indicate that the PDM is fit-for-purpose for preclinical ISCT in automated weaning under the defined COU of pre-clinical testing and risk classification. \\
Overall, the approach demonstrates that general CM\&S credibility principles can be operationalized for mechanical ventilation when adapted to domain constraints. Warnaar et~al.~\cite{warnaar_computational_2023} highlight a persistent gap between model availability and comprehensive validation in ventilation, with a predominance of lung-mechanics CPMs and comparatively limited coverage of gas exchange, diaphragm function, and integrative cardiopulmonary control. Only a minority of studies meet high validation quality, and reporting is inconsistent~\cite{warnaar_computational_2023}. Within this context, the present work advances a ventilation-specific operationalization of credibility assessment under ASME V\&V~40 and FDA guidance and ISCT workflows, addressing the research gap identified in the introduction: namely, the need for a practical, domain-tailored framework that links ?oIs and COU to evidence categories and factors suitable for physiologic PDMs in ventilation~\cite{asme_2018,fda_guidance_2023,pathmanathan_credibility_2024}. The approach is analogous in spirit to precedents in other domains (e.g.\ UVA/Padova for diabetes)~\cite{cobelli_developing_2023,dalla_man_meal_2007}.\\

\subsection{Limitations, Conclusion and Outlook}
Despite the overall favorable applicability, residual gaps remain and would need to be closed to support higher-stakes decisions. Population-based validation would benefit from broader, stratified, multi-center cohorts with richer patient-level metadata to strengthen representativeness. For model calibration, the $R^2 \approx$ 94\,\%  for capnography, though higher than the gold-standard precision of laboratory blood gas analysis ($\pm2-3$~mmHg for \pCO), is considered adequate for the present preclinical assessment of automated weaning strategies. Capnography in clinical practice exhibits inherent measurement uncertainty (sensor latency $300-500$~ms, filtering artifacts, patient-ventilator asynchrony~\cite{lumb_nunns_2017}) that accounts for much of the $10\,\%-20\,\%$ observed error range. For preclinical ISCT of automated weaning protocols, model accuracy at this level is deemed sufficient to resolve protocol logic transitions (e.g.\ pressure reduction thresholds) and compare relative efficacy (weaning success, asynchrony incidence) across protocols. Plausibility is reinforced by expert endorsement. \\
In summary, the credibility activities undertaken are proportionate to a medium-low model-risk COU for preclinical ISCT in automated weaning, in line with ASME V\&V~40 and FDA guidance~\cite{asme_2018,fda_guidance_2023}. For higher model influence or greater decision consequence, escalation of evidence is warranted: expanded in vivo comparators with statistical power, external validation and stricter calibration statistics, explicit validation of measurement-chain effects and controller timing (e.g. hardware-/software-in-the-loop). Collectively, these steps would further align ventilation ISCT practice with regulator-ready credibility frameworks and help narrow the gap between general guidance~\cite{warnaar_computational_2023,pathmanathan_credibility_2024}. \\
This work presents a standards-aligned credibility assessment of a PDM for automated weaning in mechanical ventilation. Using ASME~V\&V~40 and FDA guidance as a foundation, the model was evaluated within a clearly defined context of use through a structured, multi-source evidence framework. Across ventilation-relevant quantities of interest (\vt, RR, \etco, and asynchronies, $P_\mathrm{plat}$), the collected evidence met the predefined acceptance criteria. Calibration against literature and clinical ranges, population-based assessment of physiological plausibility, and evaluation of emergent behavior consistently support the intended use. The resulting fit-to-COU score indicates good agreement, demonstrating that the model is fit for purpose for preclinical ISCTs in automated weaning.\\
Beyond the specific model, this study contributes a practical, domain-specific operationalization of general credibility frameworks that focused validation activities for physiological PDMs in mechanical ventilation — a domain where such implementations remain limited. By explicitly tailoring evidence categories, comparators, and quantities of interest to ventilation applications and documenting applicability and limitations, the presented workflow offers a reproducible and regulator-relevant pathway for credibility assessment, aligned with established practices in other physiological modeling domains.\\
A key advantage of the proposed patient-specific modeling framework is the combination of physiologically informed prior initialization with targeted
adaptation of those parameters that are identifiable from the available
measurements. Population- and patient-specific information, including external
estimates derived from demographic and clinical data, provides a plausible
initial physiological parameterization, while the available respiratory
measurements are used to adapt only those parameter directions that are
sufficiently informed by the data. The resulting individualized model demonstrated predictive capability across different patients, ventilatory inputs, operating conditions, and validation horizons. This combination of prior physiological knowledge and measurement-driven adaptation is particularly important for prospective use: the model does not need to reconstruct the complete latent physiological parameter space from the available measurements alone, but can update those components that are identifiable from incoming data. With sufficiently efficient online parameter adaptation, this framework could support real-time prediction and model-based optimization. Such developments could ultimately provide the basis for a physiological digital twin that continuously integrates prior patient information with incoming measurements, predicts short-term physiological responses, and supports clinical decision-making and medical-device automation.

\section{Framework for the Credibility Assessment of a Physiological Computational Model}
\label{sec:cred_assessment}
For the PDM to credibly support preclinical decision-making, its development and validation must follow established standards and be proportionate to its risk and context of use. This section outlines the structured, nine-step credibility assessment workflow that anchors the entire validation effort. Two complementary frameworks are employed: ASME~V\&V~40~\cite{asme_2018}, which prescribes verification, validation, and uncertainty quantification in proportion to model risk, FDA guidance~\cite{fda_guidance_2023} on computational modeling and simulation, which operationalizes credibility across evidence categories. Presenting the methodology ensures transparency and reproducibility, enabling readers to understand both the evidence categories considered and the risk-informed, COU-anchored rationale underlying their selection. \\
ASME~V\&V~40 prescribes that verification, validation, and uncertainty quantification (V\&V/UQ) be proportionate to model risk, which is determined by the model’s influence on the decision and the consequence if the decision is wrong, and that all credibility activities be anchored in a clearly stated question of interest, context of use, and quantity of interest. The FDA guidance operationalizes credibility through evidence categories that extend beyond code and calculation verification to include model plausibility, calibration, bench and in vivo comparison, population-level validation, and emergent phenomena, alongside documentation and traceability. ISCT-focused work further clarifies planning, adequacy assessment, and factor-specific considerations for patient models and virtual cohorts~\cite{pathmanathan_credibility_2024}. \\
The credibility assessment workflow proceeds in nine tests (see Figure~\ref{fig:credibility_workflow}), structured to ensure transparency and reproducibility while keeping rigor proportionate to model risk~\cite{asme_2018, fda_guidance_2023, pathmanathan_credibility_2024}:
\begin{enumerate}
	\item \textbf{State the ?oI:} Define the decision-relevant questions the ISCT using the model must answer for a defined use case (e.g.\ in mechanical ventilation: maintenance of lung-protective \vt and $P_\mathrm{plat}$, acceptable \vt, RR , and \etco, and stability of patient-ventilator interaction).
	\item \textbf{Describe the COU:} Specify the intended (pre-)clinical use, patient population stratification (e.g. ARDS/COPD, postoperative, neuromuscular), device mode and controller settings, monitoring signals (e.g. \etco, \spo), and explicit boundaries of applicability~\cite{asme_2018}.
	\item \textbf{Assess model risk:} Assess model influence (degree to which decisions depend on model outputs) and decision consequence (clinical and regulatory impact of error) to determine model risk, which guides the appropriate level of credibility goals~\cite{asme_2018}.
	\item \textbf{Identify credibility evidence:} Select relevant evidence categories (e.g.\ code and calculation verification, model plausibility, calibration, bench test validation, in vivo validation, population-based validation. Emergent behavior, documentation/traceability) based on ?oI and COU~\cite{fda_guidance_2023, pathmanathan_credibility_2024}.
	\item  \textbf{State credibility factors and gradations:} For the selected categories, specify credibility factors, state gradations and select credibility goals (levels of rigor commensurate with assessed model risk) appropriate to ventilation PDMs and ISCTs, following ISCT guidance and domain-tailored recommendations~\cite{pathmanathan_credibility_2024}.
	\item \textbf{Prospective adequacy assessment:}  Synthesize results across categories and factors, evaluate whether acceptance criteria are met for each QoI, identify any credibility gaps~\cite{pathmanathan_credibility_2024}.
	\item  \textbf{Execute studies and collect evidence:} Perform verification tests (code verification, numerical error/convergence), calibration to literature and clinical ranges, validation against comparator (population-based, in vivo when available), UQ and sensitivity analyses, and demonstrations of emergent phenomena; ensure complete documentation~\cite{fda_guidance_2023, pathmanathan_credibility_2024}.
	\item  \textbf{Post-study adequacy assessment:} Synthesize results across credibility evidence categories and credibility factors, evaluate whether acceptance criteria are met for each ?oI, identify any credibility gaps~\cite{pathmanathan_credibility_2024}.
	\item \textbf{Finalize the credibility report:}  Present a traceable account of methods, datasets, protocols, analyses, and conclusions aligned with standards and guidance~\cite{fda_guidance_2023}.
\end{enumerate}
\noindent
These nine steps — from COU definition through final credibility assessment reporting — provide a structured workflow that guards against ad hoc validation choices and ensures proportionality between evidence rigor and model risk. 

\section*{Acknowledgement}
The authors would like to thank Pras Pathmanathan (U.S. Food and Drug Administration (FDA)) for his valuable feedback and thoughtful review of an earlier version of this manuscript. His comments and suggestions greatly helped to improve the quality of this work.\\
To improve grammar, style, and overall readability of the manuscript, the authors used
the following AI-based tools: DeepL Translator (DeepL SE, Cologne, Germany) and Fh-Genie (Fraunhofer-Gesellschaft zur Förderung der angewandten Forschung e.V., Munich, Germany). The authors have carefully reviewed all generated content and take full responsibility for the content of this work.
\section*{Funding}
C. Hennigs, C. Danielson, F. Bilda, H. Selpien, and G. Männel were supported by the German Federal Ministry of Research, Technology and Space (BMFTR) through the CMS4Vent Project (FKZ: 03LW0469K, 03LW0470, 03LW0471). D. Karachalios was supported  by the German Federal Ministry of Research, Technology and Space (BMFTR) through the KiMeKo Verbundprojekt (FKZ: 01$\vert$S24056A). C. Hennigs, F. Bilda, G. Männel, F. Spitzenberger and P. Rostalski were funded by the European Union - European Regional Development Fund (ERDF), the Federal Government and Land Schleswig Holstein, Individualisierte Medizintechnik für bildgestützte, robotische Interventionen (IMTE 2) (No. 125 24 009).
\section*{Conflict of Interest}
C. Danielson is also employed at Drägerwerk AG \& Co. KGaA, Lübeck, Germany but has contributed to this projection solely in her capacity as regulatory expert as research associate at the Fraunhofer IMTE.
\section*{Authors' contributions}
Conceptualization was undertaken by C.H., C.D., F.B., G.M. and P.R.; data curation by C.H., C.D., F.B., D.K., and H.S.; formal analysis by C.H., D.K., and H.S.; funding acquisition by C.H., F.B., G.M., D.S., F.S., and P.R.; investigation by C.H., D.K. and H.S.; methodology by C.H., C.D., F.B., D.K., and G.M.; project administration by F.B., D.S., F.S., and P.R.; resources by D.S., F.S., and P.R.; software by C.H., F.B., D.K., and N.H.; supervision by G.M., D.S., F.S., and P.R.; validation by C.H., C.D., D.K., and N.H.; and visualization by C.H. and D.K. The original draft of the manuscript was prepared by C.H., C.D and D.S. All authors contributed to manuscript review and editing.
\section*{Ethical Approval and Consent to participate}
This prospective observational single-center study was conducted in three surgical intensive care units of the University Hospital Schleswig-Holstein, Campus Kiel, Kiel, Germany. The study was approved by the ethics committee of the Christian-Albrechts University Kiel, Germany (approval D411/13). Need for informed consent was waived by the ethics committee. Study data were analyzed anonymously. 

\printbibliography
\appendix

\section{Technical Description of the Patient--Device Model}
\label{app:pdm}

\subsection{Purpose, Scope, and Model Requirements}
\label{app:pdm-scope}

The patient--device model (PDM) provides a physiologically plausible
simulation environment for the preclinical in silico assessment of
automated ventilator functions. In the present study, it is used to
evaluate an automated weaning protocol (AWP) in adult, invasively
ventilated patients under CPAP/CSV-PS conditions.

The PDM integrates respiratory mechanics, pulmonary gas exchange,
respiratory control, a ventilator representation, and a capnography
measurement model. It is intended to reproduce clinically relevant
relationships between ventilator settings, patient-specific physiology,
and observable respiratory signals.

The AWP is not part of the PDM. Instead, it constitutes the external
test article assessed in the ISCT. The AWP receives
processed model outputs, including tidal volume $V_{\mathrm{T}}$,
respiratory rate $RR$, and end-tidal carbon dioxide
$et\mathrm{CO_2}$, and returns the requested inspiratory pressure
$P_{\mathrm{insp}}$ to the ventilator model.

A set of functional (F), physiological (P), and clinical (C) requirements was formulated based on the COU. The requirements are listed in Table~\ref{tab:req}.
\begin{table*}[t]

\centering
\renewcommand{\arraystretch}{1.2}
\caption{Patient--device model requirements derived from the prospective ISCT use case.}
\label{tab:req}

\begin{tabular}{>{\centering\arraybackslash}p{1.8cm} p{12.5cm}}
\toprule
\textbf{ID} & \textbf{Requirement} \\
\midrule
\textcolor{darkblue}{\textbf{AWP--F1}} & The model \textit{shall} take the airway pressure \paw, the patient's weight, and COPD diagnosis as input. \\
\textcolor{darkblue}{\textbf{AWP--F2}} & The model \textit{shall} output the airway flow measurement $\dot{V}$. \\
\textcolor{darkblue}{\textbf{AWP--F3}} & The model \textit{shall} output a mainstream capnogram measurement $F_{\mathrm{CO_2}}$. \\
\textcolor{darkblue}{\textbf{AWP--F4}} & The model \textit{shall} implement a closed-loop respiratory control system linking neural oscillation, mechanoreceptor feedback, chemoreceptor feedback, respiratory muscle activation, and ventilator triggering. \\
\addlinespace
\textcolor{darkblue}{\textbf{AWP--P1}} & The model \textit{shall} reproduce normal lung mechanics of a healthy adult. \\
\textcolor{darkblue}{\textbf{AWP--P2}} & The model \textit{shall} be configurable to reproduce characteristic lung mechanics of COPD, including increased expiratory resistance. \\
\textcolor{darkblue}{\textbf{AWP--P3}} & The model \textit{shall} be configurable to reproduce characteristic lung mechanics of ARDS, including reduced compliance and nonlinear pressure–volume relationships. \\
\textcolor{darkblue}{\textbf{AWP--P4}} & The model \textit{shall} be configurable to simulate V/Q mismatch. \\
\textcolor{darkblue}{\textbf{AWP--P5}} & The model \textit{shall} reproduce ventilatory responses to CO$_2$ and O$_2$ changes. \\
\textcolor{darkblue}{\textbf{AWP--P6}} & The model \textit{shall} be configurable to reproduce different sedation levels. \\
\addlinespace
\textcolor{darkblue}{\textbf{AWP--C1}} & The model \textit{shall} keep respiratory rate RR, tidal volume \vt, and minute ventilation MV within clinical ranges under pressure support levels of 0 -- 25 mbar. \\
\textcolor{darkblue}{\textbf{AWP--C2}} & The incidence of double triggering, missed triggering, double effort, delayed triggering, and synchronous triggering \textit{shall} match clinical data under pressure support ventilation. \\
\bottomrule
\end{tabular}
\end{table*}

\subsection{Overall Architecture and Signal Flow}
\label{app:pdm-architecture}

\begin{figure}[t]
\centering
\resizebox{0.49\textwidth}{!}{%
\begin{tikzpicture}[
    x=1cm,
    y=1cm,
    line cap=round,
    line join=round,
    pdm patient box/.style={
        vv box,
        minimum width=3.30cm,
        minimum height=0.70cm,
        fill=modelGreenLight,
        draw=modelGreenDark,
        text=modelGreenDark
    },
    pdm device box/.style={
        vv box,
        minimum width=3.30cm,
        minimum height=0.70cm,
        fill=white,
        draw=vvNavy,
        text=vvNavy
    },
    pdm signal/.style={
        font=\sffamily\fontsize{5.4}{6.0}\selectfont,
        text=vvNavy,
        fill=vvPale!55,
        inner sep=1.0pt,
        align=center
    }
]

\path[fill=vvPale!55] (0.00,0.20) rectangle (7.60,4.10);
\path[fill=modelGreenDark] (0.00,3.68) rectangle (7.60,4.10);

\draw[phase frame] (0.00,0.20) rectangle (7.60,4.10);
\draw[phase frame] (0.00,3.68) -- (7.60,3.68);

\node[phase title] at (3.90,3.89) {
    Patient--device model architecture
};

\node[pdm patient box] (rc) at (1.80,2.90) {
    \workflowlabel{Respiratory center}
};

\node[pdm patient box] (lm) at (1.80,1.78) {
    \workflowlabel{Lung mechanics}
};

\node[pdm patient box] (ge) at (5.80,1.78) {
    \workflowlabel{Gas exchange}
};

\node[pdm device box] (vent) at (1.80,0.68) {
    \workflowlabel{Ventilator model}
};

\node[pdm device box] (cap) at (5.80,0.68) {
    \workflowlabel{Capnogram model}
};


\draw[subflow]
    (rc.south) --
    node[pdm signal, right, pos=0.50] {$P_{\mathrm{mus}}$}
    (lm.north);

\draw[subflow]
    (lm.east) --
    node[pdm signal, above, pos=0.50] {$V,\,\dot{V}$}
    (ge.west);

\draw[subflow]
    (ge.north)
    to[out=90, in=0, looseness=1.00]
    node[pdm signal, below, pos=0.48]
        {$P_{\mathrm{O_2}},\,P_{\mathrm{CO_2}}$}
    (rc.east);


\draw[flow]
    ([xshift=-4pt]vent.north) --
    node[pdm signal, left, pos=0.48] {$P_{\mathrm{aw}}$}
    ([xshift=-4pt]lm.south);

\draw[flow]
    ([xshift=4pt]lm.south) --
    node[pdm signal, right, pos=0.48] {$\dot{V}$}
    ([xshift=4pt]vent.north);

\draw[flow]
    (ge.south) --
    node[pdm signal, right, pos=0.50] {$f_{\mathrm{CO_2}}$}
    (cap.north);

\draw[flow]
    (cap.west) --
    node[pdm signal, above, pos=0.50] {$F_{\mathrm{CO_2}}$}
    (vent.east);

\end{tikzpicture}%
}
\caption{
Detailed architecture and signal flow of the coupled patient--device model.
Green boxes denote physiological patient submodels, whereas navy boxes denote
device and measurement components. $P_{\mathrm{mus}}$: respiratory muscle
pressure; $P_{\mathrm{aw}}$: airway pressure; $V$: lung volume;
$\dot{V}$: airway flow; $P_{\mathrm{O_2}}$ and $P_{\mathrm{CO_2}}$:
oxygen and carbon dioxide partial pressures; $f_{\mathrm{CO_2}}$:
alveolar carbon dioxide fraction; $F_{\mathrm{CO_2}}$: mainstream
capnogram signal.
}
\label{fig:pdm-architecture}
\end{figure}
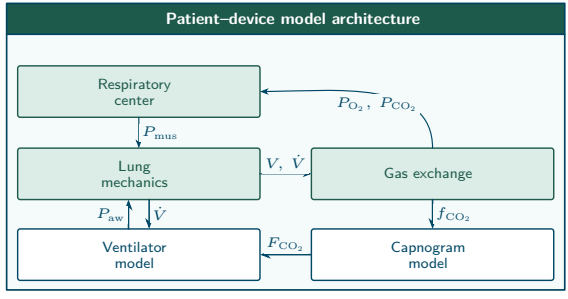

The PDM consists of five coupled components: lung mechanics, gas
exchange, respiratory control, ventilator dynamics, and measurement
models. The resulting model architecture and signal flow are summarized in Figure~\ref{fig:pdm-architecture}. The respiratory center generates neural drive and respiratory
muscle pressure. Respiratory muscle pressure and ventilator-generated
airway pressure determine lung volume and airway flow. Gas exchange
depends on the resulting ventilation and pulmonary blood flow, while
blood-gas values feed back to the respiratory center through chemical
control loops.

\subsection{Respiratory Mechanics}
\label{app:pdm-lung-mechanics}

The respiratory mechanics submodel represents the mechanical interaction between the ventilator, the respiratory muscles, the airways, the lungs,
and the chest wall~\cite{bates_lung_2009, hennigs_EFL_2024}. It describes airway flow $\dot{V}$, lung volume $V$, airway resistance, lung elastance, chest-wall elastance, and respiratory
muscle pressure $P_{\mathrm{mus}}$.

The model supports healthy respiratory mechanics as well as
pathophysiological configurations. Increased expiratory airway resistance can be used to represent obstructive disease patterns such as chronic obstructive pulmonary disease (COPD)~\cite{hennigs_copd_2023}. Reduced compliance and nonlinear pressure--volume relationships can be configured to represent restrictive or ARDS-like conditions~\cite{bates_lung_2009}.

The respiratory mechanics component is dynamically coupled to the ventilator model through airway pressure $P_{\mathrm{aw}}$ and to the gas-exchange model through lung volume and alveolar ventilation.

\subsection{Gas Exchange and Acid--Base Homeostasis}
\label{app:pdm-gas-exchange}

The gas-exchange submodel represents oxygen and carbon dioxide transport between the lungs, pulmonary circulation, systemic tissues, and the brain~\cite{ben-tal_simplified_2006, chiari_comprehensive_1997, hennigs_mathematical_2022}.
It includes lumped representations of alveolar gas, anatomical shunt, systemic tissue compartments, and a cerebral compartment.

The model accounts for anatomical and functional dead space, pulmonary shunt, diffusion limitation, metabolic oxygen consumption, carbon dioxide
production, and blood transport. Acid--base homeostasis is represented by simplified buffering relations and carbon dioxide kinetics~\cite{chiari_comprehensive_1997, loeppky_relationship_1993}. The resulting partial pressures of oxygen and carbon dioxide provide feedback to the respiratory control model~\cite{hennigs_respcenter_2025}.

Cardiac output and heart rate are fixed within an individual simulation scenario because a complete cardiovascular model is outside the present context of use.

\subsection{Respiratory Control and Respiratory Muscle Model}
\label{app:pdm-respiratory-control}

The respiratory control model consists of chemical and mechanical feedback
loops, a neural oscillator generating the breathing rhythm, and a respiratory muscle model~\cite{hennigs_respcenter_2025}. Central and peripheral chemoreflex pathways respond to changes in carbon dioxide, oxygen, and pH. Mechanical feedback
includes lung-volume-related and flow-related reflex pathways.

The resulting neural drive determines respiratory muscle activation and therefore $P_{\mathrm{mus}}$. This allows spontaneous breathing and patient--ventilator interaction to emerge from the coupling between neural drive, respiratory mechanics, and ventilator support.

The model can be configured to represent altered respiratory drive, sedation, respiratory failure severity, and respiratory muscle weakness.

\subsection{Ventilator, Measurement, and Controller Interfaces}
\label{app:pdm-ventilator-interface}

The ventilator model represents CPAP/CSV-PS ventilation with flow triggering. It receives the inspiratory pressure command $P_{\mathrm{insp}}$ from the AWP and generates the airway pressure $P_{\mathrm{aw}}$ applied to the patient model. Relevant ventilator settings include pressure support, PEEP, trigger sensitivity, rise time, and cycling-off criteria.

The model provides airway flow $\dot{V}$ and a mainstream capnogram signal $F_{\mathrm{CO_2}}$. A data-processing layer derives clinically relevant variables, including $V_{\mathrm{T}}$, $RR$, and $et\mathrm{CO_2}$, which are supplied to the AWP.

Unless measurement-noise scenarios are explicitly activated, the reported simulation outputs are noise-free. Capnography dynamics can nevertheless be represented through inspiratory and expiratory response time constants.

\subsection{Patient Parameterization and Represented Conditions}
\label{app:pdm-parameterization}

Each simulation uses a patient-specific parameter set containing anatomical, physiological, pathophysiological, ventilator, and measurement parameters.
The parameterization includes respiratory mechanics, gas-exchange, metabolic, respiratory-control, and ventilator settings.

The model can represent healthy respiratory mechanics as well as clinically relevant deviations, including COPD-related airway obstruction, ARDS-like
reduced compliance, ventilation--perfusion mismatch, increased shunt, increased dead space, sedation, and altered respiratory drive.

Patient-specific parameters may be selected from literature-derived values, estimated from available clinical data, or varied systematically to generate simulation scenarios and virtual patient profiles.

\subsection{Numerical Formulation and Implementation}
\label{app:pdm-implementation}

The PDM is a mechanistic, physics-based, nonlinear dynamical model. The
physiological submodels are formulated as coupled ordinary differential equations describing respiratory mechanics, gas exchange, transport delays, and respiratory-control dynamics.

The model is implemented in Python using NumPy and SciPy. Numerical integration is performed using an adaptive Runge--Kutta 2(3) method with variable step size. Model states and output signals are stored at 100~Hz to provide sufficient temporal resolution for analyses of respiratory mechanics and patient--ventilator interaction.

The model is predominantly deterministic. A stochastic component may be used in breath generation to reproduce physiological breath-to-breath variability when required by the simulation scenario.

\subsection{Assumptions and Known Limitations}
\label{app:pdm-limitations}
The PDM is a deterministic, mechanistic, physics-based model that represents human respiratory mechanics, gas-exchange and respiratory center physiology and dynamically couples them to a device-specific ventilator control model. It is implemented in Python (Version 3.11.11) with NumPy (Version 2.2.2) and SciPy (Version 1.15.1) libraries. A patient-specific class, parameterized according to anatomical and physiological attributes following Bilda et al.~\cite{bilda_unified_2024}, organizes all state variables and derived quantities. \\
The physiological submodel is expressed as a set of ordinary differential equations (ODEs) describing lung mechanics, tissue compartments, transport delays, and gas exchange processes. Numerical integration is performed using the adaptive Runge-Kutta 2(3) (RK23) method with variable step size, which balances computational efficiency with local error control. Model states and outputs are recorded at 100~Hz to ensure sufficient temporal resolution for analyses of ventilatory mechanics and patient-ventilator interaction.\\
The model assumes an intubated patient with no airway leakage, and a compartmental lung structure in which pressure, volume, and gas concentrations are homogeneous within each compartment. Tissue mechanics are represented by either constant or simplified nonlinear relationships. Gas concentrations in alveolar and blood compartments are assumed uniform, and spontaneous breathing is included, as the patient is not fully sedated and maintains intrinsic respiratory drive. Neurological control is assumed normal, excluding disorders that would alter the neural drive to breathe. Ventilator dynamics are idealized, representing CPAP/CSV-PS mode with flow triggering. The ventilator parameters can be fixed or varied depending on the simulation scenario. Gas exchange is modeled using lumped parameters, capturing essential shunt, V/Q mismatch, and diffusion limitations, while acid-base balance is approximated using simplified buffering equations. The model is deterministic, except for a stochastic component in breath generation to reproduce physiological variability, and all outputs are noise-free.\\
Several limitations of the model should be noted. The compartmental lung representation used in this in silico model cannot capture regional heterogeneity in ventilation or perfusion, and dynamic airway collapse, mucus, or nonlinear flow phenomena are not included. Gas exchange and acid-base models are simplified, which may reduce accuracy in pathological conditions such as severe shunt or impaired diffusion. Ventilator control is idealized, and patient-ventilator asynchrony due to fatigue, behavioral changes, or neural variability is not fully represented. Model parameterization relies on literature values or nominal anatomy, limiting the ability to reproduce individual patient variability. 

\subsection{Model Inputs, Outputs, States, and Parameters}
\label{app:pdm-inputs-outputs}
The complete model parameter set, including nominal values, units, and literature sources, is provided in Table~\ref{tab:model_input_parameter}.



\subsection{Governing Equations}
\label{app:pdm-equations}

The governing equations for lung mechanics, gas exchange, acid--base
homeostasis, respiratory control, ventilator dynamics, and capnography are
provided in Table~\ref{tab:governing_eq_results}. Equation-specific literature sources and evidence grades are reported to ensure traceability of the model formulation.

\onecolumn
\small
\begin{longtable}{p{1.1cm} p{7cm} >{\centering\arraybackslash}p{1.6cm} >{\centering\arraybackslash}p{2.2cm} >{\centering\arraybackslash}p{1.2cm}}
	\caption{Baseline patient model input parameters.}\label{tab:model_input_parameter}\\
	\toprule
	\textbf{Param}. & \textbf{Description} & \textbf{Value} & \textbf{Unit} & \textbf{Literatur} \\
	\midrule
	\endfirsthead
	
	\multicolumn{5}{c}{\tablename\ \thetable\ -- continued from previous page} \\
	\toprule
	\textbf{Param}. & \textbf{Description} & \textbf{Value} & \textbf{Unit} & \textbf{Literatur} \\
	\midrule
	\endhead
	
	\bottomrule
	\multicolumn{5}{r}{continued on next page} \\
	\endfoot
	
	\bottomrule
	\endlastfoot
	
	\multicolumn{5}{@{}l}{\textbf{Lung Mechanics (LM)}} \\ \midrule
	$E_L$ & Lung Elastance & 13 & mmHg/L & \cite{merrath_lungmechanics_2023} \\
	$E_\mathrm{CW}$ & Chest wall elastance & 8 & mmHg/L & \cite{merrath_lungmechanics_2023} \\
	$R$ & Airway Resistance & 2 & mmHg/L/s & \cite{merrath_lungmechanics_2023} \\
	$R_\mathrm{ua}$ & Upper airway resistance & 1 & mmHg/L/s &  \\
	$V_\mathrm{crit}$ & Critical volume during expiration for flow limitations & $0.5\cdot \vt$ \\
	\midrule
	\multicolumn{5}{@{}l}{\textbf{Gas Exchange (GE)}} \\
	\midrule
	$Q$ & Cardiac output & 5 & L/min & \cite{merrath_gasexchange_2023} \\
	$Q_\mathrm{b}$ & Brain blood flow & 0.7 & L/min & \cite{chiari_comprehensive_1997} \\
	$Q_\mathrm{t}$ & Tissue blood flow & 4.3 & L/min & \cite{chiari_comprehensive_1997} \\
	$\mathrm{SV_r}$ & Pulmonary stroke volume & 0.07 & L & \cite{ben-tal_simplified_2006} \\
	$V_D$ & Anatomical dead space & 0.15 & L & \cite{lumb_nunns_2017} \\
	FRC & Functional residual capacity & 2.5 & L & \cite{lumb_nunns_2017} \\
	$f_s$& Anatomical shunt & 0.04 & & \cite{lumb_nunns_2017}\\
	$f_\mathrm{V_D}$ & Functional dead space & $0 - 0.5$ & \\
	$f_\mathrm{fs}$& Functional shunt & $0 - 0.5$ &\\
	$V_\mathrm{B}$ & Body volume& 38.74 & L& \cite{chiari_comprehensive_1997}\\
	$V_\mathrm{b}$ & Brain tissue volume & 0.90 & L& \cite{chiari_comprehensive_1997}\\
	$\mathrm{MP_{t, O_2}}$ & Metabolic rate O$_2$ & 0.25 & L/min & \cite{merrath_gasexchange_2023, lumb_nunns_2017} \\
	$\mathrm{MP_{t, CO_2}}$ & Metabolic rate CO$_2$ & 0.21 & L/min & \cite{merrath_gasexchange_2023, lumb_nunns_2017} \\
	$\mathrm{MP_{{b,O_2}}}$ & Brain metabolic rate O$_2$ & 0.05 & L/min& \cite{chiari_comprehensive_1997} \\
	$\mathrm{MP_{{b,CO_2}}}$ & Brain metabolic rate CO$_2$& 0.05 & L/min & \cite{chiari_comprehensive_1997}\\
	$D_\mathrm{L, O_2}$ & Diffusion capacity of O$_2$ & 0.00056  & L/(s$\cdot$mmHg)& \cite{merrath_gasexchange_2023}\\
	$D_\mathrm{L, CO_2}$ & Diffusion capacity of CO$_2$ & 7.08$ \cdot 10^{-4}$ & L/(s$\cdot$mmHg)& \cite{ben-tal_simplified_2006}\\
	$\sigma_\mathrm{O_2}$ & Solubility of O$_2$ in plasma & 1.4$\cdot 10^{-6}$ & mol/(L$\cdot$mmHg) &  \cite{ben-tal_simplified_2006}\\
	$\sigma_\mathrm{CO_2}$ & Solubility of CO$_2$ in plasma & 3.3$\cdot 10^{-5}$ & mol/(L$\cdot$mmHg) &  \cite{ben-tal_simplified_2006}\\
	$P_\mathrm{atm}$ &  Atmospheric pressure & 760 & mmHg \\
	$p_w$ & Vapor pressure of Water at $37$°C & 47 & mmHg&  \cite{ben-tal_simplified_2006}\\
	$h$ & Concentration of H$^+$ & $10^{-7.4}$ &  mol/L&  \cite{ben-tal_simplified_2006}\\
	$r_2$& Dehydration reaction rate & 0.12 &  1/s&  \cite{ben-tal_simplified_2006}\\
	$l_2$ & Hydration reaction rate &  164$ \cdot 10^3$ & L/s $\cdot$ mol&  \cite{ben-tal_simplified_2006}\\
	$\delta$ &  Acceleration rate & $10^{1.9}$ & & \cite{ben-tal_simplified_2006}\\
	$T_h$ & Concentration of hemoglobin &  2 $\cdot 10^{-3}$ & mol/L &  \cite{ben-tal_simplified_2006}\\
	\fio & Inspired concentration of O$_2$ & $0.21 - 1$ &  \\
	\fico & Inspired concentration of CO$_2$ & $\approx 0~$& \\
	$\tau_\mathrm{CSF}$ & Time constant cerebral spine fluid compartment & 320 & s & \cite{chiari_comprehensive_1997} \\
	$k_1$ & O$_2$-carrying capacity of haemoglobin & 1.32 $\cdot 10^{-3}$ & L/g & \cite{chiari_comprehensive_1997}\\
	$k_2$ & O$_2$ solubility coefficient &  3.03 $\cdot 10^{-5}$ & 1/mmHg & \cite{chiari_comprehensive_1997}\\
	$k_3$ & P$_{50}$-like parameter & 26.6 & mmHg & \cite{chiari_comprehensive_1997}\\
	$n$ & Hill coefficient & 2.6 & & \cite{chiari_comprehensive_1997} \\
	$K_T$ & Equilibrium constants in the saturation function of hemoglobin & 10 $\cdot 10^{3}$ & L/mol & \cite{unzai_rate_1998}\\
	$K_R$ & Equilibrium constants in the saturation function of hemoglobin & 3.6 $\cdot 10^{6}$ & L/mol& \cite{unzai_rate_1998}\\
	$L$ & Equilibrium constants in the saturation function of hemoglobin & 171.2 $\cdot 10^{6}$ & & \cite{roughton_accurate_1973, severinghaus_simple_1979}\\	
	\midrule
	\multicolumn{5}{@{}l}{\textbf{Respiratory Center (RC)}} \\
	\midrule
	$a_\mathrm{ch}$   & Sensitivity of total chemoreflex drive & 1  & & \cite{botros_neural_1990} \\
	$b_\mathrm{ch}$   & Total chemoreflex drive constant & -4 &  & \cite{botros_neural_1990}\\
	$G_C$  & Central chemoreflex gain  & 27.72 & exc/$\Delta$pH  \\
	$G_\mathrm{CO_2}$   & Peripheral chemoreflex gain CO$_2$ & 0.08 & exc/kPa \\
	$G_\mathrm{O_2}$   & Peripheral chemoreflex gain O$_2$  & 0.2 & exc/kPa \\
	$T_P$   & Peripheral chemoreceptor time constant & 10 & s & \cite{botros_neural_1990}\\
	$T_C$   & Central chemoreceptor time constant  & 100 & s & \cite{botros_neural_1990}\\
	$K_I$  & Inspiratory muscle neural gain  & 1.87  & mmHg & \\
	$K_E$  & Expiratory muscle neural gain  & 1.87 & mmHg & \\
	$K_{\mathrm{PRS}}$  & Pulmonary stretch receptor gain  & 7.5 & exc/L & \cite{botros_neural_1990} \\
	$K_{\mathrm{insp}}$ & Inspiratory flow reflex neural gain  &  0.1 & exc$\cdot$s/L \\
	$K_{\mathrm{exp}}$ & Expiratory reflex neural gain  & 0.7 & exc$\cdot$L/(mmHg$\cdot$s) \\
	$K_{\mathrm{ua}}$ & Upper airway reflex neural gain & 0.1 & exc/mmHg \\
	$\sigma_\mathrm{chem}$ & Noise to chemoreceptors & 0.3 & mmHg & \cite{JAWORSKI2019148}\\
	$\sigma_\mathrm{Dt}$ & Noise to chemical drive & 0.1 &  & \cite{JAWORSKI2019148} \\        
	$R_\mathrm{th}$& Upper airway threshold resistance &  20 - 50 & mmHg$\cdot$s/L\\
	$\rho_{\mathrm{resp}}$ & Respiratory failure scaling factor & 0 - 1 & \\
	$\rho_{\mathrm{sed}}$ & Sedation scaling factor & 0 - 1 & \\
	$\tau_\mathrm{ua}$ & Time constant for upper airway resistance change & 1 & s\\
	$E_\mathrm{exp}$ & Threshold value for active expiration & 5.2 &  exc & \\
	$\theta_{\mathrm{chem}}$ & Set point of respiratory center for \paco & 40 & mmHg & \cite{lumb_nunns_2017}\\
	$\theta_\mathrm{pH_m}$ & Set point of respiratory center for pH$_m$ & 7.32 & & \cite{lumb_nunns_2017} \\
	$\theta_\mathrm{pH_a}$ & Set point of respiratory center for pH$_a$ & 7.4 & & \cite{lumb_nunns_2017}\\
	$P_\mathrm{a}\mathrm{O_{2,th}}$ & Threshold partial pressure of arterial O$_2$ & 75 & mmHg & \cite{lumb_nunns_2017} \\
	$k_\mathrm{cr}$ & Constant characterizing the central receptors &  1.139 & $\mathrm{min}^{1/2}\cdot \mathrm{L}^{1/2}$ & \cite{chiari_comprehensive_1997} \\
	$\mathrm{RR}_{\mathrm{base}}$& Baseline respiratory rate at zero chemical drive & 10 & min$^{-1}$ & \cite{reynolds_transient_1972,reynolds1973transient} \\
	$\mathrm{RR}_{\mathrm{amp}}$ & Amplitude factor of RR increase with relative chemical drive & 2.1 & min$^{-1}$ & \cite{reynolds_transient_1972,reynolds1973transient} \\
	$k_{\mathrm{RR,sens}}$ & Sensitivity of RR to relative chemical drive in exponential fit & 1.8 & & \cite{reynolds_transient_1972,reynolds1973transient} \\
	$\mathrm{RR}_{\mathrm{off}}$ & Offset used in normalization of RR to $K_\mathrm{RR}$ & 0.1212 & min$^{-1}$ & \cite{reynolds_transient_1972,reynolds1973transient} \\
	$\mathrm{RR}_{\mathrm{sc}}$ & Scaling factor used in normalization of RR to $K_\mathrm{RR}$ & 6.545 & min$^{-1}$ & \cite{reynolds_transient_1972,reynolds1973transient} \\
	$V_\mathrm{HB}$ & Threshold volume for Hering-Breuer-Reflex & $1-1.5$ & L & \cite{takashi_nishino_physiological_2000} \\
	$f_{\mathrm{P,sub}}$ & Fraction of maximum inspiratory muscle pressure reached during normal resting breathing and moderate chemostimulation & 0.25 	&  &\cite{fogarty_breathing_2018} \\
	$N_\mathrm{exp}$   & Threshold neural activity of expiratory group for onset of active expiration & 5 & – & \\
	\midrule
	\multicolumn{5}{@{}l}{\textbf{Ventilator (V)}} \\
	\midrule
	$K_p$ & Proportional gain of the ventilator PI controller & $1.0-5.0$ & L/s/mmHg & \cite{leonhardt_2016} \\
	$K_i$ & Integral gain of the ventilator PI controller & $0.1-1.0$ & L/s/mmHg/s & \cite{leonhardt_2016} \\
	$C_h$ & Compliance of ventilator tubing and patient circuit & $0.002-0.005$ & L/cmH$_2$O & \cite{leonhardt_2016} \\
	$t_\mathrm{insp}$ & Set inspiratory time in pressure-controlled breaths &  & s &  \\
	$t_\mathrm{exp}$ & Set expiratory time in pressure-controlled breaths &  & s &  \\
	$t_\mathrm{sl}$ & Pressure rise (slope) time from PEEP to $P_\mathrm{insp}$ & $0.1-0.3$ & s & \\
	$\alpha$ & Cycling-off threshold as fraction of peak inspiratory flow & $0.2-0.5$ & -- &  \\
	$\beta$ & Flow-trigger sensitivity for assisted breaths & $0.5-2.0$ & L/min &  \\
	$t_\mathrm{delay}$ & Maximum acceptable trigger delay between effort and support & $0.25$ & s & \\
	$t_\mathrm{sample}$ & Sampling interval of the ventilator state machine & $0.01$ & s & -- \\
    \midrule
	\multicolumn{5}{@{}l}{\textbf{Measurements (M)}} \\
	\midrule
	$\Delta t_d$          & Lung-to-finger circulation time for \spo measurement          & $20$          & s          & \cite{herrmann_virtual_2025} \\
	$\sigma_{\text{SpO}_2}$ & Standard deviation of \spo measurement noise                 & $1$           & \%         & \cite{herrmann_virtual_2025} \\
	$\tau_\mathrm{insp}$  & Time constant of capnography response during inspiration         & $0.05$        & s          & -- \\
	$\tau_\mathrm{exp}$   & Time constant of capnography response during expiration          & $0.15$        & s          & -- \\
	$\sigma_\mathrm{noise}$ & Standard deviation of esophageal pressure measurement noise     & $0.5$         & mmHg       & -- \\
   $Q_{\mathrm{switch}}$ & Capnogram flow-switch threshold & 0.2 & L/s & -- \\
\end{longtable}
\normalsize

\begin{longtable}{>{\raggedright\arraybackslash}p{0.7cm} 
			>{\raggedright\arraybackslash}p{3cm} 
			>{\raggedright\arraybackslash}p{8cm} 
			>{\raggedright\arraybackslash}p{1.2cm}
			>{\raggedright\arraybackslash}p{1cm}}
		\caption{Model governing equations of the respiratory system model with literature reference and grading.} \\
		\label{tab:governing_eq_results} \\
		\toprule
		\textbf{ID} & \textbf{Description} & \textbf{Equation} & \textbf{Lit.}  & \textbf{Grade}\\
		\midrule
		\endfirsthead		
		\multicolumn{5}{c}{{\tablename\ \thetable{} -- Model governing equations.}} \\
		\toprule
		\textbf{ID} & \textbf{Description} & \textbf{Equation} & \textbf{Lit.} & \textbf{Grade}\\
		\midrule
		\endhead		
		\bottomrule
		\multicolumn{5}{r}{{...}} \\
		\endfoot		
		\bottomrule
		\endlastfoot		
		LM1 & 1- Comp. Lung mechanics model for COPD & $\paw(t) = R(V) \cdot \dot{V}(t) + E \cdot V(t) - \pmus(t)$, \begin{numcases}{R(V)=}
			\nonumber R_2, & if $\dot{V}(t) \leq 0$ and $V(t) \leq \vcrit$\\
			\nonumber R_1, & else.
		\end{numcases} & \mbox{\cite{Hennigs_COPD_2024, suki_mathematical_2025}}  & 7\\
		\midrule
		GE1 & Gas exchange CO$_2$ in lungs & $	\frac{d \fco(t)}{d t} = \frac{1}{V(t)}\Bigl(D_\mathrm{L,CO_2}\bigl(\pcco(t) - \palvco\bigr) + \bigl(f_\mathrm{insp,CO_2} - \fco(t)\bigr)\,\dot{Q}_{A_i} - \fco(t)\bigl(D_\mathrm{L,O_2}\bigl(\pco(t) - \palvo\bigr) + D_\mathrm{L,CO_2}\bigl(\pcco(t) - \palvco\bigr)\bigr)	\Bigr),$ & \mbox{\cite{ben-tal_simplified_2006}} & 5\\[3pt]
		GE2 & Gas exchange O$_2$ in lungs &  $	\frac{d \fo(t)}{d t} = \frac{1}{V(t)}
		\Bigl(	D_\mathrm{L,O_2}\bigl(\pco(t) - \palvo\bigr) + \bigl(f_\mathrm{insp,O_2} - \fo(t)\bigr)\,\dot{Q}_{A_i}	- \fo(t)\bigl(D_\mathrm{L,CO_2}\bigl(\pcco(t) - \palvco\bigr) + D_\mathrm{L,O_2}\bigl(\pco(t) - \palvo\bigr)	\bigr)	\Bigr)$  & \mbox{\cite{ben-tal_simplified_2006}}  & 5\\[3pt]
		GE3 & Gas exchange CO$_2$ in capillaries & $\frac{d \pcco(t)}{d t} = \frac{D_\mathrm{L,CO_2}}{\sigma_\mathrm{CO_2} \cdot V_c} (\palvco(t) - \pcco(t)) + \frac{d \cdot l_2}{\sigma_\mathrm{CO_2}} h \cdot [\mathrm{HCO_3^-}](t) - \delta \cdot r_2 \cdot \pcco(t)$  & \mbox{\cite{ben-tal_simplified_2006}}  & 5\\[3pt]
		GE4 & Gas exchange z in capillaries & $	\frac{d [\mathrm{HCO_3^-}](t)}{d t} = \delta \cdot r_2 \cdot \sigma_\mathrm{CO_2} \cdot \pcco(t) - \delta \cdot l_2 \cdot h \cdot [\mathrm{HCO_3^-}](t)$  & \mbox{\cite{ben-tal_simplified_2006}} & 5\\[3pt]
		GE5 & Gas exchange O$_2$ in capillaries & $	\frac{d \pco(t)}{d t} =
		\frac{D_\mathrm{L,O_2}}{\sigma_\mathrm{O_2} \cdot V_c} 
		\left[	1 + \frac{4 \cdot T_h}{\sigma_\mathrm{O_2}} \frac{d \tilde{f}(\pco(t))}{d \pco}
		\right]^{-1} \Bigl( \palvo(t) - \pco(t) \Bigr).$  & \mbox{\cite{ben-tal_simplified_2006}} & 5\\[3pt]
		GE6 & Gas exchange in tissue (O$_2$ and CO$_2$)& $\frac{d C_{t}(t)}{d t} = \frac{Q_t \left( C_{\mathrm{a}}(t) - C_{t}(t) \right) + \mathrm{MP}_{t, i}}{V_t}$ with $i = O_2, CO_2$ & \cite{chiari_comprehensive_1997} &  7\\[3pt]
		GE7 & Gas exchange in brain (O$_2$ and CO$_2$) & $\frac{d C_{b}(t)}{d t} = \frac{Q_b \left( C_{\mathrm{a}}(t) - C_{b}(t) \right) + \mathrm{MP}_{b, i}}{V_b}$ with $i = O_2, CO_2$ & \cite{chiari_comprehensive_1997} &  7\\[3pt]
		GE8 & Gas exchange in cerebral fluid (CO$_2$) & $\frac{d \pcsf(t)}{d t} = \frac{1}{\tau_\mathrm{CSF}} (\pbco(t) - \pcsf(t))$ & \cite{chiari_comprehensive_1997}&  7\\[3pt]
		GE9 & Medullary partial pressure CO$_2$ & $	P_\mathrm{m}\mathrm{CO_2} = P_\mathrm{b}\mathrm{CO_2} + (\pcsf - P_\mathrm{b}\mathrm{CO_2}) \cdot e^{-k_{cr} \cdot \sqrt {Q_\mathrm{b}}}$ & \cite{chiari_comprehensive_1997} &  7 \\[3pt]
		\midrule
		BL1 & Oxygen dissociation curve & $\tilde{f}(\pO) = \frac{L \cdot K_T \cdot \sigma_\mathrm{O_2} \cdot \pO (1 + K_T \cdot \sigma_\mathrm{O_2}  \cdot \pO)^3 + K_R \cdot \sigma_\mathrm{O_2} \cdot \pO (1 + K_R \cdot \sigma_\mathrm{O_2} \cdot \pO)^3}{L(1+K_T \cdot \sigma_\mathrm{O_2} \cdot \pO)^4 + (1 + K_R \cdot \sigma_\mathrm{O_2} \cdot \pO)^4}$ & \cite{ben-tal_simplified_2006} & 5 \\[3pt]
		BL2 & Henderson-Hasselbalch formula & $\mathrm{pH} = 6.1 + \log \left(
		\frac{[\mathrm{HCO_3^-}]}{0.03~\mathrm{mmol/(L \cdot mmHg)} \cdot \pCO}
		\right)$ & \cite{hills_ph_1973} & 7\\[3pt]
		BL3 & Conversion of partial pressure to concentration (CO$_2$) & $	c_{\mathrm{CO_2}} =
		\Bigg(
		1 - \frac{0.02924~\mathrm{L/g} \cdot T_h}%
		{2.244 - 0.422 \cdot S_{\mathrm{O_2,vir}} \cdot (8.74 - \mathrm{pH})}
		\Bigg)
		\cdot 0.0301~\mathrm{L/mmHg} \cdot \pCO
		\cdot \Big( 1 + 10^{\mathrm{pH} - 6.1} \Big)
		\cdot 2.226$ & \cite{loeppky_relationship_1993} & 6\\
		BL4 & Conversion of partial pressure to concentration (O$_2$) & $p\mathrm{O}_{\mathrm{2,vir}} = \pO \left( \frac{40~\mathrm{mmHg}}{\pCO} \right)^{0.3}$ & \cite{chiari_comprehensive_1997} & 7 \\ [3pt]
		&  & $c_\mathrm{O_2} = k_1 \cdot S_\mathrm{O_2,vir}+ k_2 \cdot p_\mathrm{O_2} $ & \cite{chiari_comprehensive_1997} & 7 \\
		\midrule
		RC1 & Central chemical drive of pH level in medullar & $\frac{dD_C(t)}{dt}
		= \frac{1}{\tau_C}
		\bigl(G_C (\theta_\mathrm{pH,m} - \mathrm{pH}_\mathrm{m}) - D_C(t)\bigr)$  & modified from~\cite{JAWORSKI2019148, botros_neural_1990} & 5\\[3pt]
		& Peripheral chemical drive of \pao & $\frac{dD_\mathrm{O_2}(t)}{dt}
		= \frac{1}{\tau_P}
		\bigl(G_\mathrm{O_2} (P_\mathrm{a}\mathrm{O_{2,th}} - \pao) - D_\mathrm{O_2}(t)\bigr)$ & modified from~\cite{JAWORSKI2019148, botros_neural_1990} & 5\\[3pt]
		& Peripheral chemical drive of \paco and pH & $\frac{dD_\mathrm{CO_2}(t)}{dt}
		= \frac{1}{\tau_P}
		\bigl(G_\mathrm{CO_2} \paco
		+ G_\mathrm{CO_2,\mathrm{pH}} (\theta_\mathrm{pH,a} - \mathrm{pH_a})
		- D_\mathrm{CO_2}(t)\bigr)$ & modified from~\cite{JAWORSKI2019148, botros_neural_1990} & 5\\[3pt]
		RC2 & Total chemical drive & $D_\mathrm{total} = D_C + D_\mathrm{CO_2} + D_\mathrm{O_2}$ &  modified from~\cite{JAWORSKI2019148, botros_neural_1990} & 5\\[3pt]
		& Relative total chemical drive & $	D_\mathrm{rel} = a_\mathrm{ch} \cdot D_\mathrm{total} + b_\mathrm{ch} + \sigma_{D_\mathrm{total}}$ & \cite{JAWORSKI2019148, botros_neural_1990} & 5\\[3pt]
		RC3 & Respiratory Rate based on chemical drive & $\mathrm{RR} = \mathrm{RR}_{\mathrm{base}} + e^{\bigl(k_{\mathrm{RR,sens}} \cdot D_{\mathrm{rel}}\bigr)}\, \mathrm{RR}_{\mathrm{amp}},$ & \cite{reynolds_transient_1972, reynolds1973transient}  &  3 \\
		& Scaling factor for RR &  $K_\mathrm{RR} = \frac{\mathrm{RR} - \mathrm{RR}_{\mathrm{off}}}{\mathrm{RR}_{\mathrm{sc}}}$ &  &  3\\[3pt]
		RC4 & Hering-Breuer-Reflex & $D_{\mathrm{HB}} = S[K_{\mathrm{PRS}} \cdot (V(t) - V_0 - V_\mathrm{HB})]$ & modified from~\cite{JAWORSKI2019148} &5\\[3pt]
		RC5 & Inspiratory flow reflex & $D_{\mathrm{insp}} = S[K_{\mathrm{insp}} \cdot \dot{V}_{\mathrm{insp}}(t)]$ &  & 1\\[3pt]
		RC6 & Negative pressure reflex & $	P_{\mathrm{ua}} = R_{\mathrm{ua}} \cdot \dot{V}(t) - P_\mathrm{mus}$&  & 1 \\[3pt]
		& & $	D_\mathrm{neg} = S[K_{\mathrm{ua}} \cdot P_{\mathrm{ua}}]$ &  & 1\\[3pt]
		RC7 & Expiratory reflex & $D_\mathrm{exp} = S[K_{\mathrm{exp}} \cdot (R_{\mathrm{th}} - R_{\mathrm{ua}})]$ &  & 1\\[3pt]
		RC8 & Neural oscillator for late inspiratory neuron  & $\frac{dN_R(t)}{dt} = -a_R \cdot N_R(t) + W_{IR} \cdot S[N_I(t)] + W_{PR} \cdot S[N_P(t)] + W_{RR} \cdot S[N_R(t)] + B_R$ & ~\cite{JAWORSKI2019148, botros_neural_1990}  & 5\\[3pt]
		& Neural oscillator for inspiratory neuron& $	\frac{dN_I(t)}{dt} = -a_I \cdot N_I(t) + W_{EI} \cdot S[N_E(t)] + W_{PI} \cdot S[N_P(t)] + W_{LI} \cdot S[N_L(t)] + W_{II} \cdot S[I] + B_I$ & ~\cite{JAWORSKI2019148, botros_neural_1990} & 5\\  [3pt]
		& Neural oscillator for early inspiratory neuron&  $	\frac{dN_L(t)}{dt} = -a_L \cdot N_L(t) + W_{IL} \cdot S[N_I(t)] + W_{EL} \cdot S[N_E(t)] + W_{RL} \cdot S[N_R(t)] + W_{LL} \cdot S[L] + B_L$ & ~\cite{JAWORSKI2019148, botros_neural_1990} & 5\\[3pt]
		& Neural oscillator for post inspiratory neuron& $\frac{dN_P(t)}{dt} = -a_P \cdot N_P(t) + W_{EP} \cdot S[N_E(t)] + W_{RP} \cdot S[N_R(t)] + W_{PP} \cdot S[N_P(t)] + B_P$ &  ~\cite{JAWORSKI2019148, botros_neural_1990} & 5\\[3pt]
		& Neural oscillator for expiratory neuron&  $\frac{dN_E(t)}{dt} = -a_E \cdot N_E(t) + W_{IE} \cdot S[N_I(t)] + W_{RE} \cdot S[N_R(t)] + W_{EE} \cdot S[N_E(t)] + B_E$ & ~\cite{JAWORSKI2019148, botros_neural_1990} & 5\\ [3pt]
		\midrule
       CP1 & Mainstream capnogram &
{\scriptsize
$\displaystyle
\frac{dF_{\mathrm{CO_2}}(t)}{dt}
=\begin{cases}
\dfrac{
\fco(1-f_{\mathrm{V_D}})
+\fico f_{\mathrm{V_D}}
-F_{\mathrm{CO_2}}(t)}
{\tau_{\mathrm{exp}}},
&
\text{if }\;
\begin{array}{@{}c@{}}
\dot V(t)\leq Q_{\mathrm{switch}},\\[0mm]
V(t)\geq \vd,
\end{array}
\\[1mm]
\dfrac{\fico-F_{\mathrm{CO_2}}(t)}
{\tau_{\mathrm{insp}}},
&
\text{otherwise.}
\end{cases}
$}
& & 5 \\[-2pt]
\end{longtable}
\newpage
\twocolumn
\section{Model Credibility Assessment }
\label{appen:model_val}
\subsection{Sensitivity analysis and uncertainty quantification}
Table~\ref{tab:sa_uq_input} summarizes the input parameters of the sensitivity analysis and uncertainty quantification.
\subsection{Model Calibration}
\label{append:model_cal}
\subsubsection{Method: Test S--2}
The system formulation integrates lung mechanics and gas exchange dynamics under the influence of cardiac activity within a nonlinear state-space representation. The model is derived from first physiological principles, ensuring that all components are physiologically interpretable and that the relationships between states and parameters correspond directly to measurable respiratory and cardiovascular phenomena. \\
A central challenge addressed in this work is the presence of non-smooth or discontinuous operators arising from physiological switching and cardiac resetting mechanisms. To enable stable numerical simulation and the application of optimization-based inference and estimation methods, these operators are reformulated using smooth sigmoid-based approximations. The resulting formulation yields a globally differentiable dynamical system between heartbeats, retaining the qualitative physiological behavior while maintaining full interpretability and compatibility with observer design and parameter identification algorithms for nonlinear inference. \\
An iterative sliding-window formulation can subsequently be used to update patient-specific parameters online as physiology evolves over time. In the present validation analysis, however, parameters were identified within a defined calibration interval and subsequently kept fixed during the temporally separated validation interval. \\
Within this dynamical system, model parameters define patient-specific physiology. Variations in parameter values correspond to differences in respiratory mechanics, gas exchange efficiency, and cardio-respiratory interactions across individuals. \\
From a systems-theoretic perspective, nonlinear local observability analysis based on Lie algebraic methods indicates that the available measurement set (flow and end-tidal CO$_2$) does not fully satisfy the observability conditions required for complete state reconstruction and validation. Enhancement of the measurement setup may therefore improve validation of the involved dynamics. Embedding the system within a nonlinear observer framework enables smooth estimation of hidden states, which supports iterative parameter refinement despite partial observability. \\
\subsubsection{Results: Test S--2}
Lung volume was reconstructed by temporal integration of the measured airway-flow signal after offline leakage correction, following the preprocessing approach introduced and subsequently applied in related model-based respiratory-effort studies~\cite{grashoff_model-based_2023, globig_estimating_2026}, and was subsequently used for comparison with the simulated volume trajectory.








\paragraph{Patient 1: PEEP intervention.}
Patient~1 was used to assess frozen prediction across a change in ventilatory
operating conditions. The model was calibrated before the intervention and
all patient-specific parameters were subsequently kept fixed. During
validation, PEEP changed from approximately $6.0$ to $4.5$~mmHg, while
FiO$_2$ remained constant at $0.299$ and heart rate at $60$~bpm. Thus, the
validation interval represents a genuine change in ventilatory input rather
than a repetition of the calibration condition.

\begin{figure*}[h!]
    \centering
    \includegraphics[width=\textwidth]{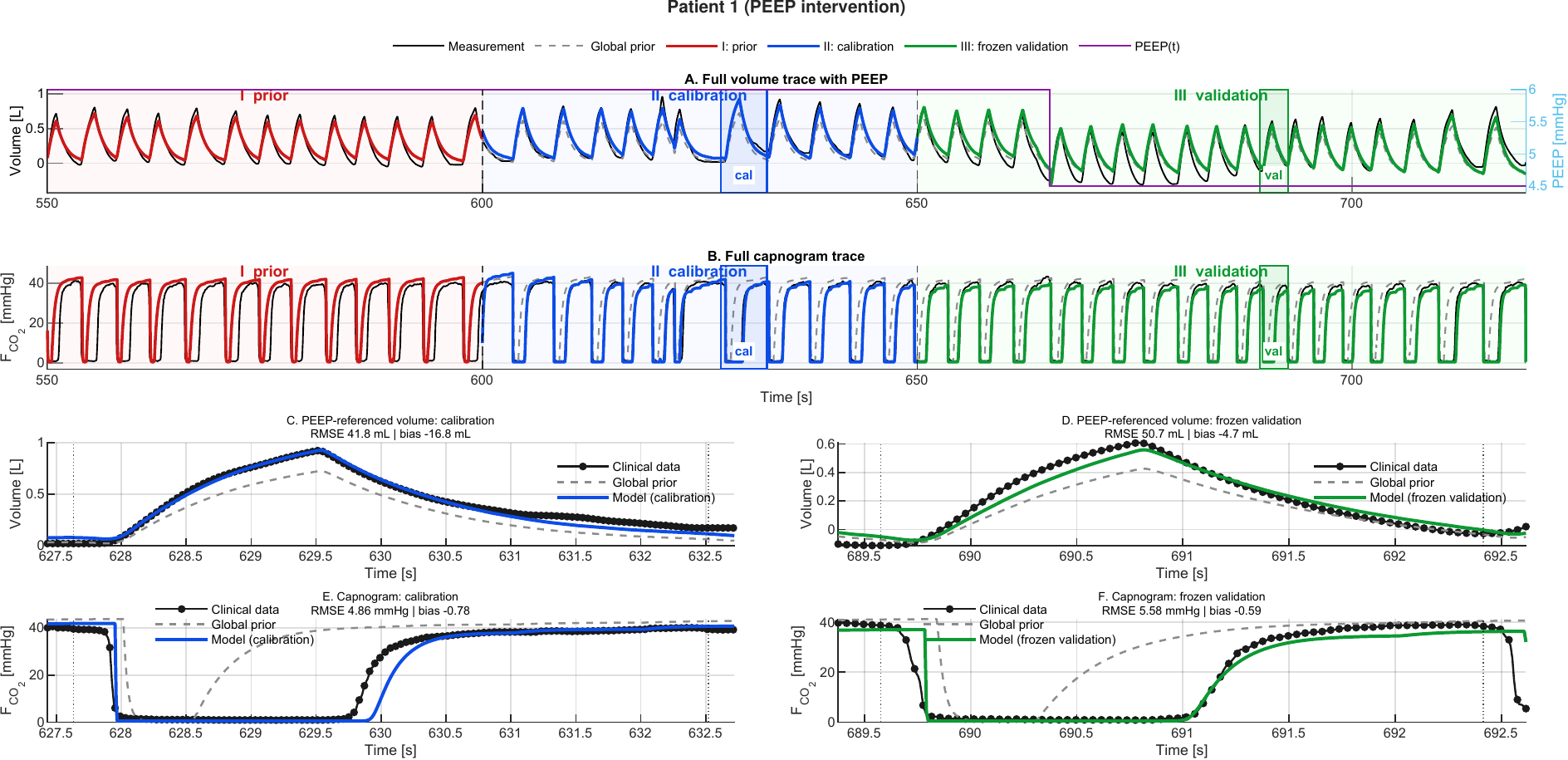}
   \caption{Patient~1: patient-specific calibration and frozen temporal validation spanning a PEEP intervention. The model was calibrated before validation and all patient-specific parameters were subsequently kept fixed throughout the validation interval. During validation, PEEP decreased from approximately $6.0$ to $4.5$~mmHg, while $F_{\mathrm{iO_2}}$ remained constant at $0.299$ and heart rate at $60$~bpm. Panels A--B show the complete PEEP-referenced relative lung-volume and capnogram trajectories together with the PEEP input, while panels C--F show representative breaths during calibration and frozen validation. The plotted volume signal is expressed relative to the calibration baseline. Changes in this baseline associated with external PEEP are represented by $\Delta V_{\mathrm{PEEP}}(t)=[PEEP(t)-PEEP_{\mathrm{ref}}]/E_{\mathrm{RS}}$, where $PEEP_{\mathrm{ref}}$ denotes the reference PEEP of the calibration interval and $E_{\mathrm{RS}}$ is the fitted respiratory-system elastance reported in Table~\ref{tab:model_parameters_detailed}. Consequently, negative values of the plotted volume indicate excursions below the selected baseline and do not represent negative absolute lung volume. The case therefore assesses frozen prediction across a change in ventilatory operating conditions, without parameter re-identification. Despite the intervention, the model retained the principal volume and capnogram dynamics, although the overall lung-volume agreement during frozen validation was lower than during calibration.}
    \label{fig:patient1_validation}
\end{figure*}

As shown in Figure~\ref{fig:patient1_validation}, the model achieved good
agreement during calibration ($R^2=0.936$, RMSE $=64.1$~mL) and retained the
principal respiratory dynamics during frozen validation across the PEEP
intervention. The lower whole-window volume agreement
($R^2=0.811$, RMSE $=117.6$~mL) was associated mainly with a slow,
localized volume-reference discrepancy around the intervention rather than
with a persistent loss of breath-scale dynamics. Importantly, prediction of
the directly measured respiratory signals remained strong during frozen
validation, with $R^2=0.942$ for flow and $R^2=0.950$
(RMSE $=3.93$~mmHg) for the capnogram.

This discrepancy should also be interpreted in the context of the
measurement preprocessing. Lung volume was not measured directly but was
obtained by integration of the offline leakage-corrected airway-flow signal,
using the preprocessing approach described in~\cite{grashoff_model-based_2023, globig_estimating_2026}. Such leakage correction is necessary
to prevent cumulative integration drift, but abrupt changes in ventilatory
operating conditions, such as the PEEP intervention considered here, can
affect the slowly varying flow balance used for volume reconstruction.
Consequently, localized transients in the reconstructed volume around the
intervention may partly reflect preprocessing effects rather than an
equivalent change in the underlying respiratory mechanics. Consistent with
this interpretation, breath-scale dynamics remained well preserved and the
volume agreement improved again later in the post-intervention interval
without parameter re-identification. The preserved flow and capnogram
predictions further indicate that the reduced whole-window volume metric does
not represent a general deterioration of the coupled model. Thus, Patient~1
provides a stringent frozen-prediction test while also illustrating the
importance of accounting for measurement-chain and preprocessing uncertainty
when evaluating derived physiological signals.

\paragraph{Patient 2: long-duration validation.}
Patient~2 was used to assess the persistence of the individualized model over a substantially longer prediction horizon. The model was calibrated over approximately $120$~s and subsequently propagated with all patient-specific parameters fixed over approximately $473$~s of frozen temporal validation. Ventilatory conditions remained essentially unchanged over this interval, allowing the analysis to focus on whether the identified physiological parameterization remained representative over time rather than on adaptation to an external intervention.

Relative to the population-level prior, the individualized parameter set (Table~\ref{tab:model_parameters_detailed}) indicated moderate respiratory-system elastance and resistance together with a comparatively larger dead-space contribution. For this patient, the standard parameter subset was augmented by the effective metabolic CO$_2$ production parameter $\mathrm{MP}_{t,\mathrm{CO_2}}/V_B$ because the nominal population-level value resulted in a persistent mismatch of the capnogram level. This parameter should be interpreted as an effective patient-specific description of the coupled CO$_2$ production and transport balance rather than as a direct measurement of whole-body metabolism.

\begin{figure*}[t]
    \centering
    \includegraphics[width=\textwidth]{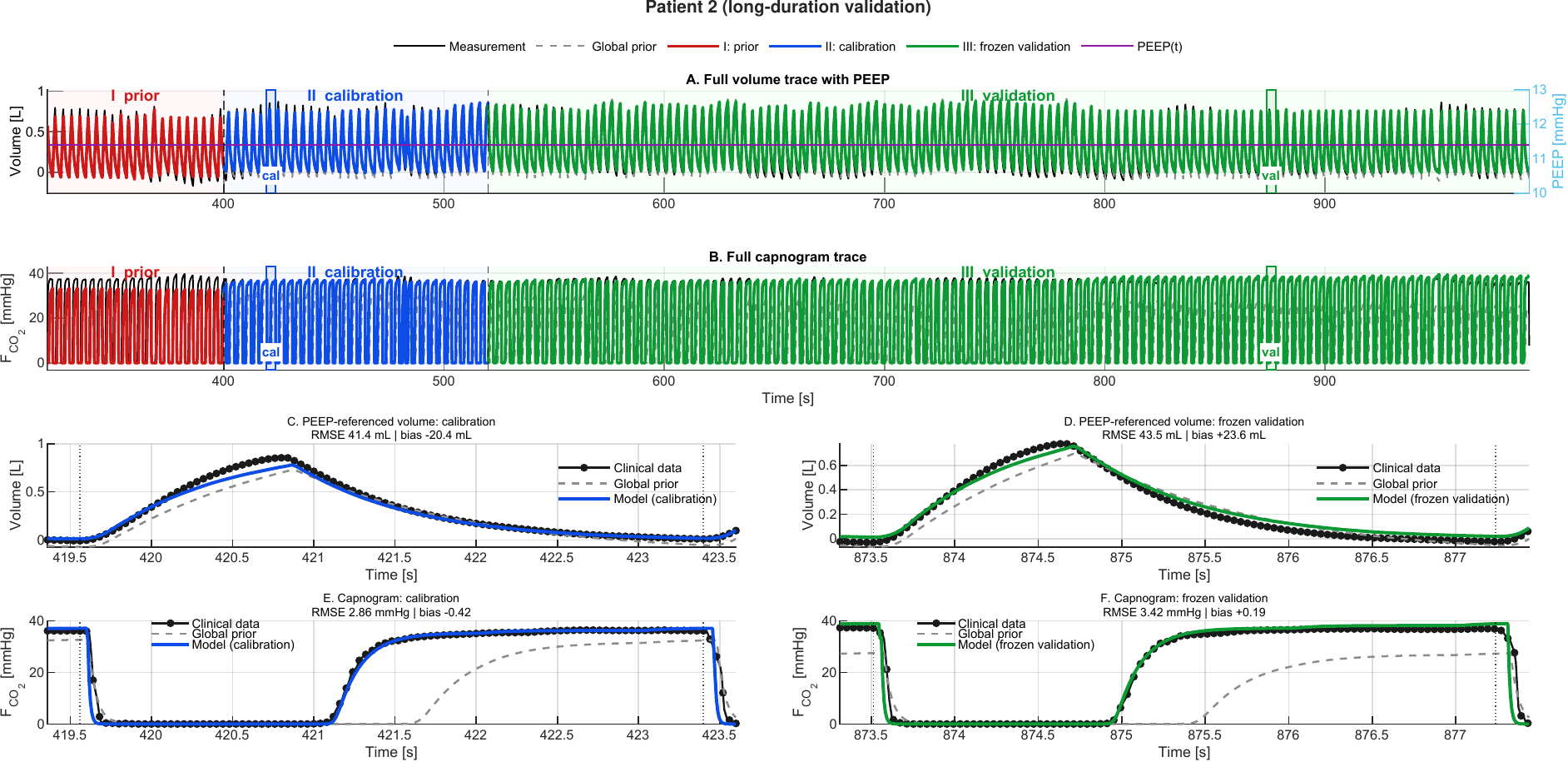}
    \caption{Patient~2: patient-specific calibration and long-duration frozen temporal validation.
    The model was calibrated over approximately $120$~s and all individualized parameters were subsequently kept fixed during approximately $473$~s of temporal validation.
    Panels A--B show the complete lung-volume and capnogram trajectories across calibration and frozen validation, while panels C--F show representative breaths from the respective intervals. In contrast to Patient~1, no relevant ventilatory intervention occurred during validation; the case therefore primarily evaluates persistence of the individualized physiological parameterization over an extended time horizon. For Patient~2, the effective metabolic CO$_2$ production parameter $\mathrm{MP}_{t,\mathrm{CO_2}}/V_B$ was additionally individualized to account for the persistent patient-specific CO$_2$ level. Despite the substantially longer validation interval, the frozen model retained the principal volume and capnogram dynamics, supporting a quasi-static approximation of the identified parameters over the investigated time scale.}
    \label{fig:patient2_validation}
\end{figure*}

As shown in Figure~\ref{fig:patient2_validation}, the frozen model retained the main respiratory-volume and capnogram dynamics throughout the extended validation interval. Lung-volume prediction achieved an RMSE of $81.2$~mL ($R^2=0.913$), while capnogram prediction yielded an RMSE of $4.51$~mmHg ($R^2=0.932$). The preservation of both waveform morphology and CO$_2$ level over several minutes without parameter updating supports the use of a quasi-static patient-specific parameterization over this time scale, while longer observation periods would naturally motivate subsequent online adaptation as physiology evolves.

\paragraph{Patient 3: reference case.}
Patient~3 served as the reference case for patient-specific model calibration and frozen temporal validation. Relative to the population-level prior parameter ranges summarized in Table~\ref{tab:model_input_parameter}, the individualized parameter set (Table~\ref{tab:model_parameters_detailed}) indicated relatively stiff respiratory mechanics, reflected by an elevated elastance, together with moderate airway resistance. The identified dead-space-related parameters remained physiologically plausible and allowed the model to reproduce the measured capnogram morphology.

The calibrated model reproduced the respiratory volume trajectory with an RMSE of $42.7$~mL and maintained comparable performance during frozen validation (RMSE $41.5$~mL, $R^2=0.967$). Given tidal-volume excursions of several hundred millilitres, this error represents only a small fraction of the respiratory amplitude. Capnogram prediction was similarly preserved, with a validation RMSE of $2.81$~mmHg and $R^2=0.977$. The complete calibration and validation trajectories are shown in Figure~\ref{fig:model_calibration}.

\paragraph{Patient 4: robustness to capnogram artefact.} Patient~4 was used to assess frozen model prediction in the presence of a localized distortion of the measured capnogram. Relative to the population-level prior, the individualized parameter set indicated comparatively high
respiratory-system elastance and resistance. The standard capnogram-related parameter subset was retained for patient-specific adaptation. The functional dead-space fraction $f_{VD}$ converged to a value effectively equal to zero,
indicating that no additional functional dead-space contribution was required to reproduce the calibration data for this patient. All patient-specific parameters were subsequently kept fixed during temporal validation.

Respiratory-volume prediction remained accurate throughout frozen validation, with an RMSE of $21.4$~mL and $R^2=0.975$. In contrast, the measured expiratory
CO$_2$ signal developed a marked plateau distortion during validation that was not accompanied by a corresponding change in the respiratory-volume waveform or in the recorded ventilatory inputs. The complete frozen-validation capnogram nevertheless retained an $R^2$ of $0.921$ with an RMSE of $8.13$~mmHg.

\begin{figure*}[t]
    \centering
    \includegraphics[width=\textwidth]{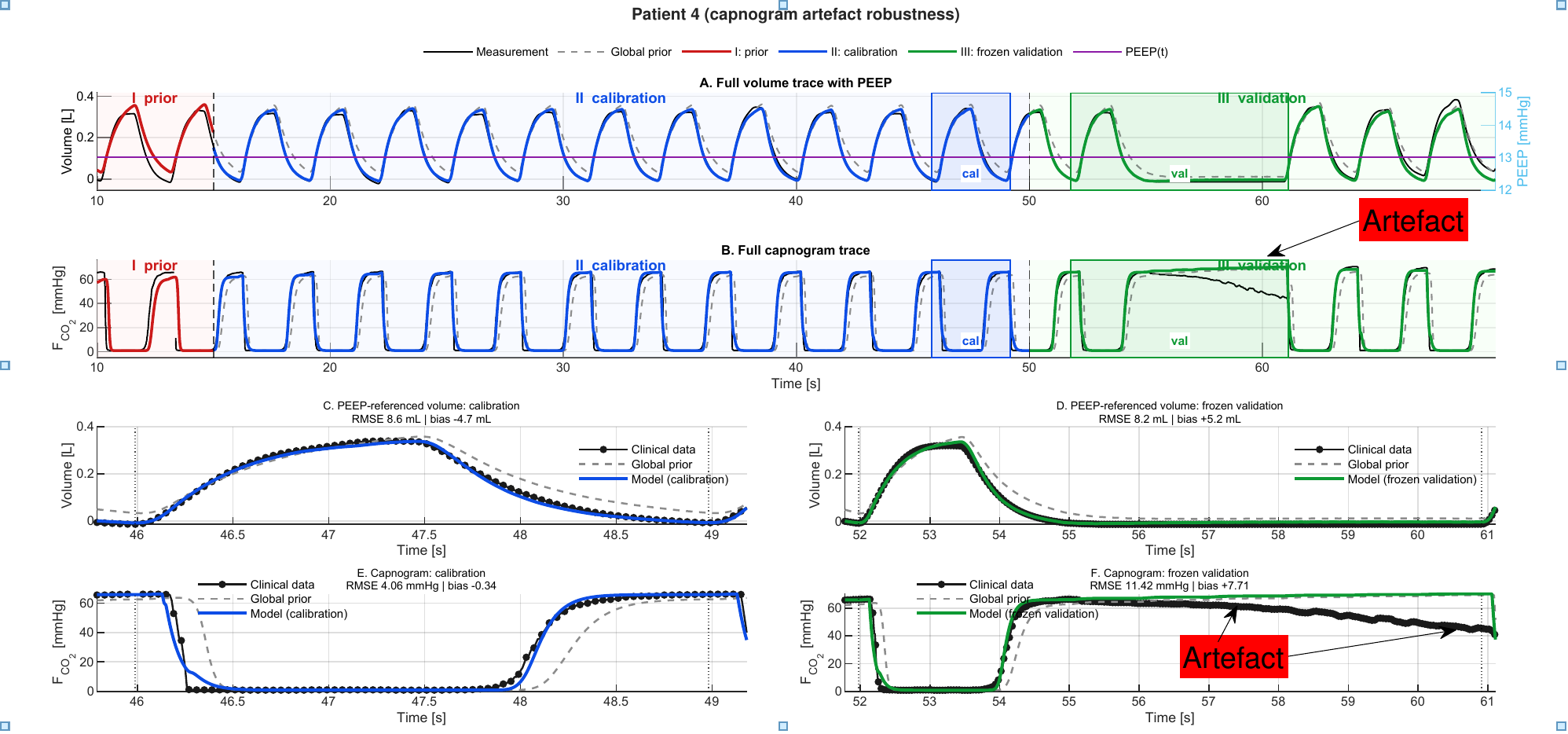}
    \caption{Patient~4: patient-specific calibration and frozen temporal validation in the presence of a capnogram distortion interpreted as a likely measurement artefact. Panels A--B show the complete lung-volume and capnogram trajectories across calibration and frozen validation, while panels C--F show representative
    breaths from the respective intervals. Respiratory-volume prediction remains accurate during frozen validation,
    whereas the measured expiratory CO$_2$ plateau develops a localized drift that is not reproduced to the same extent by the frozen mechanistic model. The measured early-to-late plateau changed by approximately $1.68$~mmHg, compared with approximately $0.43$~mmHg for the corresponding frozen-model prediction. The preserved respiratory-volume dynamics and the substantially smaller change in the model-predicted CO$_2$ plateau are consistent with a localized measurement-chain disturbance rather than a corresponding change in
    respiratory mechanics or ventilatory input. Potential mechanisms include leakage, suction, sampling-line disturbance, or transient disconnection; the available data do not permit attribution to a specific cause.}
    \label{fig:patient4_validation}
\end{figure*}

As illustrated in Figure~\ref{fig:patient4_validation}, the measured
early-to-late expiratory plateau changed by approximately $1.68$~mmHg, whereas the corresponding change of the frozen model prediction was only approximately $0.43$~mmHg. The mechanistic prediction therefore remained substantially more stable than the locally distorted measurement. We interpret this discrepancy as being consistent with a measurement-chain artefact rather than forcing the model to reproduce it as a physiological change. Possible explanations include transient leakage, suction, sampling-line disturbance, or temporary disconnection, although the available recordings do not permit attribution to a specific mechanism.

This case illustrates a potentially useful property of the physiological model: a localized disagreement between a dynamically consistent frozen prediction and an otherwise unexplained measurement distortion may provide an indication of measurement artefact rather than necessarily implying a change in the underlying patient physiology.

\paragraph{Parameter identifiability.}
Only parameters sufficiently informed by the available measurements were individualized. The remaining model parameters correspond to weakly observable directions of the parameter space under the present measurement configuration and were therefore retained at physiologically plausible population-informed prior values. Exploratory fits allowing separate inspiratory and expiratory resistances yielded ratios $R_{\mathrm{insp}}/R_{\mathrm{exp}}$ close to unity across all investigated patients, providing no evidence that two distinct resistance parameters were required by the available data. This avoids over-parameterization and non-unique parameter combinations while preserving a consistent physiological baseline for the latent model dynamics. This combination of physiologically informed prior initialization and selective adaptation of measurement-identifiable parameters forms the basis for subsequent online model updating and patient-specific prediction.

\begin{table*}[t]
\centering
\footnotesize
\setlength{\tabcolsep}{3.2pt}
\renewcommand{\arraystretch}{1.16}

\newcommand{\upd}[1]{\mathbf{#1}}

\caption{Patient-specific model parameterization for the four investigated
cases in Test~S--2. Parameters not individualized during calibration were
retained at physiologically plausible prior values. Bold values indicate
parameters identified or derived from patient-specific measurements.}
\label{tab:model_parameters_detailed}

\begin{tabular}{
p{1.6cm}
p{3.7cm}
p{1.8cm}
>{\centering\arraybackslash}p{1.40cm}
>{\centering\arraybackslash}p{1.40cm}
>{\centering\arraybackslash}p{1.40cm}
>{\centering\arraybackslash}p{1.40cm}}
\toprule
\textbf{Param.} & \textbf{Description} & \textbf{Unit} &
\textbf{Pat. 1} & \textbf{Pat. 2} & \textbf{Pat. 3} & \textbf{Pat. 4} \\
\midrule

\multicolumn{7}{l}{\textbf{Lung Mechanics}} \\ \midrule

$E_{\mathrm{RS}}$ &
Respiratory-system elastance &
mmHg/L &
$\upd{6.479}$ & $\upd{9.433}$ & $\upd{22.813}$ & $\upd{27.542}$ \\

$R$ &
Airway resistance &
mmHg\,L$^{-1}$\,s &
$\upd{6.572}$ & $\upd{6.286}$ & $\upd{7.868}$ & $\upd{11.402}$ \\

$\mathrm{EELV}$ &
End-expiratory lung volume &
L &
3.000 & 3.500 & 4.000 & 1.500 \\

$q$ &
Volume offset &
L &
$\upd{-0.0369}$ & $\upd{0.00218}$ & $\upd{0.0775}$ & $\upd{0.0457}$ \\

$V_T$ &
Tidal-volume reference &
L &
$\upd{0.762}$ & $\upd{0.804}$ & $\upd{0.716}$ & $\upd{0.329}$ \\

$V_D$ &
Anatomical dead space &
L &
$\upd{0.110}$ & $\upd{0.238}$ & $\upd{0.209}$ & $\upd{0.187}$ \\

\midrule
\multicolumn{7}{l}{\textbf{Gas Exchange}} \\ \midrule

$D_{\mathrm{L,O_2}}$ &
O$_2$ diffusion capacity &
L\,s$^{-1}$\,mmHg$^{-1}$ &
$9.62{\times}10^{-5}$ &
$1.03{\times}10^{-4}$ &
$9.62{\times}10^{-5}$ &
$5.60{\times}10^{-4}$ \\

$D_{\mathrm{L,CO_2}}$ &
CO$_2$ diffusion capacity &
L\,s$^{-1}$\,mmHg$^{-1}$ &
$1.92{\times}10^{-3}$ &
$2.06{\times}10^{-3}$ &
$1.92{\times}10^{-3}$ &
$7.08{\times}10^{-4}$ \\

$V_c$ &
Capillary volume &
L &
0.0673 & 0.0778 & 0.0673 & 0.0778 \\

$T_h$ &
Hemoglobin concentration &
mol\,L$^{-1}$ &
$2.30{\times}10^{-3}$ &
$2.30{\times}10^{-3}$ &
$2.30{\times}10^{-3}$ &
$2.30{\times}10^{-3}$ \\

$(1-f_s)/V_B$ &
No-Shunt fraction/tissue vol. &
L$^{-1}$ &
0.0159 & 0.0261 & 0.0159 & 0.0261 \\

$f_{V_D}$ &
Functional dead-space fraction &
-- &
$\upd{0.151}$ &
$\upd{0.281}$ &
$\upd{0.0224}$ &
$\upd{7.81{\times}10^{-8}}$ \\

\midrule
\multicolumn{7}{l}{\textbf{Metabolism \& Acid--Base}} \\ \midrule

$\mathrm{MP}_{t,O_2}/V_B$ &
O$_2$ consumption/tissue vol. &
mol\,L$^{-1}$ &
$-9.14{\times}10^{-5}$ &
$-1.04{\times}10^{-4}$ &
$-9.14{\times}10^{-5}$ &
$-1.00{\times}10^{-3}$ \\

$\mathrm{MP}_{t,CO_2}/V_B$ &
CO$_2$ production/tissue vol. &
mol\,L$^{-1}$ &
$5.97{\times}10^{-5}$ &
$\upd{2.56{\times}10^{-4}}$ &
$5.97{\times}10^{-5}$ &
$8.00{\times}10^{-5}$ \\

$h$ &
H$^+$ concentration &
mol\,L$^{-1}$ &
$3.50{\times}10^{-8}$ &
$3.52{\times}10^{-8}$ &
$3.50{\times}10^{-8}$ &
$3.52{\times}10^{-8}$ \\

$r_2$ &
Dehydration rate &
s$^{-1}$ &
0.0800 & 0.0700 & 0.0800 & 0.0700 \\

\midrule
\multicolumn{7}{l}{\textbf{Capnogram Dynamics}} \\ \midrule

$Q_{\mathrm{switch}}$ &
Flow-switch threshold &
L/s &
$\upd{0.234}$ &
$\upd{0.139}$ &
$\upd{0.184}$ &
$\upd{0.536}$ \\

$\tau_{\mathrm{exp}}$ &
Expiratory time constant &
s &
$\upd{0.185}$ &
$\upd{0.152}$ &
$\upd{0.168}$ &
$\upd{0.0876}$ \\

$\tau_{\mathrm{insp}}$ &
Inspiratory time constant &
s &
$\upd{0.00815}$ &
$\upd{0.0166}$ &
$\upd{0.00907}$ &
$\upd{0.0806}$ \\

\bottomrule
\end{tabular}

\end{table*}

\subsection{Emergent Model Behavior}
\label{sec:Emergent model behavior}
Emergent model behavior is evidence that demonstrates that the finalized computational model reproduces known physiological phenomena under specified conditions, even though these behaviors were not explicitly implemented or programmed in the governing equations. Instead, such behaviors arise from the interaction of physiological subsystems — such as chemoreflex control, neuronal oscillators, lung mechanics, and gas exchange — and are assessed based on established scientific knowledge and qualitative consistency with clinical observations. These emergent behaviors strengthen the model’s credibility, particularly in simulating complex physiological and pathophysiological states.
\subsubsection{Lung mechanics}
\begin{itemize}
	\item \textbf{Auto-PEEP:} In modeling respiratory mechanics, a computational lung model is developed including airway resistance and lung compliance. It is demonstrated that under conditions of high respiratory rate and increased resistance, the model produces intrinsic positive end-expiratory pressure (auto-PEEP) due to incomplete exhalation. This behavior is not explicitly coded but emerges from the interaction of flow resistance and timing. A developer using the model for ventilator strategy optimization could include this as credibility evidence.
	\item \textbf{Recruitment and derecruitment with varying PEEP:} A computational lung model with recruitable alveolar units is developed. It is demonstrated that increasing positive end-expiratory pressure (PEEP) leads to a gradual increase in recruited volume, while decreasing PEEP results in derecruitment with hysteresis. This behavior is not pre-specified but arises from threshold distributions and time-dependent opening/closing dynamics. This could be used as credibility evidence in the study of lung-protective ventilation strategies.
\end{itemize}

\subsubsection{Gas Exchange}
\begin{itemize}
	\item \textbf{Emergent Hypoxia or Hypercapnia via Structural Pathology:} The gas exchange model  demonstrates that modifying parameters such as dead space, shunt fraction, or cardiac output leads to emergent changes in arterial O$_2$ and CO$_2$ partial pressures, consistent with hypoxemia or hypercapnia. These changes are not directly programmed but arise from the altered balance between ventilation and perfusion. A user modeling pathological conditions like ARDS could include this as credibility evidence. 
	\item \textbf{Effect of Lung Mechanics on Gas Exchange:} The computational model of the respiratory system is developed with variable compliance and resistance. It is shown that modifying these parameters alters tidal volume and minute ventilation, which in turn indirectly affects alveolar and arterial gas levels. These changes in gas exchange are not directly specified but emerge through mechanical-chemical interaction. This could be used as credibility evidence when modeling conditions such as restrictive or obstructive lung disease.
	\item \textbf{Temperature-Dependent O$_2$ Dissociation Effects:} The gas exchange model includes temperature dependence of the hemoglobin-O$_2$ binding curve. When body temperature is increased, the model shows a rightward shift in the dissociation curve, resulting in decreased hemoglobin saturation at a given \pao. This behavior emerges from physiological equations and is not manually enforced. A user modeling hyperthermia or febrile illness could cite this as credibility evidence for thermodynamic realism.
\end{itemize}

\subsubsection{Respiratory Center}
\begin{itemize}
	\item \textbf{Asynchrony Emergence:} Evidence is collected to demonstrate that, depending on ventilator settings and patient model parameters, the model exhibits patient-ventilator asynchrony events such as double triggering or ineffective efforts. These asynchronies are not explicitly coded but emerge from the interaction between ventilator support and simulated respiratory drive. A researcher studying ventilator control algorithms could include this information as credibility evidence for model realism under clinical conditions.
	\item \textbf{Response to Different Pressure Support Levels:} It is demonstrated that increasing levels of PSV lead to a reduction in respiratory muscle activity and breathing frequency, as the model's chemoreflex reduces neural drive due to increased alveolar ventilation. This behavior is not hardcoded but emerges from the physiological feedback mechanisms. A user applying this model to simulate ventilator adaptation could cite this as credibility evidence.
	\item \textbf{Setpoint Emergence of Blood Gases:} It is demonstrated that under normal physiological conditions, the model settles to a PaCO$_2$ of approximately 40 mmHg and a blood pH of 7.4, without explicitly targeting these values. The emergent setpoint arises from the balance of feedback gains and thresholds. This behavior can be modified to simulate pathological states such as COPD, where higher \paco setpoints are observed. A developer using this model for disease simulation could include this behavior as credibility evidence.
	\item \textbf{Biphasic Ventilatory Response to Hypoxia:} The model shows a characteristic biphasic ventilatory response to sustained hypoxia — an initial increase in ventilation followed by decline — which is a known neurophysiological phenomenon~\cite{duffin_measuring_2007, MILHORN196527}. This occurs without being directly implemented, and arises from oxygen-sensitive drive components and their adaptation dynamics.
	\item  \textbf{CO$_2$ Apnea Threshold Behavior:} When CO$_2$ falls below a certain threshold, the model spontaneously ceases respiratory rhythm, resembling CO$_2$ apnea — a known phenomenon during hyperventilation. This threshold arises from the parameterization of chemoreceptor activity.
	\item \textbf{Emergence of optimal ventilation settings:} In the computational model of ventilated patients, it is shown that certain combinations of PEEP, pressure support, and trigger sensitivity result in synchronous breathing patterns with minimal respiratory muscle load. These optimal conditions are not explicitly defined but emerge from feedback-regulated interaction between ventilator and simulated patient. This behavior can support credibility for use in ventilator setting optimization.
	\item \textbf{Cycle shortening through premature cycling:} It is demonstrated that with high cycling-off thresholds, the ventilator prematurely ends inspiration while the respiratory drive is still active. This results in premature cycling, a type of timing asynchrony, which arises naturally from model dynamics and ventilator settings. This behavior can be cited as credibility evidence for simulating trigger-timing interactions.
	\item \textbf{Ineffective inspiration attempts with trigger delay:} The computational model simulating patient-triggered mechanical ventilation is developed. It is shown that under conditions of low respiratory drive or high trigger sensitivity thresholds, the model generates inspiratory efforts that fail to trigger the ventilator — resulting in ineffective efforts. This behavior arises from the interaction of respiratory muscle pressure and ventilator settings and is not explicitly defined. This emergent behavior can support credibility in simulating patient-ventilator interaction.
	\item \textbf{Double triggering with high spontaneous activity:} In a simulation of pressure support ventilation, it is demonstrated that when the neural inspiratory time of the patient exceeds the duration of ventilator-delivered support, the model exhibits double triggering — a known form of patient-ventilator asynchrony. This phenomenon arises from mismatches in timing between spontaneous effort and ventilator cycling-off criteria and is not directly coded. This can be used as credibility evidence for temporal realism in support ventilation.
\end{itemize}

\subsection{Model Plausibility}
Table~\ref{tab:lit_grading} summarizes the quality grading of the governing equations, ranging from LOW to HIGH.
\subsection{Credibility Assessment}
\label{appen:credibility}
Table~\ref{tab:assess_credibility} provides a detailed overview of the credibility assessment, rating each validation step as a (weak), b (moderate), or c (good).

\onecolumn
\small
\begin{longtable}{p{1.5cm} p{5cm} >{\centering\arraybackslash}p{2cm} >{\centering\arraybackslash}p{2.5cm} >{\centering\arraybackslash}p{2cm}}

\caption[Baseline patient model input parameters for global sensitivity analysis and uncertainty quantification]{Model input parameters for global sensitivity analysis and uncertainty quantification with mean, standard deviation, and distribution.}
\label{tab:sa_uq_input} \\

\toprule
\textbf{Param.} & \textbf{Description} & \textbf{Mean} & \textbf{Std.}  & \textbf{Distribution} \\
\midrule
\endfirsthead

\multicolumn{5}{c}{\tablename\ \thetable\ -- continued from previous page} \\
\toprule
\textbf{Param.} & \textbf{Description} & \textbf{Mean} & \textbf{Std.}  & \textbf{Distribution} \\
\midrule
\endhead

\midrule
\multicolumn{5}{r}{continued on next page} \\
\endfoot

\bottomrule
\endlastfoot

\multicolumn{5}{l}{\textbf{Lung mechanics (LM)}} \\
\midrule

    $R$ & Airway resistance & 3 & 0.3 & lognormal \\
	$E_L$   & Lung elastance & 10 & 1 & lognormal \\
	$E_\mathrm{CW}$  & Chest wall elastance & 5  & 0.5 & normal \\
	\midrule
	\multicolumn{3}{@{}l}{\textbf{Gas Exchange (GE)}} \\
	\midrule
	$f_\mathrm{V_D}$      & Pulmonary dead space & 0.1 & 0.06 &lognormal \\
	\vd       & Anatomical dead space &  0.15  & 0.06 & lognormal \\
	\fs       & Shunt fraction & 0.1 & 0.06 & lognormal \\
	$Q$        & Cardiac output & 5.5 & 0.55 & normal \\
	$D_\mathrm{L,O_2}$       & O$_2$ diffusing capacity & 3.5e-4  & 3.5e-5 & normal \\
	$D_\mathrm{L,CO_2}$       & CO$_2$ diffusing capacity & 2.52e-5  & 2.52e-6 & normal \\
	$V_c$       & Pulmonary capillary volume & 0.08  & 0.008 & lognormal \\
	$\mathrm{MP_{t, O_2}}$     & Metabolic rate O$_2$ & 0.2 & 0.02 & lognormal \\
	$\mathrm{MP_{t, CO_2}}$    & Metabolic rate CO$_2$  & 0.2  & 0.02 & lognormal \\
	$V_B$       & Body volume & 38  & 3.8 &  normal \\
	$T_h$ & Concentration of hemoglobin & 20 g & 2 & normal \\
	$\delta$ & Acceleration rate & 79.433 & 8 & normal \\
	$\sigma_\mathrm{O_2}$ & Solubility of O$_2$ in plasma & 1.34e-6 & 1.34e-7 & normal \\
	$\sigma_\mathrm{CO_2}$ & Solubility of CO$_2$ in plasma & 3.3e-5 & 3.3e-6 & normal \\
	FRC & Functional residual capacity & 3 L & 0.3 & normal \\
	\midrule
	\multicolumn{3}{@{}l}{\textbf{Respiratory Center (RC)}} \\
	\midrule
	$G_C$ & Central chemoreflex gain & 35 & 3 & normal \\
	$G_\mathrm{CO_2}$ & Peripheral chemoreflex gain CO$_2$ & 0.168 & 0.03 & normal \\
	$G_\mathrm{O_2}$ & Peripheral chemoreflex gain O$_2$ & 0.105 & 0.03 & normal \\
	$\rho_{\mathrm{sed}}$ & Sedation scaling factor & 0.5 & 0.25 & uniform \\
	$\rho_{\mathrm{resp}}$ & Respiratory failure scaling factor & 0.5 & 0.25 & uniform \\
	$\theta_{\mathrm{chem}}$ & Set point \paco & 42.5 & 4 & normal \\
	$\theta_\mathrm{pH_m}$ & Set point pH$_m$  & 7.32 & 0.7 & normal \\
	$\theta_\mathrm{pH_a}$ & Set point pH$_a$  & 7.4 & 0.7 & normal \\
	\pao & Arterial partial pressure O$_2$ & 82.5 & 8 & normal \\
	\paco & Arterial partial pressure CO$_2$ & 42.5 & 5.0 & normal \\
	\pbco & Brain partial pressure CO$_2$ & 90 & 9.0 & normal \\

\end{longtable}
\normalsize

\small
\begin{table*}[h!]
	\centering
	\caption{Governing equation quality grading from LOW to HIGH.}
	\begin{tabular}{lp{13cm}}
		\toprule
		& \textbf{Literature quality (LQ) based on citation count and author recognition} \\
		\midrule
		LOW    & The literature has fewer than 10 citations and the authors are not well known in the community. \\
		MED & The literature has at least 10 citations, or fewer citations but authors well known in the community. \\
		HIGH   & The literature has more than 50 citations, or at least 10 citations and authors well known in the community. \\
		\midrule 
		& \textbf{Data quality if physics based (DQP)} \\
		\midrule
		LOW    & The equation is neither based on physical principles nor physiological assumptions. \\
		MED & The equation is largely derived from physiological assumptions or physical principles. \\
		HIGH   & The equation is based on fundamental physical principles or conclusively derived from physiological assumptions. \\
		\midrule
		& \textbf{Data quality if study based (DQS)} \\
		\midrule
		LOW  & Small number of subjects; very narrow range of intended population (e.g. average); key patient data missing; single-center study. \\
		MED  & Multiple subjects, but not statistically representative; partial range of intended population covered; most key patient data available; single- or limited multi-center study. \\
		HIGH & Statistically relevant number of subjects ($> 100$); full range of characteristics of intended population represented; all key patient data available; multi-center study or multiple studies. \\
		\midrule
		& \textbf{Model validation (V)} \\
		\midrule
		LOW & The model is checked for plausibility. \\
		MED & Results are compared to other models or literature data. \\
		HIGH & Results are compared to clinical data. \\
		\bottomrule
	\end{tabular}
	
	\label{tab:lit_grading}
\end{table*}
\begin{table}[h]
	\centering
	\small
	\caption{Credibility assessment detailed overview with rating each validations step based on gradations a (weak), b (moderate) and c (good).}
	\label{tab:assess_credibility}
	
	\rotatebox{90}{%
	\begin{tabular}{l>{\centering\arraybackslash}p{1.2cm}>{\centering\arraybackslash}p{1.2cm}>{\centering\arraybackslash}p{1.2cm}>{\centering\arraybackslash}p{1.2cm}>{\centering\arraybackslash}p{1.2cm}>{\centering\arraybackslash}p{1.2cm}>{\centering\arraybackslash}p{1.2cm}>{\centering\arraybackslash}p{1.2cm}>{\centering\arraybackslash}p{1.2cm}>{\centering\arraybackslash}p{1.2cm}>{\centering\arraybackslash}p{1.2cm}}
		\toprule
		
		\multicolumn{11}{l}{\textbf{Population-Based Validation (5)}}\\
		\midrule
        & Number of subjects & Character-istics & Patient-level data & Source & Quantity & Equiva-lency  & Output comp. & Rigor of comparison & Agree-ment & Relevance of QoI & Relevance to COU \\
		\cmidrule(lr){1-12}
		\textbf{V--1} & \gC & \gA & \gA & \gB & \gB & \gC & \gC & \gB & \gC & \gA & \gB \\
		\textbf{V--2} & \gC & \gA & \gA & \gB & \gA & \gC & \gC & \gB & \gC & \gA & \gB \\
		\textbf{V--3} & \gA & \gA & \gA & \gA & \gB & \gC & \gC & \gB & \gC & \gB & \gB \\
		\textbf{V--4} & \gB & \gB & \gB & \gA & \gB & \gB & \gC & \gB & \gB & \gB & \gB \\
		\textbf{V--5} & \gB & \gB & \gC & \gA & \gB & \gB & \gC & \gB & \gC & \gA & \gB \\
		\midrule
		\multicolumn{11}{l}{\textbf{Computational Model Calibration (2)}}\\
		\midrule
		& Quality of data & Quantity of data & Relevance to COU & Inputs vs. params & Goodness of fit \\
		\cmidrule(lr){1-12}
		\textbf{S--1} & \gC & \gB & \gC & \gB & \gB \\
		\textbf{S--2} & \gB & \gA  & \gC & \gA & \gB \\
		\midrule
        \multicolumn{11}{l}{\textbf{In vivo validation (4)}}\\
		\midrule
		& Number &  Range & Patient data & Measurements \\
		\cmidrule(lr){1-12}
		\textbf{S--2} & \gA & \gC  & \gB & \gA &  \\
		\midrule
		\multicolumn{11}{l}{\textbf{Computational Model Plausibility (7)}}\\
		\midrule
		& Rationale & Justification  & Input param. & Relevance to QoI &  Consis-tency & Expert endorsement & Limitations & & & \\
		\cmidrule(lr){1-12}
		\textbf{S--3} & \gC & \gC & \gD & \gD & \gD & \gB & \gC \\
		\midrule
		\multicolumn{11}{l}{\textbf{Emergent Model Behavior (6)}}\\
		\midrule
		& Identi-fication & Relevance to COU & Reprodu-cibility & Confir-mation & Consis-tency & Impact & & & & \\
		\cmidrule(lr){1-12}
		\textbf{S--4} & \gC & \gB & \gC & \gB  & \gD & \gB \\
		\bottomrule
	\end{tabular}
}
\end{table}

\end{document}